\documentclass[]{aa}

\usepackage{graphicx}
\usepackage{enumitem}
\usepackage{color}
\usepackage{txfonts}
\usepackage{bbm}
\newcommand{\kms}{{ km~s$^{-1}$}}

\newcommand{\hMpc}{{ \textit{h}$^{-1}$~Mpc}}

\begin{document} 

\title{Statistically bias-minimized peculiar velocity catalogs from Gibbs point processes and Bayesian inference}
 
   \author{Sorce Jenny G.\inst{1,2,3}\fnmsep\thanks{jenny.sorce@univ-lille.fr / jenny.sorce@universite-paris-saclay.fr}
          \and
          Stoica Radu S.\inst{4}
          \and
          Tempel Elmo\inst{5,6}
          }
          
   \institute{Univ. Lille, CNRS, Centrale Lille, UMR 9189 CRIStAL, F-59000 Lille, France
         \and
            Universit\'e Paris-Saclay, CNRS, Institut d'Astrophysique Spatiale, 91405, Orsay, France
         \and
             Leibniz-Institut f\"{u}r Astrophysik, An der Sternwarte 16, 14482 Potsdam, Germany
           \and
             Universit\'e de Lorraine, CNRS, IECL, Inria, F-54000 Nancy, France
             \and
             Tartu Observatory, University of Tartu, Observatooriumi 1, 61602 T\~oravere, Estonia
             \and
              Estonian Academy of Sciences, Kohtu 6, 10130 Tallinn, Estonia
             }
             
   \date{Received XX XX, 2023; accepted XX XX, XXXX}

 
  \abstract
   {Galaxy peculiar velocities are excellent cosmological probes provided that biases inherent to their measurements are contained before any study. This paper proposes a new algorithm based on an object point process model whose probability density is built to statistically reduce the effects of Malmquist biases and uncertainties due to lognormal errors in radial peculiar velocity catalogs. More precisely, a simulated annealing algorithm permits maximizing the probability density describing the point process model. The resulting configurations are bias-minimized catalogs. Tests are conducted on synthetic catalogs mimicking the second and third distance modulus catalogs of the Cosmicflows project from which peculiar velocity catalogs are derived. By reducing the local peculiar velocity variance in catalogs by an order of magnitude, the algorithm permits recovering the expected one while preserving the small-scale velocity correlation. It also permits retrieving the expected clustering. The algorithm is then applied to the observational  catalogs. The large-scale structure reconstructed with the Wiener-filter technique applied to the bias-minimized observational catalogs matches with great success the local cosmic web as depicted by redshift surveys of local galaxies. These new bias-minimized versions of peculiar velocity catalogs can be used as a starting point for several studies from {possibly estimating} the most probable Hubble constant, H$_0$, value to the production of simulations constrained to reproduce the local Universe.}
 

   \maketitle


\section{Introduction}

The peculiar velocities of galaxies result from the action of the entire underlying gravitational field. Additionally, they are linear and correlated on large scales. As such they are excellent cosmological probes to study the dark side of the Universe. However, peculiar velocity catalogs are also grandly affected by different sources of biases. Some are known, some are not. In any case accounting for their effects is not completely mastered. Disentangling the true underlying signal from noises in radial peculiar velocity catalogs became a major issue within the last decade with the advent of using them to derive cosmological parameters \citep[e.g.][]{2011ApJ...736...93N,2017MNRAS.468.1420F,2017MNRAS.464.2517H,2017MNRAS.470..445N,2018PhRvD..98f3503W}, to map the local distribution of matter \citep[e.g.][]{2014Natur.513...71T,2018NatAs...2..680H} and to constrain initial conditions that evolve into our local neighborhood, the local Universe \citep[e.g.][]{2010arXiv1005.2687G,2014MNRAS.437.3586S,2016MNRAS.455.2078S}.\\

\citet{2013AJ....146...86T} then \citet{2016AJ....152...50T} released two large peculiar velocity catalogs. The second improves on the first one with additional major contributions from the 6dF Galaxy Survey \citep[e.g.][]{2003ASPC..289...97W,2014MNRAS.443.1231C} and two Spitzer surveys: CosmicFlows with Spitzer \citep[CFS,][]{2014MNRAS.444..527S} and the Spitzer Survey of Stellar Structure in Galaxies \citep[S$^{4}$G,][]{2010PASP..122.1397S}. However, with the increasing distance coverage, the impact of biases affecting the catalogs has grown stronger. 
\citet{2015MNRAS.450.2644S} proposed a technique to minimize the biases in such peculiar velocity catalogs. It permitted erasing their effects such as a spurious strong infall onto the reconstructed local volume. Later it allowed simulating the local clusters such as our closest neighbor: the Virgo cluster of galaxies \citep{2016MNRAS.460.2015S,2019MNRAS.486.3951S,2021MNRAS.504.2998S}, but also other local clusters \citep[such as Centaurus, Coma and Perseus,][]{2018MNRAS.478.5199S,2023arXiv230101305S}. Even more recently, \citet{2019MNRAS.488.5438G,2022MNRAS.517.4529B} and 
\citet{2022MNRAS.513.5148V}, borrowing from \citet{2016MNRAS.457..172L}, proposed Bayesian techniques that sensibly reduced the infall onto the reconstructed local volume. Like any Bayesian techniques, they rely on theoretical expectations, in that case mostly coming from the $\Lambda$CDM cosmological paradigm. However, they also rely (heavily) on prior knowledge of the dataset. They invoke multiple functions making it difficult to disentangle what can really be deduced from the data from what is included as a prior to correct the data. On the other hand, although \citet{2015MNRAS.450.2644S} rely only on the expected radial peculiar velocity 1D distribution, they forget about the 3D {small-scale} correlations related to the 3D spatial distributions of galaxies.  \\

In this paper, we propose to emphasize on the 3D {small-scale} correlations using a probabilistic approach. More precisely, considering the finite size of the peculiar velocity catalogs of galaxies in a finite region (the local Universe), we base our approach on an object point process model built such that its probability density tends to be maximal when bias effects are minimal. The configurations (or realizations) maximizing the probability density are then bias-minimized radial peculiar velocity catalogs. From a broad point of view, it is a typical inverse problem regularly solved in various fields, including in astrophysics, for instance for image restoration \citep[e.g.][]{vanlieshout1994,2013EAS....59..265B}. 

In our case, functions used within the core of the algorithm should rely heavily neither on the cosmological model nor on the dataset configuration. It makes it easier to:
\begin{enumerate}
\item change for another cosmological model,
\item change for another dataset and
\item disentangle the underlying signal in the data from the signal that is induced by priors.
\end{enumerate}
More specifically, galaxy radial peculiar velocities are derived via a cosmological model from galaxy distance moduli and observational redshifts. The latter being far more precisely determined than the former, uncertainties on the latter are usually considered as negligible. Galaxy observational redshifts are thus assumed to be fixed. The algorithm should then output the most probable location of the galaxies given the uncertainty on their distance modulus measurements and the associated derived radial peculiar velocities. This kind of configurations should be likely to maximize the proposed probability density. {Note that one may advocate for using observational redshifts as proxies for distances. Radial peculiar velocities derived from such distances would all be zero. This is unrealistic. The resulting configuration should thus not maximize the proposed probability density}. \\

To include 3D {small-scale} correlations, for a set of galaxy distance moduli and uncertainties ($\mu, \sigma$), the algorithm should consider for each galaxy a pre-determined spatial region that characterize a zone of interactions. This zone should depend on the galaxy distance modulus and uncertainty and, by extension, on its radial peculiar velocity. The algorithm should rely on the underlying correlation, in the catalog of galaxy distance moduli, between the directly derived peculiar velocities of galaxies. Radial peculiar velocities of galaxies sharing a same local region in space are indeed linked to the local underlying gravitational field, namely the local structure. As mentioned above, given our spatial data set (a distribution of galaxies in a finite local Volume), we assume that the realization that maximizes the probability density of our point process model should permit retrieving the underlying correlation. The probability density $p$, with respect to the unit intensity Poisson reference measure, should depend on the data ($\mu, \sigma$) and the model parameters ($c$) \citep[see][and references therein for detailed explanations]{stoica2010}. {Given the intractability nature of the underlying probability density, in order to derive the configuration that maximizes it, we need instead to construct a function, $U(\mu|\textbf{d},c)$ where $\textbf{d}=(\mu_{init},\sigma_{init},z_{obs})$ and sample from it (cf. Metropolis-Hastings algorithm).} This function should reach its maximum for a realization (set of distance moduli $\mu_i$ and associated uncertainties $\sigma_i$) that minimizes the biases. Moreover, since the function is not \textit{a priori} convex (i.e. not a single maximum), the realization should be obtained with a global optimization technique through a simulated annealing to sample a probability law in the form $p(\mu|\textbf{d},c)^{1/T} \propto \mathrm{exp}(-U(\mu|\textbf{d},c)/T)$ with $T$ slowly going to zero.\\

Such techniques have been used in the past in astronomy to find, for instance, filaments and groups in redshift surveys \citep{2016A&C....16...17T,2018A&A...618A..81T} as well as to build maps of optimal tile distributions in order to observe efficiently multi-source catalogs \citep{2020MNRAS.497.4626T}. However, in these examples, celestial object (galaxy and star) distributions are considered fixed. {Considering these distributions, one of the realizations maximizing the probability gives one of the optimal arrangement of filaments, groups and tiles.} In our case, the galaxy distribution is not fixed but constitutes a realization by itself. One of the galaxy distributions that maximizes the probability is retrieved, in particular, thanks to the {local} underlying velocity correlations. Contrary to \citet{2015MNRAS.450.2644S}, bias-minimized radial peculiar velocities of galaxies are not obtained on a one-to-one basis (probability of the velocity to exist given the 1D velocity probability distribution) but collectively (probability of the velocity to exist given the 1D velocity probability distribution \emph{and} given the 3D velocity local variance probability). In both cases, the resulting bias-minimized peculiar velocity catalog should be considered as a whole. Namely, datapoints cannot be considered individually as better estimates. The realization, i.e., the full catalog, constitutes statistically a better representation of the true datapoint distribution. This concept is at the core of the algorithm we propose. \\

This paper starts with a description of the biases affecting the catalogs and the computation of radial peculiar velocities. The third section builds  the probability law as well as the algorithm and the associated inference processes. The algorithm is subsequently applied to mock catalogs mimicking observational ones. The building of the latter is also detailed. The results of the application of the algorithm to the synthetic catalogs are analyzed. For the sake of conciseness, plots are shown for one of the mock catalogs. To show one use-example of the bias-minimized catalogs, they are plugged into a Wiener-filter algorithm to recover the full 3D velocity and density fields. {This technique is chosen as a case study because of its sensitivity to biases.} The technique is then applied to the observational catalogs. Again, for conciseness, results are shown only for one of them: the third catalog of the Cosmicflows project, i.e. cosmicflows-3 \citep{2016AJ....152...50T}. {The Wiener-filter is also applied to the initial and post-treatment observational catalogs to validate further the bias-minimization algorithm.} {Before concluding, the influence of the H$_0$ parameter value on the results is quantified using both the synthetic and observational catalogs.}


 \section{Biases and Uncertainties}
 
 \subsection{Biases and effects}
Peculiar velocity measurements\footnote{derived from distances, themselves obtained from distance moduli.} and their gathering into catalogs are complicated by several biases described at length in \citet{2015MNRAS.450.2644S} and references therein. Whilst generally all gathered under the term `Malmquist bias', three types of Malmquist bias can in fact be distinguished. In addition to these biases, there is a lognormal error distribution which requires some attention. Here we present only a short description of these biases:
\begin{enumerate}[wide, labelwidth=!, labelindent=0pt, label={b\arabic*)}]
\item the Malmquist Bias, due to selection effects, is usually taken care of when calibrating distance indicators used to derive distances and then velocities  \citep[Kaptney, 1914; Malmquist, 1922;][]{1992ApJ...395...75H,1994ApJ...430....1S,1997ARA&A..35..101T,1993A&A...280..443T,1990A&A...234....1T,1994ApJ...435..515H,1994ApJS...92....1W,1995ApL&C..31..263T}. This is the case in the catalogs we use \citep[e.g.][]{2012ApJ...749...78T,2013ApJ...765...94S,2014MNRAS.444..527S}. \\
\item the Homogeneous Malmquist Bias, due to a higher probability of scattering galaxies further away closer than the opposite (increasing surface of shells centered on us with the distance), needs to be dealt with  \citep[Kaptney, 1914; Malmquist, 1920;][]{1988ApJ...326...19L,1992ApJ...395...75H,1997ARA&A..35..101T,1994ApJ...430....1S,1993A&A...280..443T,1990A&A...234....1T,1994ApJ...435..515H,1995ApL&C..31..263T,1995PhR...261..271S}. {When considering a complete up-to-a-given-distance galaxy sample}, on average, post-bias-minimized galaxies should end up with larger distance estimates than originally measured. Although it must not be an explicit requirement of the algorithm (i.e. no function should directly enforce distances to be larger), it must be checked that it ends up being the case when comparing the initial and bias-minimized catalogs {that are mainly constituted of complete galaxy surveys}\footnote{{Notably, $\sim$50\% of the third cosmicflows dataset is constituted of the 6-degree Field Galaxy Survey peculiar velocity sample \citep{2014MNRAS.445.2677S}, a complete up-to-a-given-distance survey.}}.\\
\item the Inhomogeneous Malmquist Bias, due to a higher probability of scattering galaxies from high density regions to low density regions, is not taken into account either  \citep[e.g.][]{1994ARA&A..32..371D,1994MNRAS.266..468H}. Similarly, post-bias-minimized galaxies should be more numerous in high density regions than initially. We will thus check that, although it is not a direct requirement, the algorithm tends to cluster galaxies. Note that we group galaxies gathered into one group/cluster to derive a unique distance/velocity estimate, that of the group/cluster. This permits removing non-linear virial motions from the catalog \citep{2017MNRAS.469.2859S,2018MNRAS.476.4362S}, a source of biases for the Wiener-filter technique. Clustering will thus be smoothed out on very small scales. \\
\item The logarithmic relation between distance moduli and distances, hence velocities, introduces a non-Gaussian distribution of errors on velocities, called the lognormal error distribution. Typically, an overestimated distance modulus results in a higher error on the velocity estimate than would an underestimated distance modulus \citep[e.g.][]{2016AJ....152...50T}. Distance over- and underestimates must then not be considered similarly. There exist analytical and ad-hoc solutions to take care of this bias \citep[e.g.][]{1992ApJ...391..494L,2021MNRAS.505.3380H}. Note though that unlike for distances, errors on distance moduli can be considered symmetric. We will thus ensure that distance moduli rather than distances are the starting point of the algorithm. There will be no need to deal with this bias at the distance modulus level. 
\end{enumerate}
 
 \subsection{The main source of uncertainties} 
 
To obtain the 3D galaxy distribution from observations (= initial realization) and to control any additional source of systematics when deriving peculiar velocities {(cf. bias b4 above)}, we start directly from the catalog of galaxy distance moduli ($\mu$) and observational redshifts \citep[z$_{obs}$,][]{2014MNRAS.442.1117D} to which we add Supergalactic longitude and latitude coordinates. This allows us to derive galaxy cartesian Supergalactic coordinates.  A cosmological model is then required to determine peculiar velocities. While we use $\Lambda$CDM, later on another model can easily substitute it. \\

We remind in the following the different relations between distance moduli, observational and cosmological redshifts, luminosity distances and radial peculiar velocities. From observations of all the galaxies in the catalog, we have, thanks to distance indicators and the Doppler effect, {two independent measurements}:\\
\noindent\textbf{- Distance modulus measurements, $\mu$},\\
\noindent\textbf{- Observational redshifts, z$_{obs}$}.\\
We want:\\
\noindent\textbf{- Luminosity distances, $d_{lum}$}. They are obtained via distance moduli:
 \begin{equation}
 \mu=5\mathrm{log_{10}}(d_{lum}~\mathrm{(Mpc)})+25
 \end{equation}
 
 \noindent\textbf{- Cosmological redshifts, z$_{cos}$}. They are derived with luminosity distances using the equation:  
   \begin{equation}
   d_{lum}=(1+z_{cos})\int_0^{z_{cos}}\frac{c_ldz}{H_0\sqrt{(1+z)^3\Omega_m+\Omega_\Lambda}}
   \end{equation}
   where $H_0$ is the Hubble Constant, $c_l$ is the speed of light, $\Omega_m$ and $\Omega_\Lambda$ are the cosmological parameters corresponding to the matter and the dark energy respectively.
   \vspace{0.15cm}
   
\noindent\textbf{- Radial peculiar velocity estimates, $v_{pec}$}. They are finally obtained, using the observational  z$_{obs}$ and cosmological z$_{cos}$ redshifts  with the following formula:
   \begin{equation}
   v_{pec} = c_l \frac{z_{obs}-z_{cos}}{1+z_{cos}}
   \end{equation}
 where $v_{pec}$ will always refer to the \emph{radial} peculiar velocity in this paper.\\
 
 Among all the parameters used to derive the galaxy peculiar velocities, the largest source of uncertainties comes unquestionably from their distance estimates. This reinforces our original idea to minimize biases in peculiar velocity catalogs through the minimization of biases in distance (modulus) catalogs. We reiterate that starting from distance moduli permits avoiding the lognormal error distribution, bias b4 above, in the initial setting. Considering their precision with respect to that of distance moduli, Supergalactic latitude and longitude coordinates as well as the observational redshift are considered as error free in a first approximation. Our goal is thus designed to provide a new distance modulus estimate for each galaxy of the catalog. {These new distance moduli will give new distances (cf. eq. 1) thus radial peculiar velocities (cf. eq. 2 and 3).}  \\

   \section{Model construction: a new Gibbs field model for minimizing biases}
 \begin{enumerate}[wide, labelwidth=!, labelindent=0pt, label={\textbf{{3.\arabic*}}}]
 \item The main goal of the algorithm is to find the position of galaxies that results, given their distance modulus (by extension peculiar velocity) and uncertainty on the latter, in the highest probability density of a point process.
 \end{enumerate}
 
 \noindent To reach that goal, the principle is as follows, for each galaxy:
 \begin{enumerate}[wide, labelwidth=!, labelindent=0pt, label={\textbf{{3.\arabic*}}}]
 \setcounter{enumi}{1}
  \item The distance modulus is slightly perturbed from its initial estimate inducing a new distance estimate, thus a new peculiar velocity (cf. eq. 1 to 3). {The distance modulus is modified proportionally to  its uncertainty to be more conservative towards distance modulus measurements with higher confidence level than others} (e.g. obtained with supernovae against Tully-Fisher relation). \vspace{0.15cm}
  \item The resulting new distance modulus and associated peculiar velocity of a galaxy are compared to the initial/previous distance modulus and velocity to ensure their likelihood given the uncertainty.  \vspace{0.15cm}
  \item The resulting new peculiar velocity of a galaxy has to be compared to the peculiar velocities of surrounding galaxies. To that end,  {a 3D local-region shape is required to identify which velocities should be compared. To avoid rounded structures in density fields reconstructed with catalogs,} clear signs of bias residuals as observed with other proposed bias minimizations, the shape should be extended along the line-of-sight with a short coverage tangentially.  This prevents fortuitous transversal unrealistic correlations.  This volumetric shape is essential to ensure that both the homogeneous (further to closer positions as per bias b2) and inhomogeneous (higher to lower densities as per bias b3) Malmquist effects are probed during the algorithm run. Hence, although there is an explicit requirement of the algorithm to increase neither distances on average nor clustering, it can have an impact on these two aspects. Whether this results in the desired effect constitutes a proof of concept that the algorithm reaches the bias-minimization goal.   \vspace{0.15cm}
  \item Additionally, to be able to use the newly derived distance modulus  (hence, distance and velocity as per eq. 1 to 3) for various studies, a new uncertainty must be assigned. 
  \end{enumerate}

In the following, the first subsection sets the basis of the point process model, i.e. its density probability and the technique to find a realization that corresponds to a maximum of the latter. The next subsections define the different terms required to propose new configurations. They also build the terms of the density probability function.

    \subsection{Maximized probability density and bias-minimized catalog}

The algorithm should give the most probable location of a galaxy within a pre-determined spatial region given the uncertainty on its distance modulus measurement and, by extension, the associated radial peculiar velocity with respect to the entire catalog. Because Supergalactic longitude and latitude coordinates are considered error free in a first approximation, the spatial region is shaped along the line-of-sight to retrieve the distances. In future developments, they could be relaxed alongside observational redshift measurements and the Hubble constant, H$_0$ or more generally the cosmology. In other words, considering a finite volume (the local Universe), the algorithm, input with a given set of $n$ ($\mu_{init,\, i}$,$\sigma_{init,\, i}$) with $n$ fixed (= initial configuration),  will result in a new set of $n$ ($\tilde\mu_i$,$\tilde\sigma_i$) (one realization maximizing the probability density). Since our number of galaxies is fixed, we construct the probability density $p$ for a Gibbs point process \citep{chiu2013} with the objective that this probability density is maximal when the effects of the biases are minimal. The realization corresponding to the bias-minimized catalog is thus obtained by maximizing the probability density of the point process where:
   \begin{equation}
 p(\mu|\textbf{d},c) \propto \mathrm{exp}[-U(\mu|\textbf{d},c)] 
    \end{equation}
  with $\textbf{d}=(\mu_{init},\sigma_{init},z_{obs})$, $c$=\{$c_i$\}$_{i \in \mathbbm{N}}$ a set of positive parameters, $\mu$=($\mu_1$,$\mu_2$,...,$\mu_n$) the set of distance moduli, n the number of galaxies (datapoints) and $U$ the energy function.  {Again, given the intractability nature of $p$, building $U$ permits us to sample and thus to propose new sets.}\\
  
 {In addition to the standard likelihood included via a data energy term, $U_D$, the energy function must take into account the local peculiar velocity correlation via an interaction energy term, $U_I$.} The energy function, $U$ can thus be written:
   \begin{equation}
   U(\mu|\textbf{d},c)=U_D(\mu|\textbf{d},c)+U_I(\mu|\textbf{d},c)
   \end{equation}

More specifically, U$_D$ is required to control the galaxy position (distance modulus) and associated uncertainty and, by extension, velocity. It depends on the distance modulus, its uncertainty and the derived velocity. Namely, for a given galaxy, it depends only on its associated properties. U$_I$ allows controlling the probability that the object is at this position given its peculiar velocity (i.e. would it be at this distance modulus) in conjunction with its neighboring galaxies. It thus builds also upon the neighboring galaxy properties of a given galaxy. In that respect, the algorithm aims at resulting in a statistically bias-minimized catalog but there is no information on individual galaxies \textit{per se}.\\

Subsequently, the minimization of both energy terms for a given realization of the dataset corresponds to a maximum of the probability density, i.e. to a bias-minimized catalog: a new set of distance moduli and their uncertainties, $(\tilde\mu,\tilde\sigma)$. In other words, we need to minimize with a given set of parameters:
  \begin{equation}
  (\tilde\mu)=\mathrm{arg\ min}\{U_D(\mu|\textbf{d},c)+U_I(\mu|\textbf{d},c) - \mathrm{log}p(c)\}
  \end{equation}
 {This can be solved sampling the Gibbs point process \citep{chiu2013} within a simulated annealing algorithm.}

 \subsection{New distance modulus, $\tilde\mu_i$}
New distance moduli, $\tilde\mu_i$ are drawn as follows:
    \begin{equation}
   \tilde\mu_i = \mu_i + U_n[-0.5,+0.5] \times \gamma~  \sigma_i
     \end{equation}
 where $U_n[-0.5,+0.5] $ defines a random number from a uniform distribution between -0.5 and 0.5, $\gamma$ is set between 1 and 3 (see Table \ref{tbl:1}). Choosing a uniform distribution, centered on the previous distance modulus, with a variance proportional to the uncertainty, permits being computationally faster than when using a fixed step (slower convergence).  {Note that it does not affect the results.}\\

 \subsection{Data energy term, $U_D$}
 The data energy term controls that new drawn distance moduli and newly derived velocities are probable given the initial and previous distance modulus and velocity values. It can be decomposed into a first term controlling the newly assigned distance moduli, $U_{D1}$, and a second term controlling the associated newly derived velocities, $U_{D2}$:
    \begin{equation}
  U_D(\mu|\textbf{d},c) = U_{D1}(\mu|\textbf{d},c) + U_{D2}(\mu|\textbf{d},c)
     \end{equation}

  \subsubsection{Data energy term 1, $U_{D1}$}

\noindent\textbf{- The data energy term 1 for each datapoint, $e_1$}, is associated to the likelihood. It must ensure that any new drawn distance modulus, $\mu_i$,  for a galaxy is contained within a restricted range of values imposed by its uncertainty, $\sigma_{init,\, i}$. In particular, a term preserving a relative memory with respect to the initial distance modulus and its related uncertainty value is essential to prevent an infall onto the observer (at the center of the catalog by definition). Since the probability distribution of a galaxy distance modulus follows a Gaussian of variance $\sigma_{init,\, i}$, centered on  $\mu_{init,\, i}$, the term $e_1$ can be written:
    \begin{equation}
       e_1(\mu_i)=c_1 \times \frac{(\mu_i-\mu_{init,\, i})^2}{2\sigma_{init,\, i}^2}
     \end{equation} 
         where c$_1$ is a constant (see Table \ref{tbl:1} for its value), $\sigma_i$s are the uncertainties on the $\mu_i$s and the subscript ${init}$ refers to the initial values in the catalog. The derivation of new uncertainty estimates, $\sigma_i$s is detailed in section 3.5. While theoretically $\sigma_i$s can be zeros, in practice it is never the case thus, here and hereafter, we do not specify a special treatment.  {Note again that since distance moduli constitute the starting point of the algorithm, we avoid bias b4 mentioned hereabove. Namely, it is possible to use a Gaussian for the probability distribution of a galaxy distance modulus without approximation.}   \\

\noindent\textbf{- The total data energy term 1, $U_{D1}$}, is then:
    \begin{equation}
    U_{D1}(\mu|\textbf{d},c_1)=\sum_{i=1}^{n}e_1(\mu_i)
    \end{equation} 
    where n is the total number of galaxies. Note that in practice at a given time, this term is different from that of the previous time step only for the point perturbed from its previous step position. \\
    
      \subsubsection{Data energy term 2, $U_{D2}$}

A second data energy term is essential to encourage the decrease of high velocities in absolute value when initializing the Metropolis-Hastings random sampling. Indeed, initially the interaction term (cf. next subsection) can be null because galaxies are either isolated or  {clustered but biased in the same way, i.e. with matching velocities and associated uncertainties (cf. biases b2 and b3)}. It needs also to prevent points at the edge of the sample to simply flee away where they would have no interaction (cf. next subsection).

\noindent\textbf{- The data energy term 2 for each datapoint, $e_2$}, can thus be written:
           \begin{equation}
       e_2(\mu_i)=c_{2} \times |v_{pec,\, i}(\mu_i,z_{obs})|/v_{ref}
     \end{equation} 
 with $c_2$ a constant (see Table \ref{tbl:1} for its value) and $v_{ref}$ a reference velocity set to 10,000~km~s$^{-1}$.  {Note that $v_{ref}$ permits ensuring that all the constants have no physical unit. In addition, $v_{ref}$ value is chosen so that when all the constants are of the same order of magnitude, all the terms in $U$ are also of the same order of magnitude. All the terms weight similarly in $U$.}\\

\noindent\textbf{- The total data energy term 2, $U_{D2}$}, is then:
       \begin{equation}
    U_{D2}(\mu|\textbf{d},c_2)=\sum_{i=1}^{n}e_2(\mu_i)
    \end{equation} 
        
 Currently, the energy terms are simple. Later H$_0$, etc could vary to include their uncertainties. We might also consider a different cosmology. In the current paper, we consider though that H$_0$, z$_{obs}$, etc are constants. The line-of-sight position (distance modulus / radial peculiar velocity) and its uncertainty are the only measurements allowed to vary.  {Namely, we assume that distance moduli (and, by extension, peculiar velocities) are the only measurements with an error. Therefore, a likelihood term needs to be written only for them}. {Still Appendix A presents results using another set of cosmological parameters and section 6 quantifies the significance of H$_0$  value (and associated cosmological parameters) change on the results}.\\
  
  As an aside note, we define the decrease of velocity in absolute value parameter.   
 
 \noindent\textbf{- The decrease of velocity in absolute value for the ith galaxy:}
     \begin{equation}
 g_i(v_{pec,\, i},\tilde v_{pec\, i})=|\tilde v_{pec\, i}|-|v_{pec\, i}|
          \end{equation}            
    with $v_{pec\, i}$ the initial/previous peculiar velocity, $\tilde v_{pec\, i}$ the new peculiar velocity. $g_i$ will then appear when comparing the probabilities between ancient and new datapoints in the Markov chain (see below), more specifically in $\Delta U_{D2}$. Note that the symmetric uncertainty on the distance modulus propagates to an asymmetric uncertainty on the peculiar velocity. Namely, the new peculiar velocity, obtained with the distance modulus, is not strictly the mean of the peculiar velocities obtained with distance modulus plus and minus the uncertainty, $\tilde v_{pec\, i}^+$ and $\tilde v_{pec\, i}^-$. Tests conducted using $\tilde v_{pec\, i}$ or $<\tilde v_{pec\, i}^+,\tilde v_{pec\, i}^->$ reveal that results are unchanged given the precision reached when sampling. Indeed, this term mostly acts as a regulator not to have very large velocities. The sign of $g_i$ is statistically unchanged when using $<\tilde v_{pec\, i}^+,\tilde v_{pec\, i}^->$ rather than $\tilde v_{pec\, i}$. In the future, increasing precision of the algorithm might require this distinction.  \\%
      
     \subsection{Interaction energy term, $U_I$, step by step}
    
    The sole data energy term does not give much information on where it is best to locate the datapoints (i.e. their true location). The role of the interaction energy term is to favor configurations with plausible peculiar velocities in a statistical sense. It is essential to enforce the small-scale correlations. It should result in dealing with the effects of biases b2 and b3. It will be checked as a proof-of-concept of the algorithm. The interaction energy term thus compares peculiar velocities of interacting galaxies. It must permit distinguishing between `positive' and `negative' interactions. `Positive' means that one or both peculiar velocities are unlikely given the proximity of the two galaxies, i.e. one or both galaxies are likely not to be at the proper distance, `negative' means the opposite\footnote{Note that while the naming convention might seem counterintuitive, it makes sense when refereeing to the ultimate goal that is minimizing the energy function, $U$ i.e. lowering the interaction energy term, $U_I$ or looking for maximizing the number of `negative' interactions.}. {Consequently, in order to define the interaction energy term, we first need to introduce an interaction shape (subsection 3.4.1) that permits determining interacting galaxies and a function accounting for the small-scale velocity correlation (subsection 3.4.2) that permits attributing a state to the interactions.}
    
     \subsubsection{Interaction shape, $S$}
     
\begin{figure}
\includegraphics[width=0.5 \textwidth]{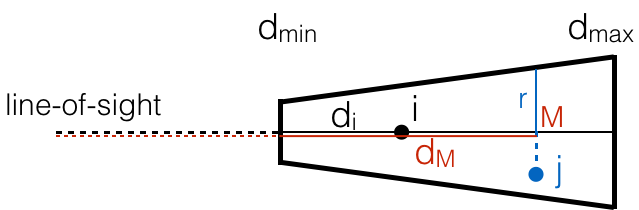}
\includegraphics[width=0.5 \textwidth]{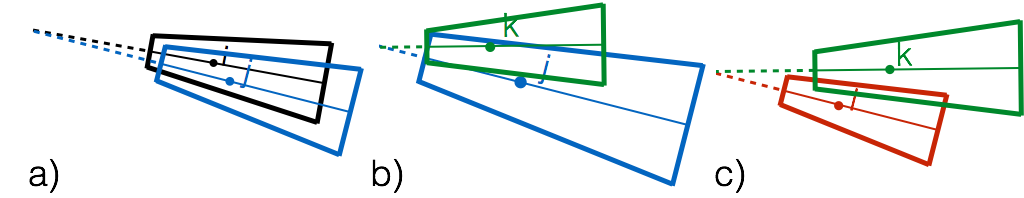}
\caption{Schematic view of the interaction shape. \textit{Top}: Interaction shape in 2D showing how i's shape is derived. j belongs to i's shape. \textit{Bottom}: Examples of interactions - a) i and j are both within one another shapes - b) k is in j's shape but j is not in k's - c) l and k are not in one another shapes.}
\label{fig:shape}
\end{figure}

The shape, $S$, permits determining whether two galaxies must be considered as interacting. The shape length and size depend on the galaxy distance and associated uncertainty. However,  its direction is always aligned with the line-of-sight. It is a 3D region shown in 2D in Figure \ref{fig:shape} top with the following properties:
\begin{itemize} 
\item d$_{min/max}$ are the distances obtained with $\mu_i\pm n_\sigma\sigma_{init,\, i}$  (see Table \ref{tbl:1} for $n_\sigma$ value) {to prevent introducing the lognormal error effect (bias b4) at this stage}. Note that to ensure a minimum size for the shape,  the initial uncertainty is used. The final uncertainty within a single run of the algorithm (see hereafter for a detailed explanation) indeed tends toward zero, thus decreasing the size of the shape and preventing any interaction.
\item M is the projection of a given point $j$ on the line-of-sight of the perturbed point $i$.
\item $r$ is equivalent to the aperture of the shape and is defined by $r~=~\alpha_{pc}~\times~d_M$ (see Table \ref{tbl:1} for $\alpha_{pc}$ value) with $d_M$ the distance of point M.
\end{itemize}
 Any point $j$ within the region of another point $i$ is then considered as interacting and their velocity are compared. The only difficulty is to check that the origin (us) is not in the shape. If this is the case, a point $j$ on the opposite side of the origin with respect to the point $i$ should not be considered as an interacting point. Consequently if the dot product between the direction of the two points is -1, there is no interaction by definition.  Figure \ref{fig:shape} bottom shows three types of situations: a) i and j are in one another shapes, b) k is in j's shape but j is not in k's, c) l and k are not in each others' shapes. \\

  \subsubsection{The {small-scale} velocity variance function, $\sigma_v$}

\begin{figure}
 \vspace{-1cm}
 \includegraphics[width=0.5 \textwidth]{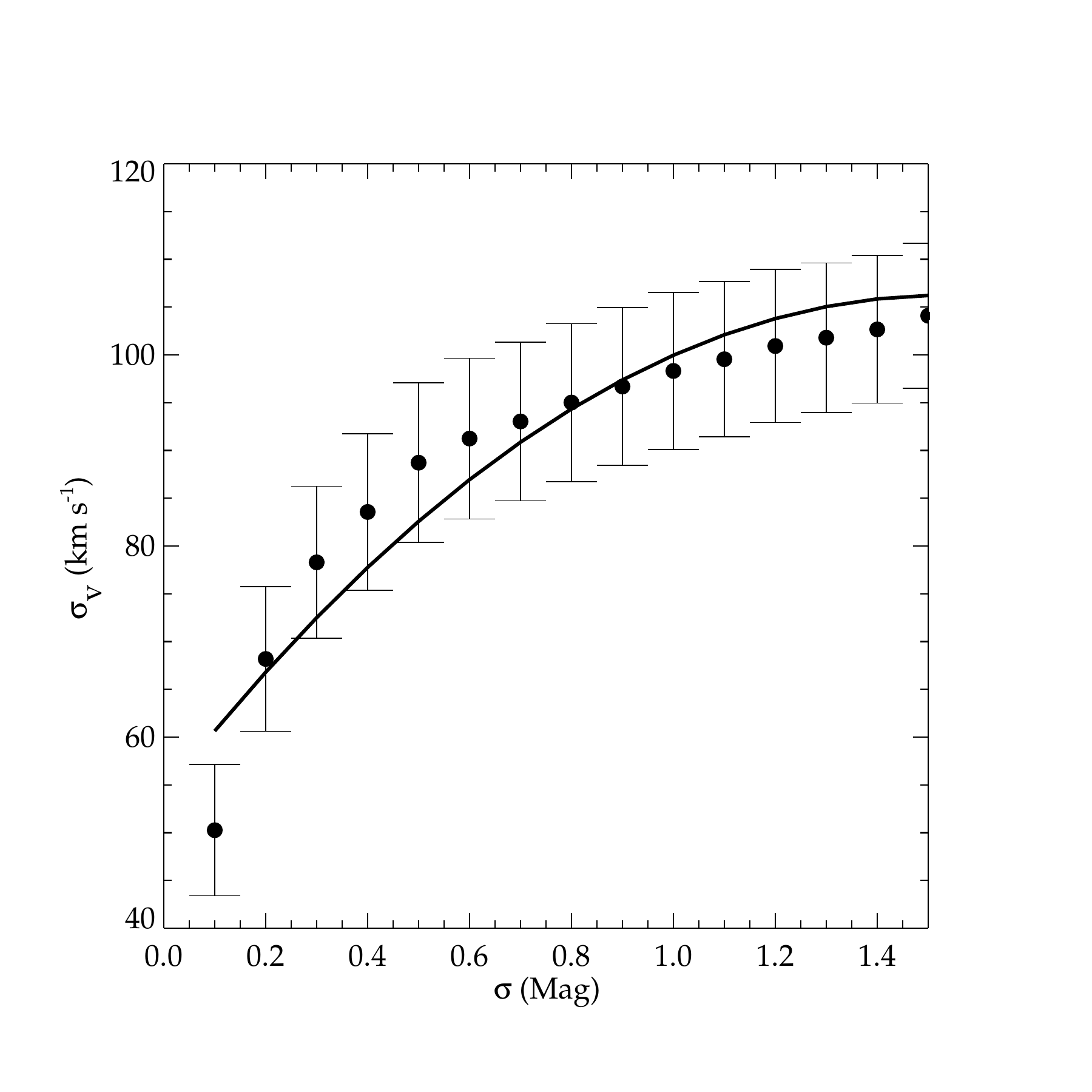}
\vspace{-1cm}
\caption{Correlation between the velocity variance and the uncertainty. More precisely, variance between velocities of objects in the same shape $S$ whose size is defined by the uncertainty. Filled circles are obtained throwing shapes, as defined in Figure \ref{fig:shape}, onto mock catalogs without errors. The solid line is the polynomial fitting a$+$b$x+$c$x^2$ with a=64, b=70, c=22~km~s$^{-1}$  assuming uncertainty magnitudes (Mag) in dex.}
\label{fig:varvelocityfit}
\end{figure}

  \noindent\textbf{- The {small-scale} velocity variance}, $\sigma_v$, and its fit are shown on Figure \ref{fig:varvelocityfit} with filled circles and a solid line respectively. {The filled circle values and their standard deviations are obtained by throwing randomly shapes, $S$, of different hypothetical sizes  onto a mock catalog of radial peculiar velocities \textit{without errors}. Each size corresponds to a given uncertainty, $\sigma$. The velocity variance $\sigma_v$ of objects within each shape is then derived. Note that changing the value of the parameters defining the shape (e.g. $\alpha_{pc}$) entails re-deriving the corresponding correlation of velocities at {small-scale}. In practice, small variations of the parameter values do not drastically modify the relation.} Note that modifying the cosmological model implies also re-deriving the small-scale velocity  variance using a mock catalog without errors from the corresponding model for consistency.\\
 
  \noindent\textbf{- The {small-scale} velocity variance function}, $\sigma_v$, is a polynomial fitting of the {small-scale} velocity variance:    
  \begin{equation}
  \sigma_v = \mathrm{a} + \mathrm{b} \times \sigma + \mathrm{c} \times \sigma^2 
  \end{equation}
   where a=64, b=70 and c=22~km~s$^{-1}$ assuming uncertainty magnitudes (Mag) in dex.  {It determines the average maximum authorized difference between velocities of galaxies belonging to the same shape as a function of the size of the shape. This shape is itself related to uncertainties on distance moduli. Note that the {small-scale} velocity correlations, obtained throwing shapes onto different synthetic catalogs (either mimicking the second or third Cosmicflows catalog distributions but without error, see description hereafter), are within their 3$\sigma$ uncertainty range.} Using fitted parameters within their 3$\sigma$ uncertainty range does not impact drastically the final output new distance modulus catalog. However, it certainly depends on the cosmological model. In the future, they will need to be relaxed together with the latter.\\
           
  \subsubsection{Interaction functions, $h$, $f$ and $q$}

The shape and the velocity correlation permits defining the interaction functions those values depend on the galaxy positions (shape membership) and velocity values (difference). Additionally, an interaction with a galaxy $j$ that has a small distance uncertainty must have a higher weight than that with a galaxy with a large distance uncertainty. The inverse of the uncertainty is thus used as a weight.

\noindent\textbf{- The interaction} between two points (galaxies) $i$ and $j$ exists,  i$\ \sim_s$ j, if:
  \begin{itemize} 
  \item $i \in S_ j$ and $j \notin S_ i$, or
  \item $j \in S_i$ and $i \notin S_ j$, or
  \item $j \in S_i$ and $i \in S_ j$\\
  \end{itemize}
  
     \noindent\textbf{- The weighted `positive' interaction function}, $h$ between the point $i$ and the other points $j$, with which it interacts, is then defined as\footnote{Note that given the precision of the small-scale velocity variance function, here again using peculiar velocities derived from distance moduli rather than the mean of the peculiar velocities derived from distance moduli plus and minus their uncertainties does not impact the output ($\mu$,$\sigma$) set.}:                 
       \begin{equation}      
        h_i=           \sum_{j=1,j\ne i}^{n}  \mathbbm{1}\{i~\sim_s\ j\} \times  \mathbbm{1}\{||v_{pec, i}|-|v_{pec, j}||>\sigma_v\}  ~ \times~ 1/\sigma_j^2
        \end{equation}  
        
           \noindent\textbf{- The weighted total -- `positive or negative' -- interaction function}, $f$ between the point $i$ and the other points $j$, with which it interacts, is then defined as:            
       \begin{equation}
       f_i=    \sum_{j=1,j\ne i}^{n}  \mathbbm{1}\{i~\sim_s\ j\}  ~ \times~ 1/\sigma_j^2 
      \end{equation}

           \noindent\textbf{- The absence of interaction function}, i.e. if i$\ \nsim_s$ j, $q$ between the point $i$ and the other points $j$, is then defined as:  
      \begin{equation}
      q_i= \mathbbm{1}\{\forall j, j\ne i, \;\; i~\nsim_s\ j\}
      \end{equation}
       
$\mathbbm{1}$ is an indicator function equal to 1 if the condition within is met and 0 otherwise. Namely, in $f$ case, $\mathbbm{1}\{i~\sim_s\ j\} $ equals 1 if the interaction between i and j exists (cf. conditions above). In $h$ case, $\mathbbm{1}\{||v_{pec, i}|-|v_{pec, j}||>\sigma_v\}$ equals 1 if i and j velocities differ by more than $\sigma_v(\sigma_i)$. Note that the $q_i=1$ choice when there is no interaction induces a constant penalization of isolated points. While values between 0.5 and 1 do not drastically change the results, we stick to $1$ to prevent potential fleeing away points. We avoid values between 0 and 0.5 to prevent ending up with an almost repulsive configuration (especially with the 0 value).\\
   
        \subsubsection{Interaction energy term, $U_I$}
   
Finally, the energy term ensures that, for objects that are close-by transversally and with position uncertainties on the line-of-sight allowing them to be within the range of distances of one another  (given d$_{min}$ and d$_{max}$), their peculiar velocities in absolute value are of the same order.\\

\noindent\textbf{- The pairwise interaction process, $e_3$}, in the interaction energy term for each point is then defined as follows:
    \begin{equation}
       e_3(\mu_i)=c_{3} ~ (h_i / f_i + q_i)
   \end{equation} 
    where $c_3$ is a constant (see Table \ref{tbl:1} for its value) and with the convention $h_i / f_i =0$ when $f_i=0$, i.e. there is no interaction.\\

\noindent\textbf{- The total interaction energy term} is:
 \begin{equation}
 U_I(\mu|\textbf{d},c_3)= \sum_{i=1}^{n} e_3(\mu_i)
 \end{equation}
 Note that this term has to be computed for every point because their interaction term can be affected by the perturbed point.  {In practice, we consider only the interaction between the perturbed point and each other point because of the required reciprocity of the interaction (see the interaction requirements with the `or' condition)}.

 \subsection{New uncertainty on distance moduli, $\tilde\sigma_i$}
 
New distance modulus uncertainties must be assigned to the datapoints. Typically, new uncertainties  should depend on the probability of the new datapoint position thus peculiar velocity with respect to the entire catalog. Since the catalog is statistically bias-minimized but not individual datapoints, new uncertainty and distance modulus cannot be used individually but within the context of the entire catalog.\\
     
\noindent We thus first define $p_v$ as the cumulative distribution function of the velocity value probability given the theory. Indeed, \citet{2001MNRAS.322..901S} proved that the distribution of radial peculiar velocities considering groups and clusters (virial motions removed) is a Gaussian. Unless the Milky Way is at a peculiar position in the Universe, the distribution of radial peculiar velocities obtained from our position should be close to a Gaussian too. This was verified by \citet{2015MNRAS.450.2644S} with constrained simulations of the local Universe.\\

\noindent\textbf{- The (cumulative distribution function of the) probability of a peculiar velocity (in absolute value)}, $p_v$, is then defined as follows: 
    \begin{equation}
    p_v=\frac{1}{\sigma_{v'}\sqrt{2\pi}}\int_{-\infty}^{-|v_{pec}|}\mathrm{exp}(-\frac{v^2}{2\sigma_{v'}^2})\mathrm{d}v
    \end{equation} 
with $\sigma_{v'}$ derived from mock catalogs mimicking the distribution of the observational catalog to be bias-minimized.  Its value is given in Table~\ref{tbl:1}.  For the sake of simplicity, in the following we will refer to $p_v$ as the probability of a given velocity value. Note that here again because of the precision with which the current algorithm computes this integral, using $v_{pec}$ instead of $<v_{pec}^+,v_{pec}^->$ does not impact the final result. In future developments, if this precision were to increase, this lognormal distribution effect on the peculiar velocity value might need to be taken into account.\\

\noindent Then, 
 \begin{enumerate}[wide, labelwidth=!, labelindent=0pt, label={\roman*)}]
\item The higher the probability is, the more the uncertainty should decrease. Thus the term $(1-p_v)$ must appear to derive the new uncertainty from the previous step uncertainty.\\
\item However, since the maximum probability $p_v$ is at a zero velocity value, the interaction term $h/f$ (see previous subsection) is also required. The text in the previous subsubsection explains this term in more detail. Briefly, the smaller $h/f$ is, the less `positive' (in the sense unrealistic) interactions with its neighbors, with respect to all its interactions, the datapoint has. Thus, the more probable the velocity is, meaning the smaller the uncertainty can be. \\
\item {Still, with the increasing probability $p_v$ of the velocity, the interaction term $h/f$ should have an increasing weight with respect to the probability term ($1-p_v$) (from (i)) and reversely.} This prevents the gathering of wrongly high absolute velocity values that have a low probability $p_v$ but few 'positive' interactions (small $h/f$)  because they form together an isolated ensemble of high velocity values. The weight on the probability (i) and interaction (ii) terms is thus simply the probability $p_v$ for the interaction term (ii) and, by extension, ($1-p_v$) for the probability term (i). {Since the individual probability of the velocity appears in the uncertainty term, it reduces sensibly the small-scale correlation of the errors on distance moduli (by extension peculiar velocities) inherent to the interaction shape. In any case, this small-scale correlation of errors has no impact on large-scale studies because of the small sizes of the shapes.}
   \end{enumerate}
   
 \noindent\textbf{- The new uncertainty, $\tilde\sigma_i$}, is thus:
    \begin{equation}
  \tilde\sigma_i= \sigma_i [  (1-p_v)^2 + p_v~(h_i/f_i+q_i) ) ] 
     \end{equation} 
     where $h_i$, $f_i$ and $q_i$ are the interaction and absence of interaction functions defined in the previous subsubsection.\\
   
       \subsection{Simulation method}
       
\begin{figure}
 \includegraphics[width=0.49 \textwidth]{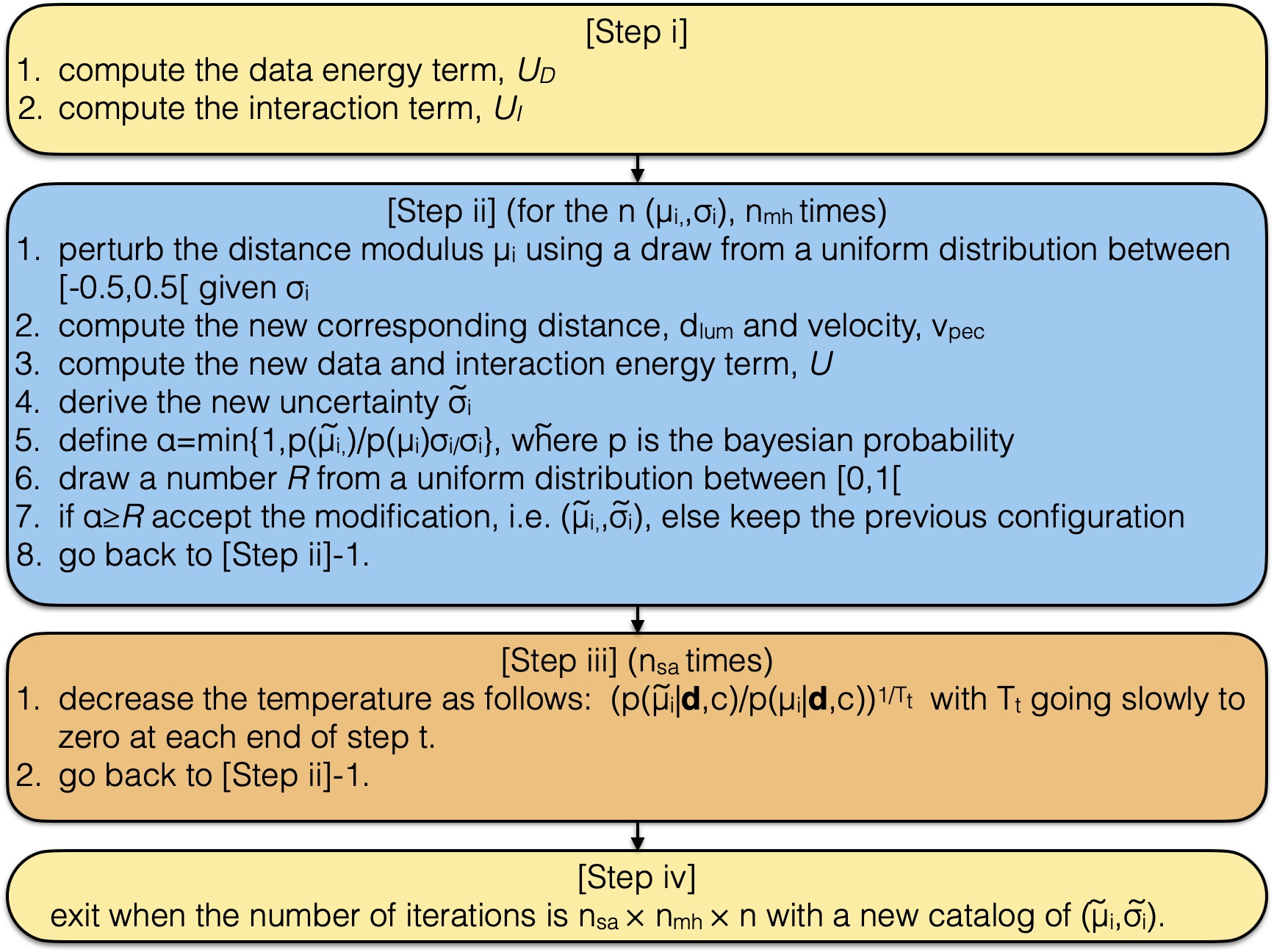}
\caption{Details of the steps to obtain the new set of ($\mu_i,\sigma_i$), namely ($\tilde\mu_i,\tilde\sigma_i$). Note that [Step ii] is done n$_{mh}$ times for the n points. [Step iii] implies fictively decreasing the temperature of the new points after and before going back to [Step ii]. There are  n$_{sa}$ timesteps, i.e. temperature decreases.}
\label{fig:schemabloc}
\end{figure}

  We use a Metropolis-Hastings sampling embedded into a simulated annealing algorithm as detailed by Figure \ref{fig:schemabloc} with blue and orange colors respectively. The first yellow panel gives the initialization with the initial realization and the last yellow panel gives the resulting realization (one maximum of the density probability).  More precisely, the steps are as follows:
  
  \begin{enumerate}[wide, labelwidth=!, labelindent=0pt, label={[Step \roman*]}]
\item Compute the data and interaction energy terms. \vspace{0.2cm}
\item Perturb the distance moduli, $\mu_i$s using a draw from a uniform distribution between [-0.5,0.5[ given the  $\sigma_i$s to get $\tilde\mu_i$s. Compute for each perturbed distance modulus, the corresponding new distance, $d_{lum}$, velocity, $v_{pec}$, energy term, $U$ and uncertainty, $\tilde\sigma_i$. Define $\alpha$=min$\{1,p_{\tilde \mu_i, \tilde \sigma_i}/p_{\mu_i, \sigma_i}\sigma_i/\tilde \sigma_i\}$, where $p$ is the Bayesian probability. Draw a number $R$ from a uniform distribution between [0,1[. if $\alpha \ge R$, accept the modification else keep the previous configuration. Repeat the process, for every single point ($n$ points), $n_{mh}$ times. 
\vspace{0.2cm}
\item decrease the probability ratio as follows:  ($p_{\tilde \mu_i, \tilde \sigma_i}/p_{\mu_i, \sigma_i}$)$^{1/T_{t}}$  with T$_t$ going slowly to zero at each end of [Step iii] according to the equation defined below.  Go back to [Step ii]. Proceed likewise $n_{sa}$ times. 
\vspace{0.2cm}
\item exit with a new catalog of datapoints with new positions along the line-of-sight, thus new peculiar velocities, and new uncertainties -- more precisely a new inseparable ($\mu$,$\sigma$) set.
\end{enumerate}
  
\noindent\textbf{- The temperature, $T_t$}, required by the simulated annealing algorithm after each loop of the Metropolis-Hastings algorithm times the number of galaxies $n$, is defined as follows:
\begin{equation}
 T_{t }=\frac{T_{0}}{1+\mathrm{ln}(t+1)} 
\end{equation}
with t the time-step. The initial temperature $T_0$  is given in Table \ref{tbl:1} together with the number of steps of the two embedded algorithms: $n_{mh}$ and $n_{sa}$.\\

 \begin{figure*}
 \vspace{-1cm}
 \includegraphics[width=0.36 \textwidth]{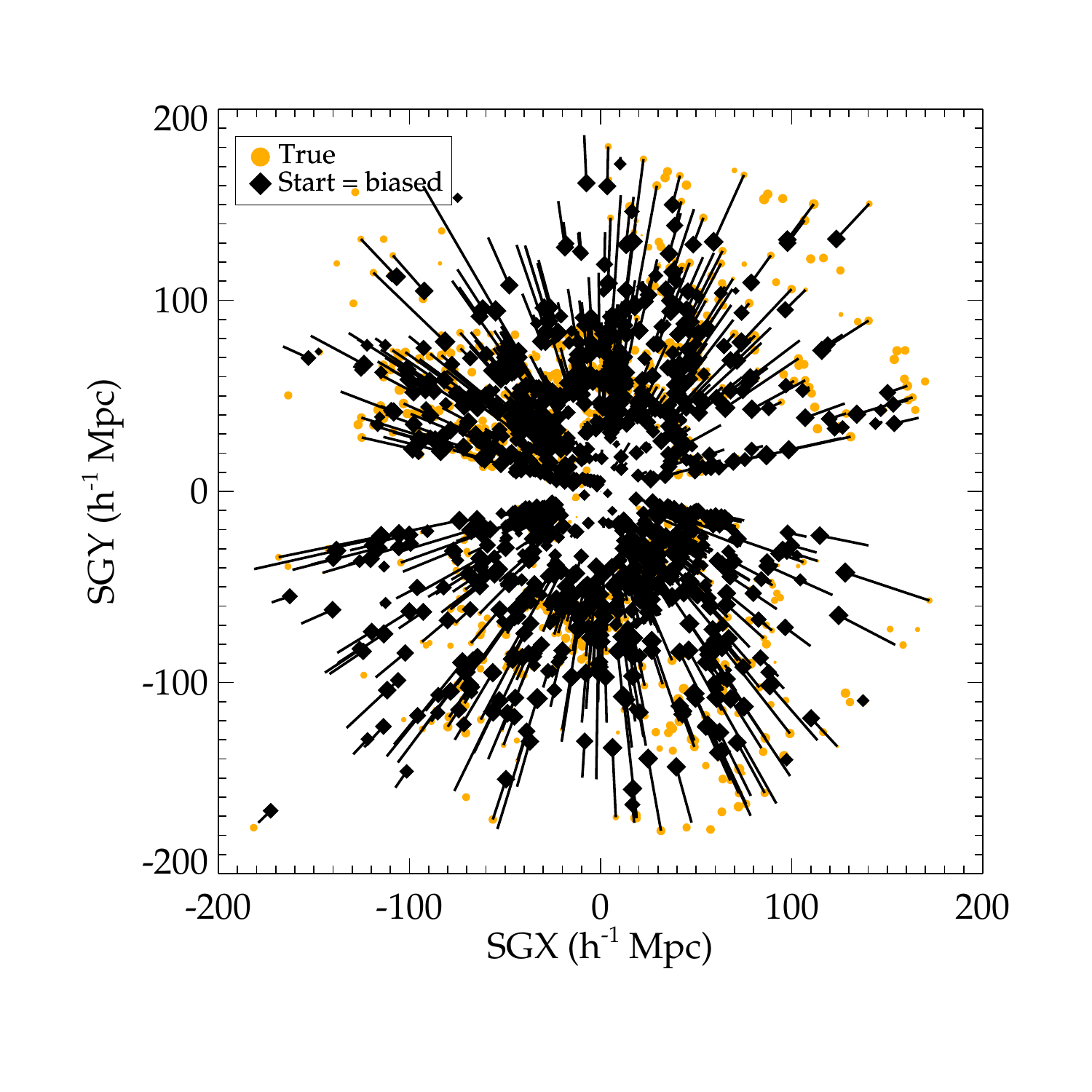}\hspace{-1cm}
  \includegraphics[width=0.36 \textwidth]{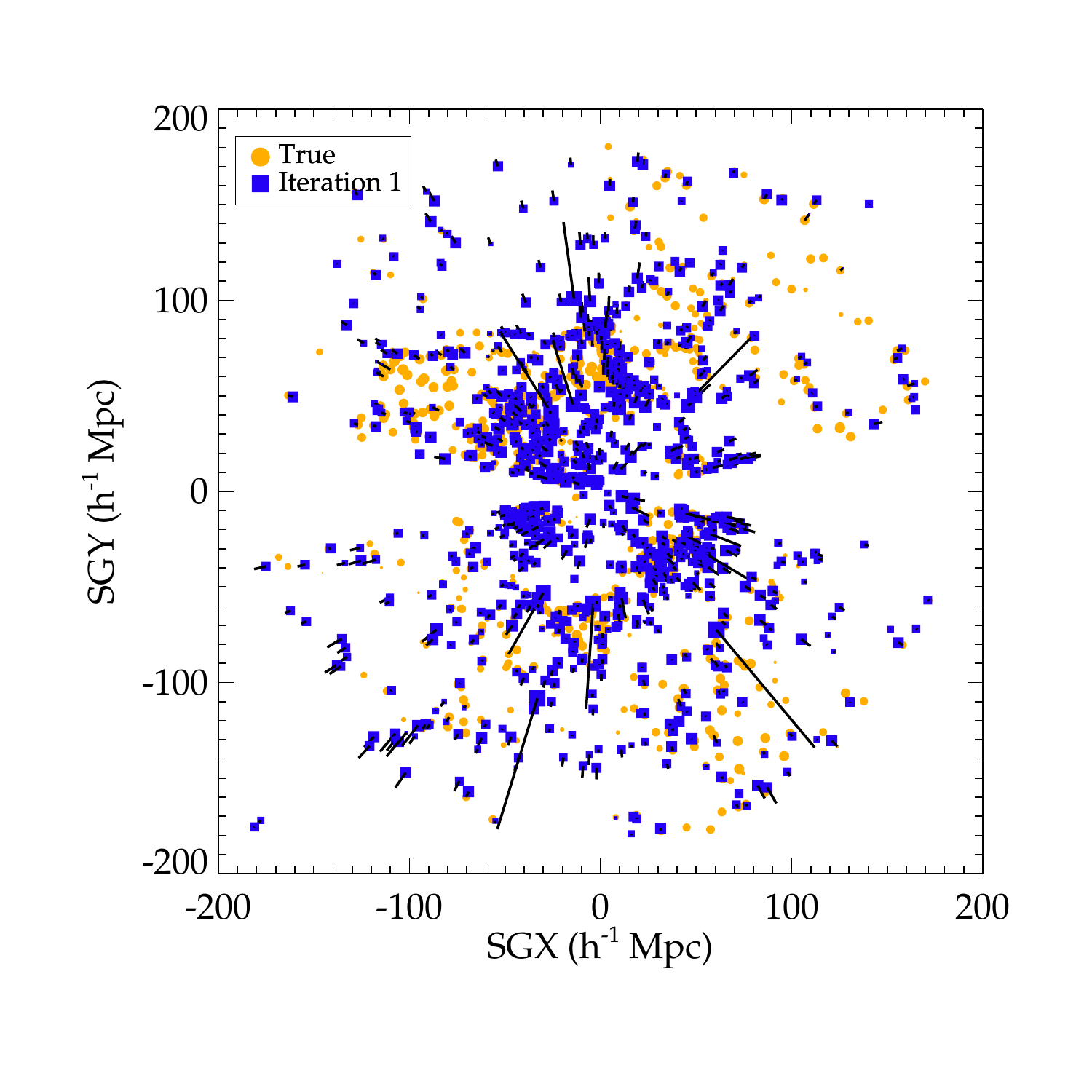}\hspace{-1cm}
\includegraphics[width=0.36 \textwidth]{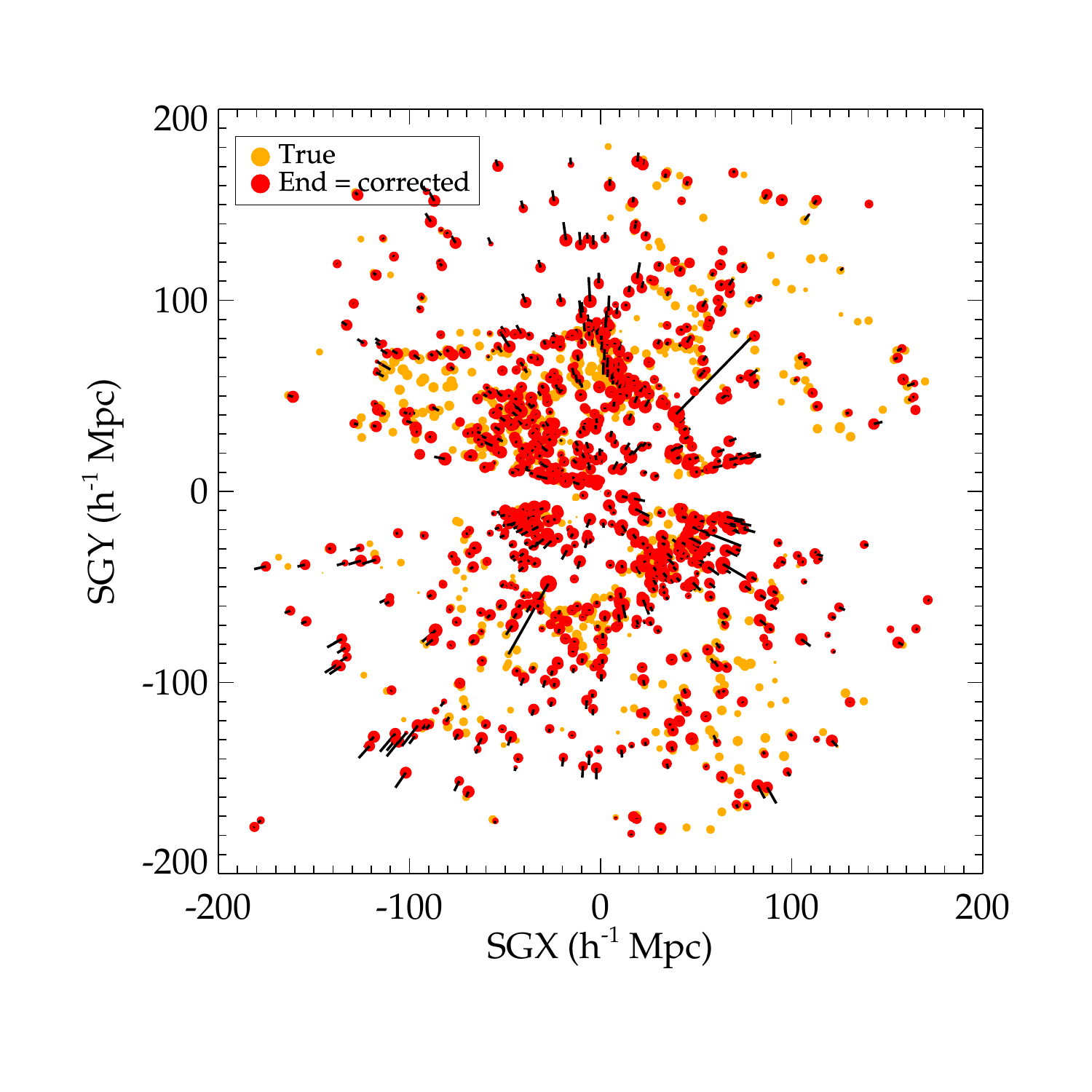}
\vspace{-0.8cm}
\caption{Galaxies in 5~\hMpc\ thick slices of the XY supergalactic plane. \textit{From left to right}: Black lines show the projected distances between true (yellow filled circles) and biased (black filled diamonds), after $n_{mh}$ Metropolis Hastings loops but no cooling for the $n$ datapoints (in the text Iteration 1, blue filled square) and, at the end of all loops (red filled circles), datapoint positions. The filled symbol sizes are proportional to velocities {on a logarithmic scale}. Note that because errors are large in the left panel, yellow filled circles are harder to distinguish. The algorithm reduces on the average errors on datapoint positions thus on their peculiar velocities: on average shorter black lines and by extension better matching filled symbol sizes of different colors.}
\label{fig:befafter}
\end{figure*}

\begin{table}
\begin{center} 
\begin{tabular}{l@{}r@{ }@{ }l}
\hline
\hline
Parameter & value & function\\
\hline
\hline
n$_{sa}$ & [5-\textbf{10}] & iteration s.a.\\
n$_{mh}$&  [2-\textbf{4}-5]$\times$1000 & iteration m.h.\\
n$_\sigma$ & 1 & shape interaction\\ 
$\gamma$ &  [1,3] & draw new $\mu$ (depends on $\sigma_i$, $m_{mh}$) \\ 
T$_0$ & 1 & initial temperature\\
$\alpha_{pc}$ & 0.05 & elongation shape \\
$\sigma_{v'}$ & 300$^{2}$ km$^2$~s$^{-2}$ & 1D radial $v_{pec}$ distribution variance\\
c$_1$ & 1 & constant in data term \\
c$_2$ & [\textbf{1}-2] & constant in data term\\
c$_3$ & [0.5-\textbf{1}] & constant in interaction term \\
\hline
\hline
\end{tabular}
\end{center}
\vspace{-0.25cm}
\caption{List of parameters used in the algorithm. When a range is given, tests with the parameter values in this range do not demonstrate any major change in the final datasets. For results presented in this paper, the bold values have been used. The constants are calibrated only once on one mock catalog and are kept unchanged for all the other synthetic catalogs and \textit{a fortiori} for the observational catalogs. Note that the $\gamma$ parameter serves the only purpose of speeding up the algorithm.}
\label{tbl:1}
\end{table}

To speed up of the process, we implemented slight modifications as follows. {We emphasize that they have no impact on the final result}. 
 \begin{itemize}[wide, labelwidth=!, labelindent=0pt]
\item We parallelized the algorithm using both MPI and openMP to run on several points from different part of the local Volume at the same time and to speed up the interaction term calculation. Namely, we perturb several points simultaneously (points sufficiently far away do not interact) and we derive their interaction energy term cumulatively respectively. We split points on processors depending on their Supergalactic longitude. \\
\item Points with an uncertainty below a certain threshold are not perturbed anymore. The threshold is determined as max$\{(\sigma_i)\}/10\times(10-m)$ where $m$ is increased by one unit each n$_{mh}$/10 iterations of the Metropolis-Hastings algorithm. Additionally, the minimum is set to 10$^{-4}$. This limit is just above the minimum change possible on the distance moduli given the precision of our algorithm. It thus prevents irrelevant iterations on datapoints those distance moduli would end up unchanged at the given precision in [Step ii].\\
\item $\gamma$ used to draw a new distance modulus decreases each n$_{mh}$/10 iterations as $\gamma=1+2/10\times(10-m)$, $m$ as defined before. {It decreases drastically the number of rejected new points in [Step ii] thus the number of irrelevant iterations}.\\
 \item If $\tilde v_{pec,\, i}$ is greater in absolute value than the largest initial peculiar velocity absolute value in the catalog, we immediately reject the new distance modulus.\\
 \item If the new distance modulus is smaller than 25 (i.e. $d_i<1$~Mpc), we immediately reject it since it is the size of the local Group (Milky Way, Andromeda and their satellites).
 \end{itemize}
 

 \section{Application to synthetic catalogs}

 In this section, the newly developed algorithm is applied and tested on synthetic catalogs. For the sake of conciseness, results being identical for all the synthetic catalogs, details are given for one of the mock catalogs.

 \subsection{Building synthetic catalogs}
 
\noindent To build mock catalogs matching our observational catalogs, we use a CLONE (Constrained LOcal \& Nesting Environment) simulation obtained with the technique described in \citet{2018MNRAS.478.5199S}. It contains 2048$^3$ particles in a $\sim$738~Mpc box and it ran from z=120 to z=0 in the Planck cosmology framework ($\Omega_m$=0.307~; $\Omega_\Lambda$=0.693 ;  H$_0$=67.77\kms~Mpc$^{-1}$).  {To obtain the different synthetic catalogs (named hereafter \textbf{true} and \textbf{biased} with the latter used as input for the algorithm), we proceed as follows}:
\begin{itemize}
\item Cut the catalog in mass and remove substructures to mimic grouping. We are not interested in testing again the grouping technique here \citep[see e.g.][for such tests]{2017MNRAS.469.2859S}. Only dark matter halos with masses greater than 10$^{12}$M$_\odot$ are preserved.
\item {Set an observer at the center of the box. From the $x$, $y$, $z$ coordinates and $v_x$, $v_y$, $v_z$ velocity components derive the distance, $d$, Supergalactic longitude and latitude, $sgl$, $sgb$ and radial peculiar velocity, $v_{pec}$ of each halo with respect to the observer.}
\item Compute cosmological redshifts, $z_{cos}$ with $d~=~\int_0^{z_{cos}}\frac{cdz}{H_0\sqrt{(1+z)^3\Omega_m+\Omega_\Lambda}}$.
\item Compute luminosity distances, $d_{lum}=(1+z_{cos})~d$, distance moduli, $\mu=5\times \mathrm{log_{10}}(d_{lum})+25$, and observational redshifts, $z_{obs}=v_{pec}/c~(1+z_{cos})+z_{cos}$.
\item Build a mock zone of avoidance by removing any halo, at more than $d_{lum}$=10~Mpc from the center of the box, within a cone which apex is the box origin and, its aperture is 0.2 radian assuming the same orientation within the XYZ simulated volume as the observational one in the Supergalactic XYZ volume.
\item For each datapoint in the observational catalog find all the halos such that $|z_{obs}-z_{obs,\ datapoint}|<0.01$ then sort these points by this value and by $|sgl-sgl_{datapoint}|$ and $|sgb-sgb_{datapoint}|$. Take the first halo of the sorted list as the mock point for the observational datapoint. \\ {$\Rightarrow$ The obtained ($sgl$,$sgb$,$z_{obs}$,$\mu$) set constitutes the \textbf{true} synthetic catalog.}
\item Add an uncertainty to the halo distance modulus as $\mu~=~\mu~+~R~\sigma_{datapoint}$. Assign $\sigma_{datapoint}$ as the uncertainty for the halo distance modulus. $R$ is drawn from a Gaussian distribution of mean zero and variance one. {Note that because the observational catalogs consist in a collection of distance moduli obtained with different indicators, coupled with the fact that several distance modulus estimates may be available for a given galaxy and \textit{a fortiori} for groups and clusters,  the $\sigma_{datapoint}$ ensemble spreads over a large range of values rather than being unique for all the datapoints.} \\ $\Rightarrow$ The obtained ($sgl$,$sgb$,$z_{obs}$,$\mu$,$\sigma_{datapoint}$) set constitutes the \textbf{biased} synthetic catalog. The goal of the algorithm is to retrieve a statistically bias-minimized synthetic catalog starting from the biased one.

\end{itemize}
  
 \subsection{Results}

  \begin{figure*}
 \vspace{-1cm}
\includegraphics[width=0.36 \textwidth]{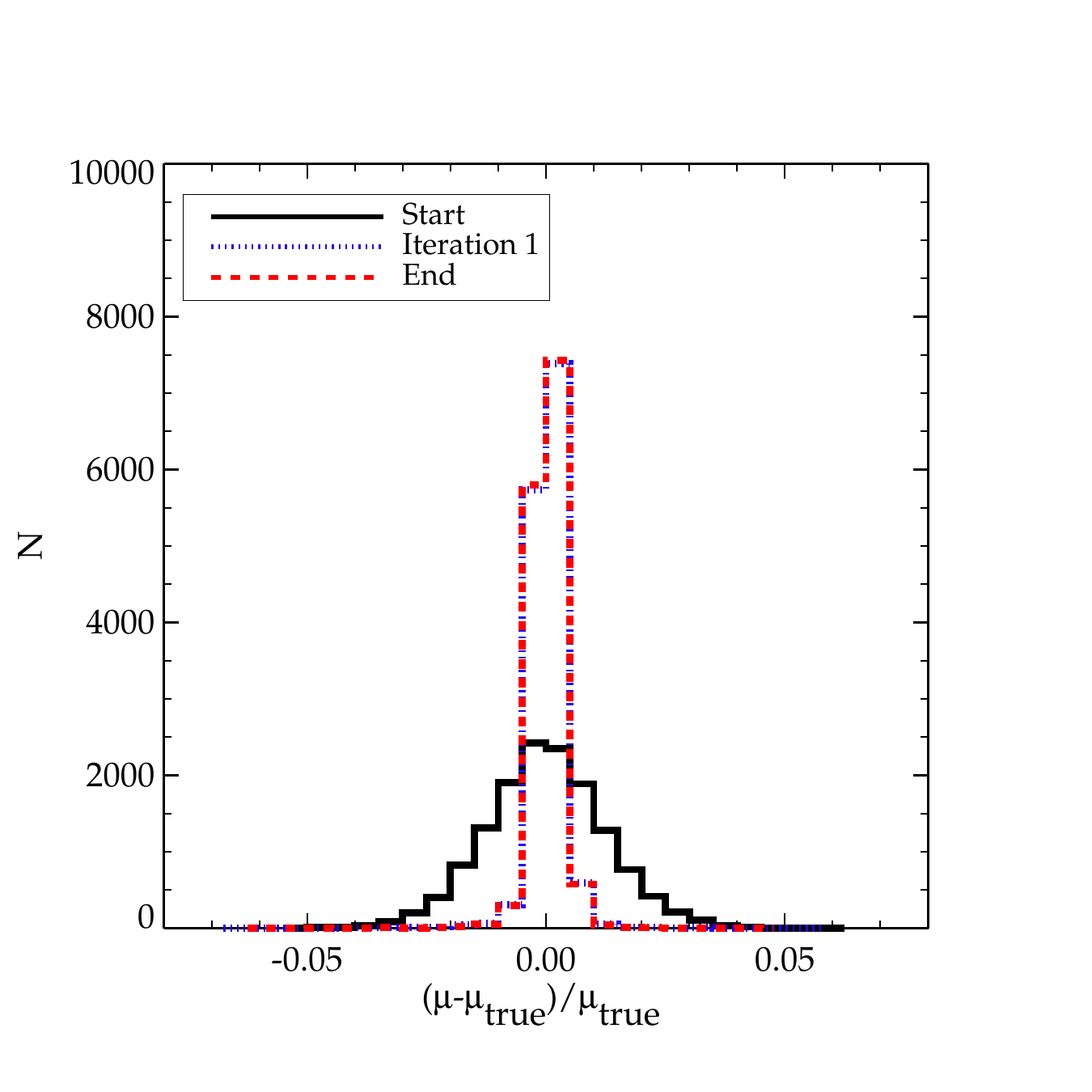}\hspace{-0.3cm}\hspace{-1cm}
\includegraphics[width=0.36 \textwidth]{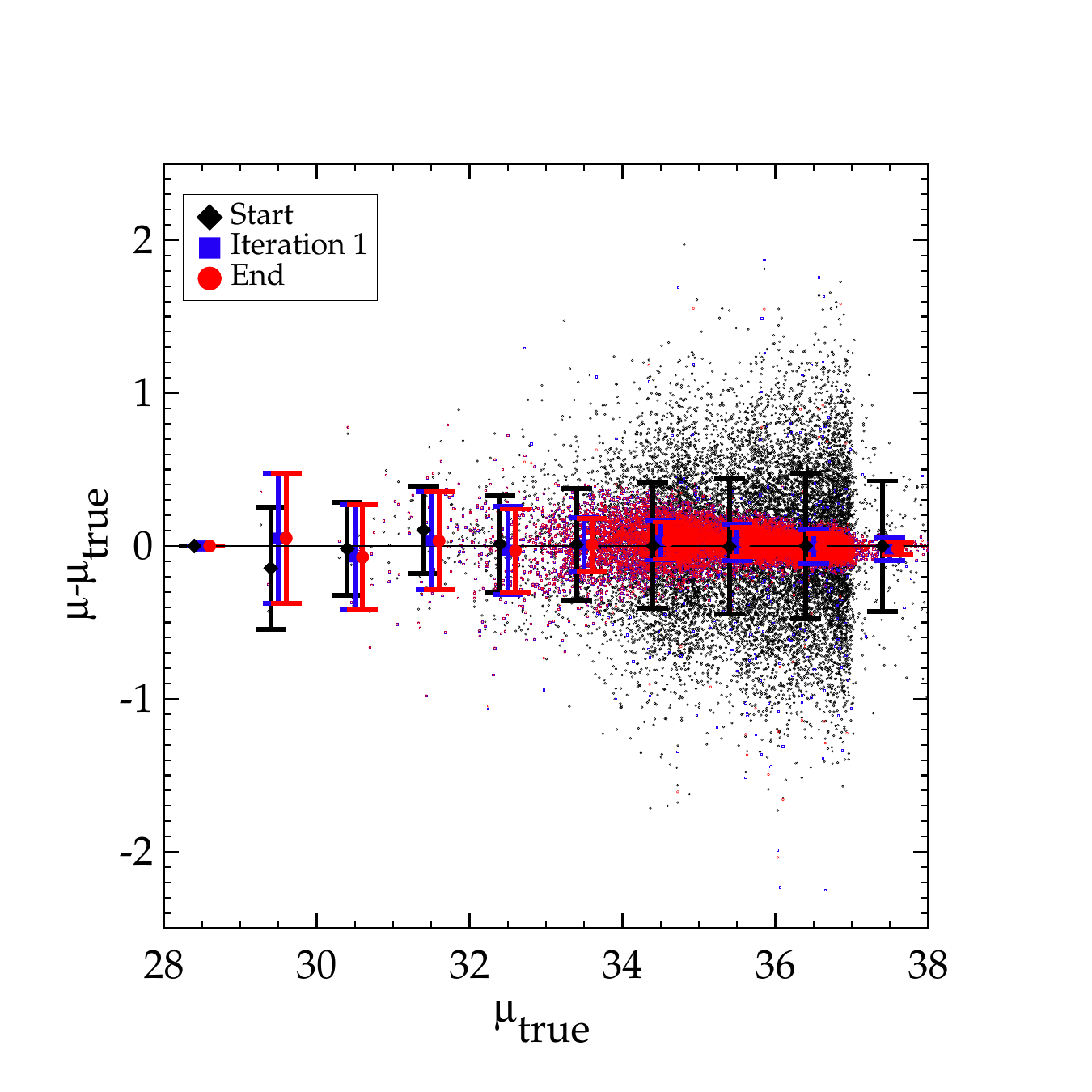}\hspace{-1cm}
\includegraphics[width=0.36 \textwidth]{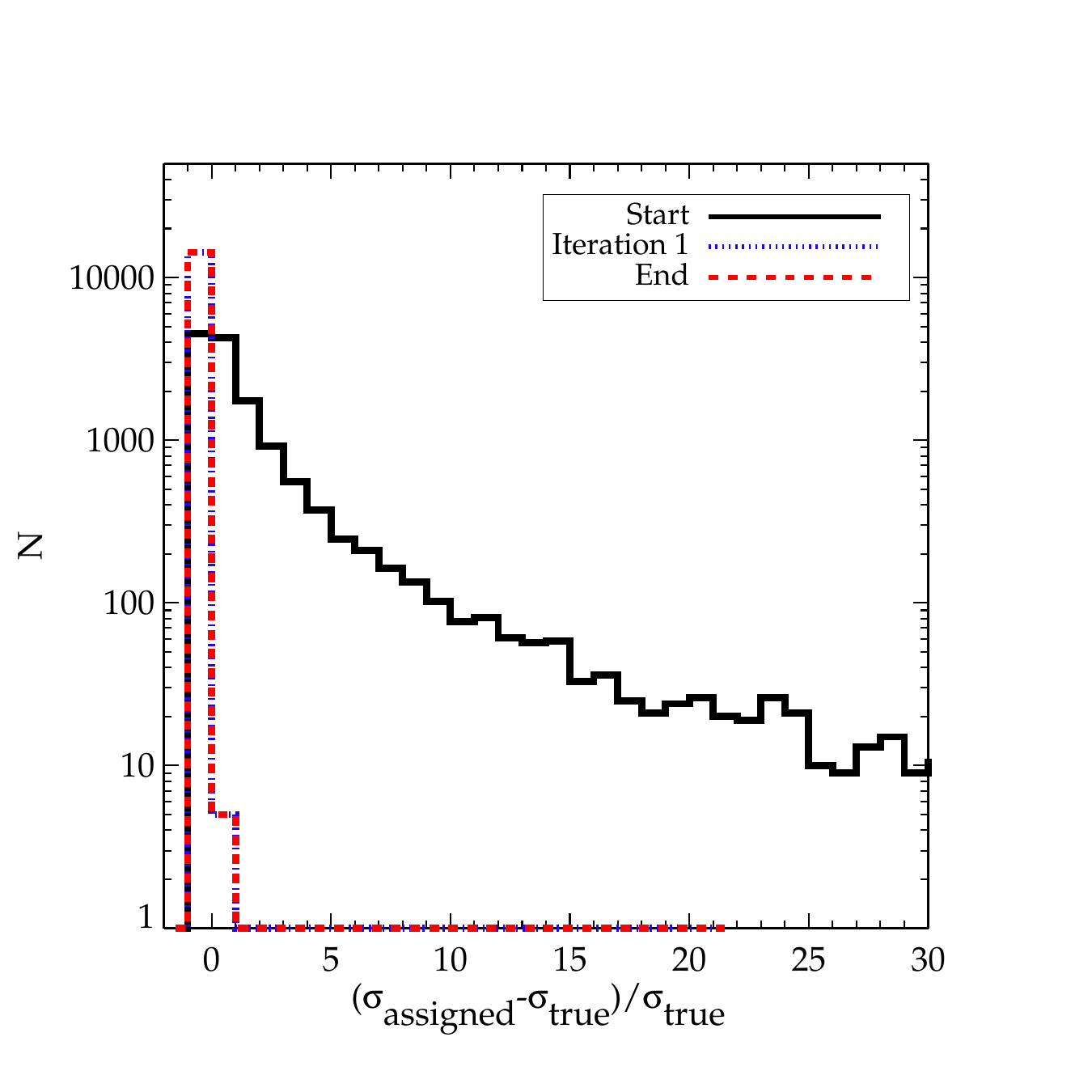}
\vspace{-0.4cm}
\caption{Comparisons of spatial distribution properties in the datasets. \textit{Left}: Histograms of the ratio of the differences between biased (black solid line), after Iteration 1 (see text for a definition, blue dotted line), corrected (red dashed line) and true distance moduli to the true distance moduli, namely errors in percent on distance moduli. \textit{Middle}: Difference between true and biased (black diamonds), after Iteration 1 (blue square) and at the end of the process or corrected (red circles) distance moduli versus true distance moduli. \textit{Right}: Ratio of the difference between assigned uncertainties and true errors to the true errors on distance moduli. Same colors and linestyles as the left panel.}
\label{fig:distmod}
\end{figure*}

 \begin{figure*}
 \vspace{-1cm}
\includegraphics[width=0.36 \textwidth]{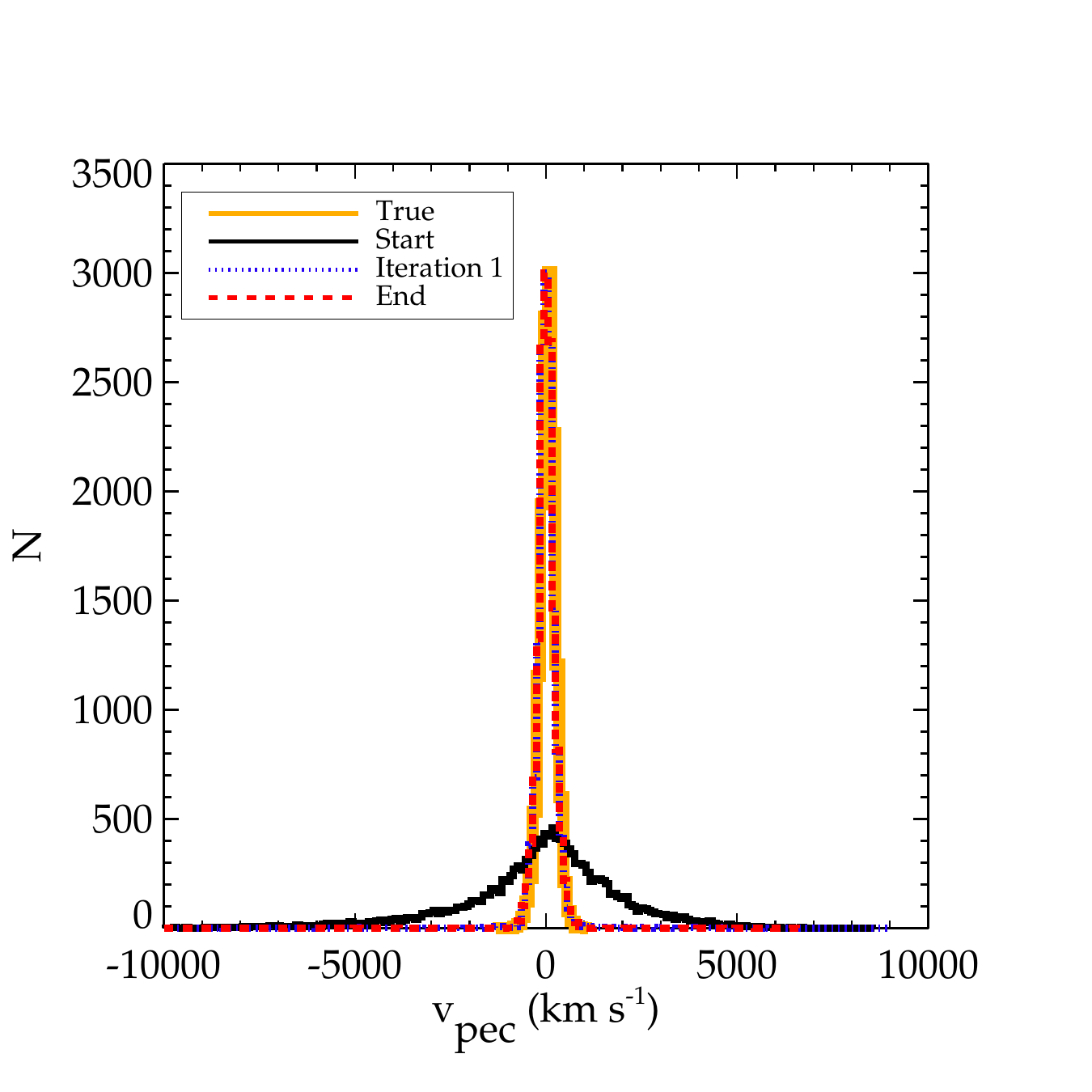}\hspace{-1cm}
\includegraphics[width=0.36 \textwidth]{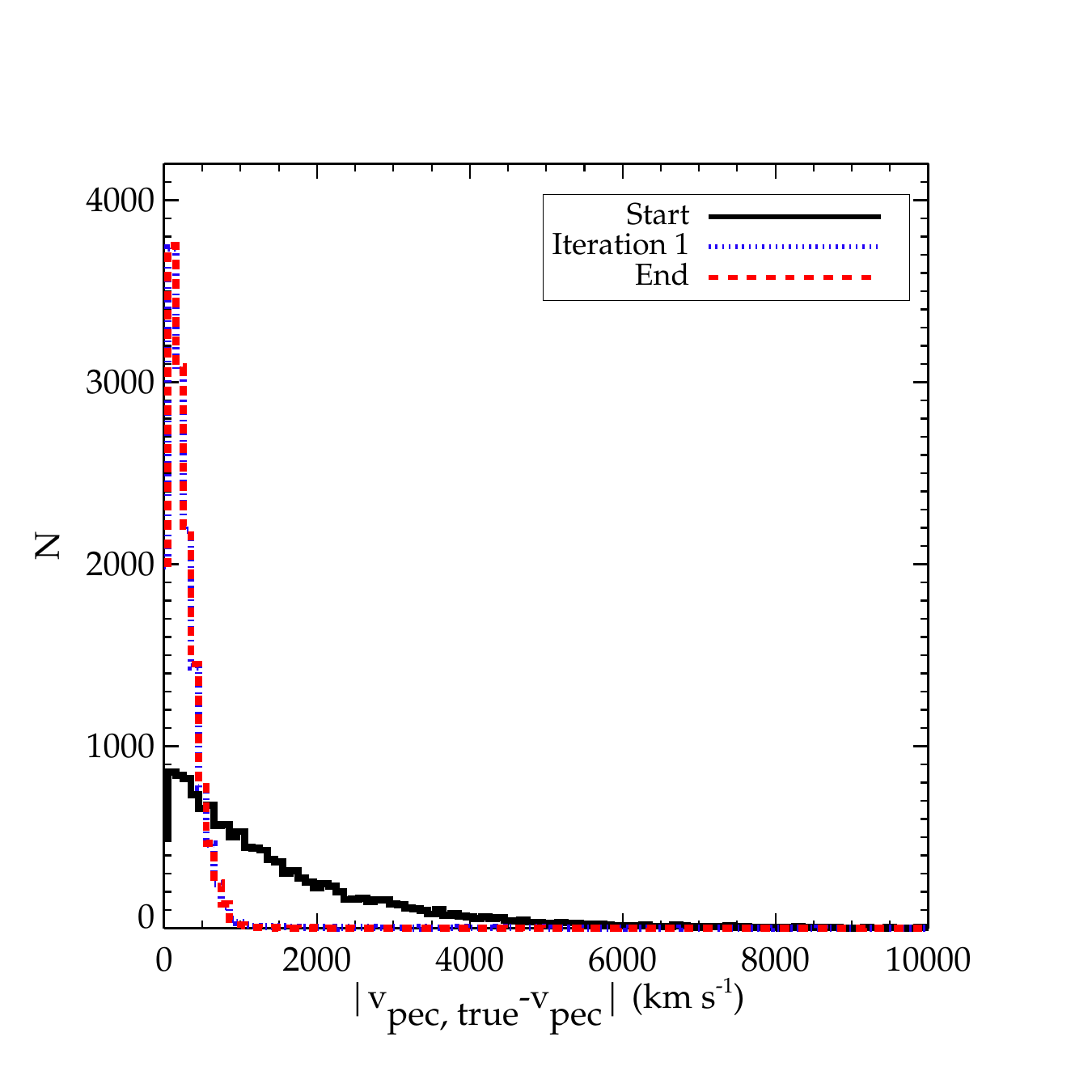}\hspace{-1cm}
 \includegraphics[width=0.36 \textwidth]{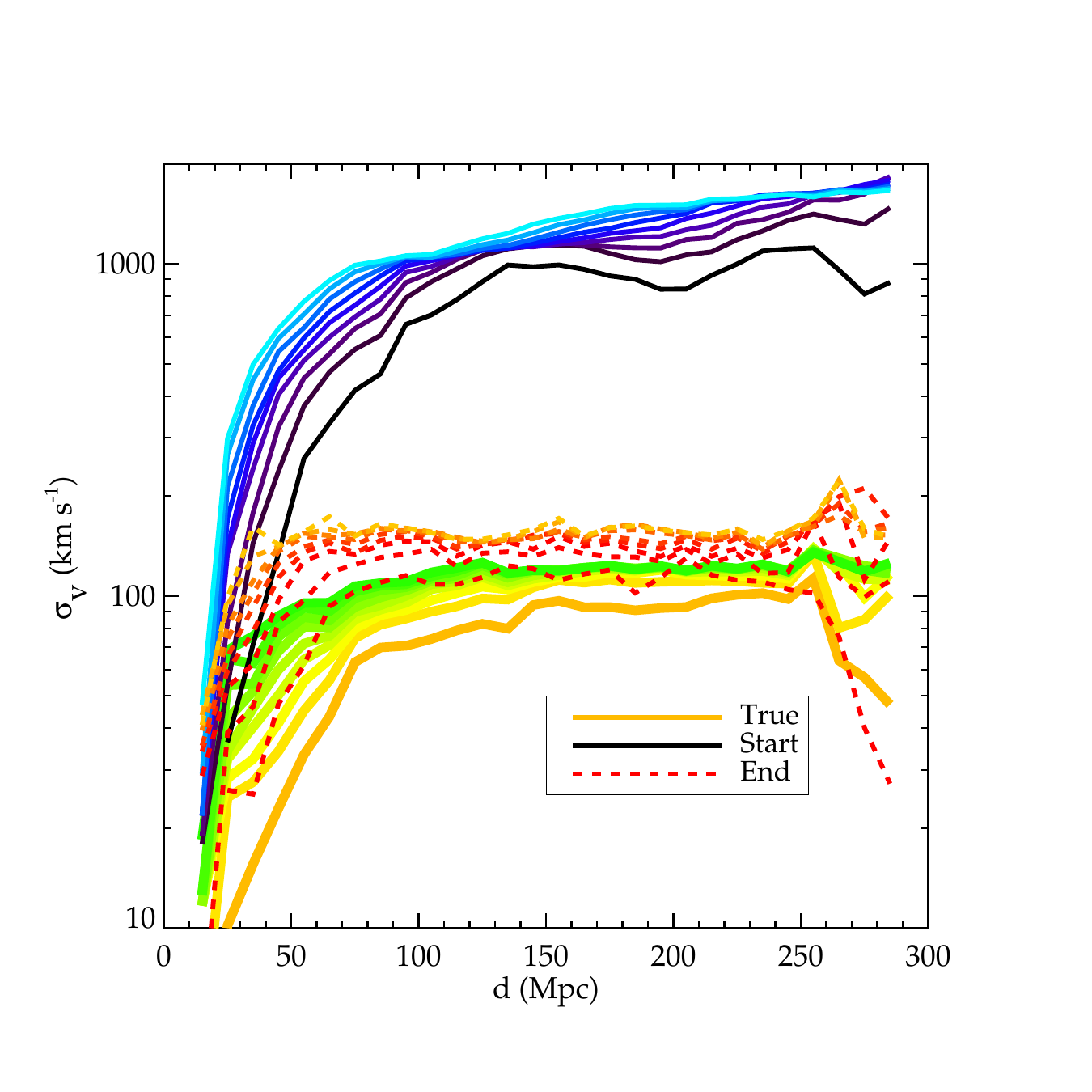}
\vspace{-0.2cm}
\caption{Comparisons of velocity properties in the datasets. \textit{Left}: True (thick yellow solid line) and biased (black solid line) peculiar velocity distributions vs. those derived from after Iteration 1 (blue dotted line) and at the end of the process or corrected (red dashed line) distance moduli.  \textit{Middle}: Histograms of errors in absolute value on peculiar velocities derived from before correction, after Iteration 1 and after correction distance moduli. Same color and linestyle codes. \textit{Right}: {Local} velocity variance in different shape $S$ elongations (one line per elongation) for a catalog free of errors (solid lines with warm colors), with errors (solid lines with cold colors), after applying the algorithm (dashed lines with warm colors). The variance is defined as the standard deviation between peculiar velocities of galaxies belonging to a same, elongated along the line-of-sight, shape $S$ (see exact definition in the text). Shape elongations at given distances are obtained ranging fictively distance modulus uncertainties from 0.2 (orange, red, black) to 1.8 (green, orange, light blue) mag.}
\label{fig:vpec}
\end{figure*}

\begin{figure*}
 \vspace{-1.5cm}
\hspace{1cm}\includegraphics[width=0.43 \textwidth]{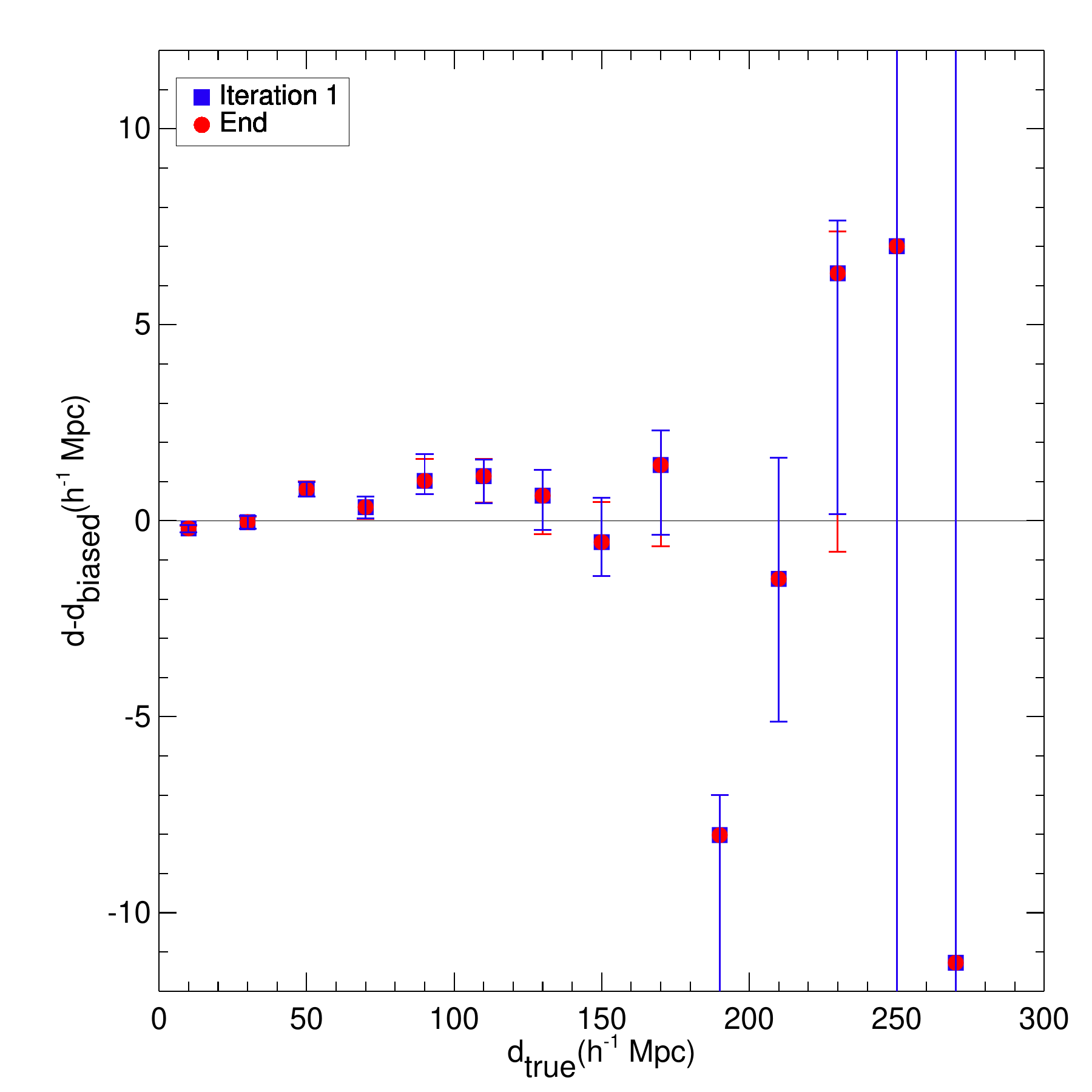}
\includegraphics[width=0.51 \textwidth]{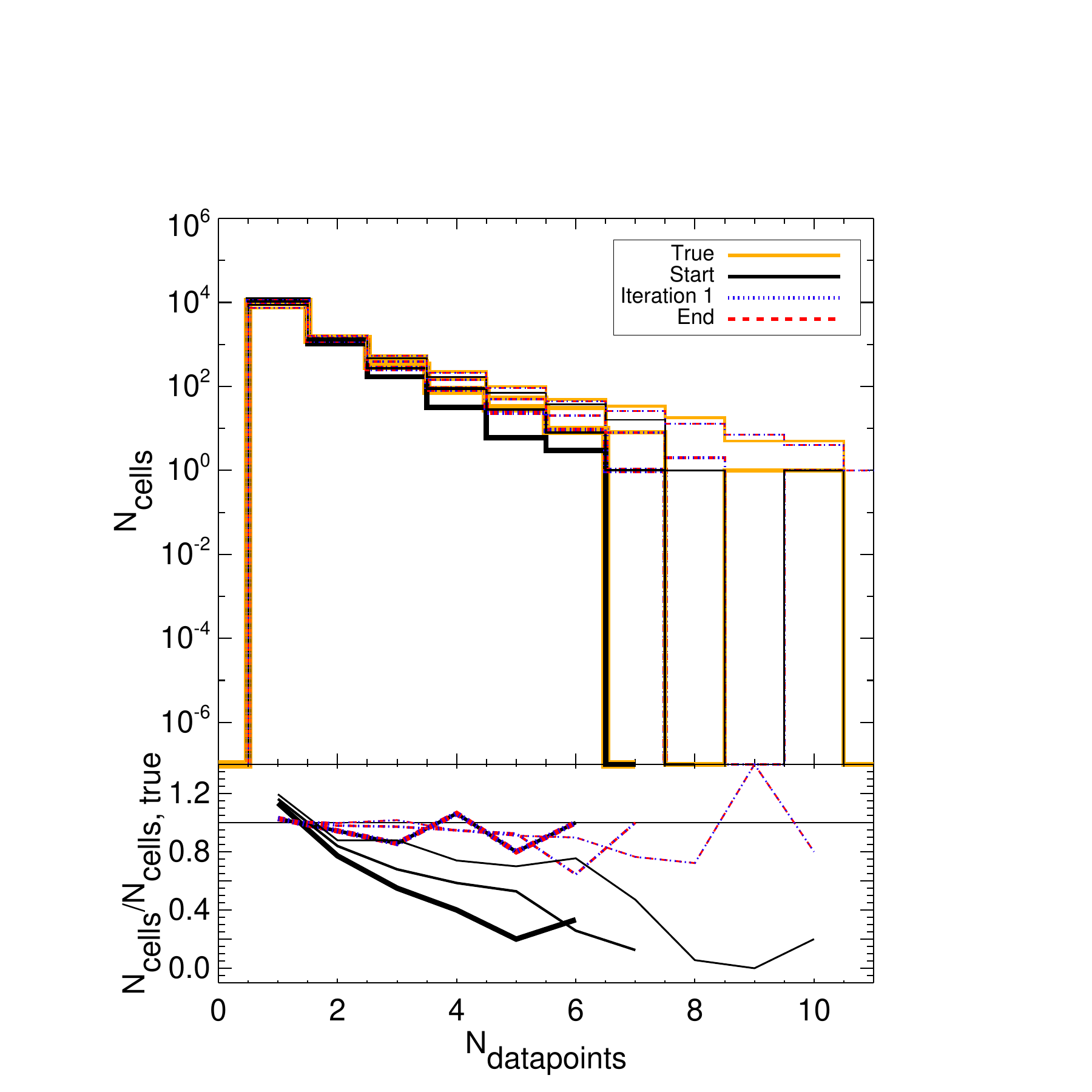}
\caption{{Comparisons of dataset properties. \textit{Left}: Median of the difference between distances obtained from after Iteration 1 (filled blue squares) and at the end of the process or corrected (filled red circles) distance moduli and those derived from biased distance moduli as a function of the true distances. Errors on the median are obtained with bootstrapping. New distances tend to be larger than initial ones. This is in agreement with the reduction of the homogeneous Malmquist bias that statistically tends to put objects closer than they are. After bias-minimization of distance modulus catalog, new distances are indeed statistically larger than initial ones derived from biased distance moduli. At large distances, effects of the catalog edges, like sharp cut-off, are preponderant. \textit{Top-right panel}: Histograms of the number of datapoints per grid cells obtained from true (thick yellow solid line), biased (black solid line), after Iteration 1 (blue dotted line) and at the end of the process or corrected (red dashed line) distance modulus catalogs. Grids are built to split the Supergalactic coordinate space uniformly. Since catalogs are not complete, cells with no datapoint have been removed. The histograms represent a measurement of the clustering. The more clustered the datapoints in a catalog are, the more datapoints there can be per grid cells. \textit{Bottom-right panel}: ratios between the biased (black solid line), after Iteration 1 (blue dotted line) and at the end of the process or corrected (red dashed line) histograms and the true one. These right panels show that datapoints in the biased catalog are statistically less clustered than in the true and corrected ones. This is in agreement with the reduction of the heterogeneous Malmquist bias. Indeed the latter tends to reduce clustering by statistically scattering objects from high density regions to low ones. The line thickness stands for the grid-cell size. From the thicker to the thinner lines, the cell sizes are $\sim$4.6, 5.5 and 6.9 \hMpc.}}
\label{fig:biascheck}
\end{figure*}

\noindent {Figures \ref{fig:befafter} to \ref{fig:biascheck} present the results of the above-described algorithm applied to one of the biased mock catalog mimicking cosmicflows-3. The yellow color stands for true datapoints, namely without error (true catalog), while the black color is used for the catalog with errors (biased catalog) used as input for the algorithm.} The blue color is used for datapoints after $n_{mh}$ Metropolis-Hastings iterations of every single one of the $n$ points (called \textbf{Iteration 1}) and the red color after $n_{mh}~\times~n_{sa}$ iterations, i.e. the Metropolis Hastings samplings embedded into the simulated annealing algorithm  (in other words, Iteration $n_{sa}$, called hereafter \textbf{corrected}). In the following, figures and the associated results are described more thoroughly. {Note that although the ultimate goal is to obtained bias-minimized peculiar velocity catalogs, since the input consists in distance moduli (that permit deriving ultimately peculiar velocities), analyses are conducted on the former as well as on the latter and on distances.} \\

From left to right, Figure \ref{fig:befafter} shows the datapoints in the mock catalog at their true positions with symbol sizes proportional, {in a logarithmic scale}, to their true peculiar velocities (yellow filled circles) alongside the initial biased positions and associated velocities (black filled diamonds), after Iteration 1 (blue filled squares) and finally the recovered or corrected positions and associated velocities (red filled circles). The solid black lines connect the true datapoint positions to their biased, after Iteration 1 and corrected positions. The mean length of these lines  and its standard deviation starts from $18\pm19$\hMpc\ to decrease to $3.1\pm5.4$~\hMpc\ and ends at $2.9\pm3.7$\hMpc. The algorithm after Iteration 1 already minimizes on the average errors on datapoint positions. According to the definitions given in eq. 1 to 3 and the propagation of uncertainty, it reduces statistically the errors on luminosity distances, cosmological redshifts and peculiar velocities. The cooling process permits small refinements on a point-to-point basis but still to be considered within the full catalog environment.\\

Figure \ref{fig:distmod} confirms that errors on distance moduli are statistically reduced. The left panel shows that true and bias-minimized distance moduli differ on average by less than 1\% (blue and red dashed and dotted lines) against 3-4\% without corrections (black solid line). More precisely, the middle panel shows that true and biased minimized distance moduli (blue and red squares and triangles) differ at most by about 0.5~mag against twice, up to four times, that value for biased distance moduli (black diamonds) especially at large distances. At small distances, the average error becomes more centered on zero, a clear indication that a systematic, galaxies too close on average as per bias b2, has been decreased. Note that the more datapoints there are in a given region, the better the algorithm performs. This is in agreement with the data interaction energy term. It proves that this term is essential and that it enforces the small-scale velocity correlation in the interaction shapes. It also confirms that datapoints cannot be considered individually but together as a whole: the bias-minimized catalog. The right panel shows for information that newly assigned uncertainties and true errors on datapoints are consistent as they differ at most by a few percent of the true error.

  \begin{figure*}
 \vspace{-0.5cm}
\includegraphics[width=0.36 \textwidth]{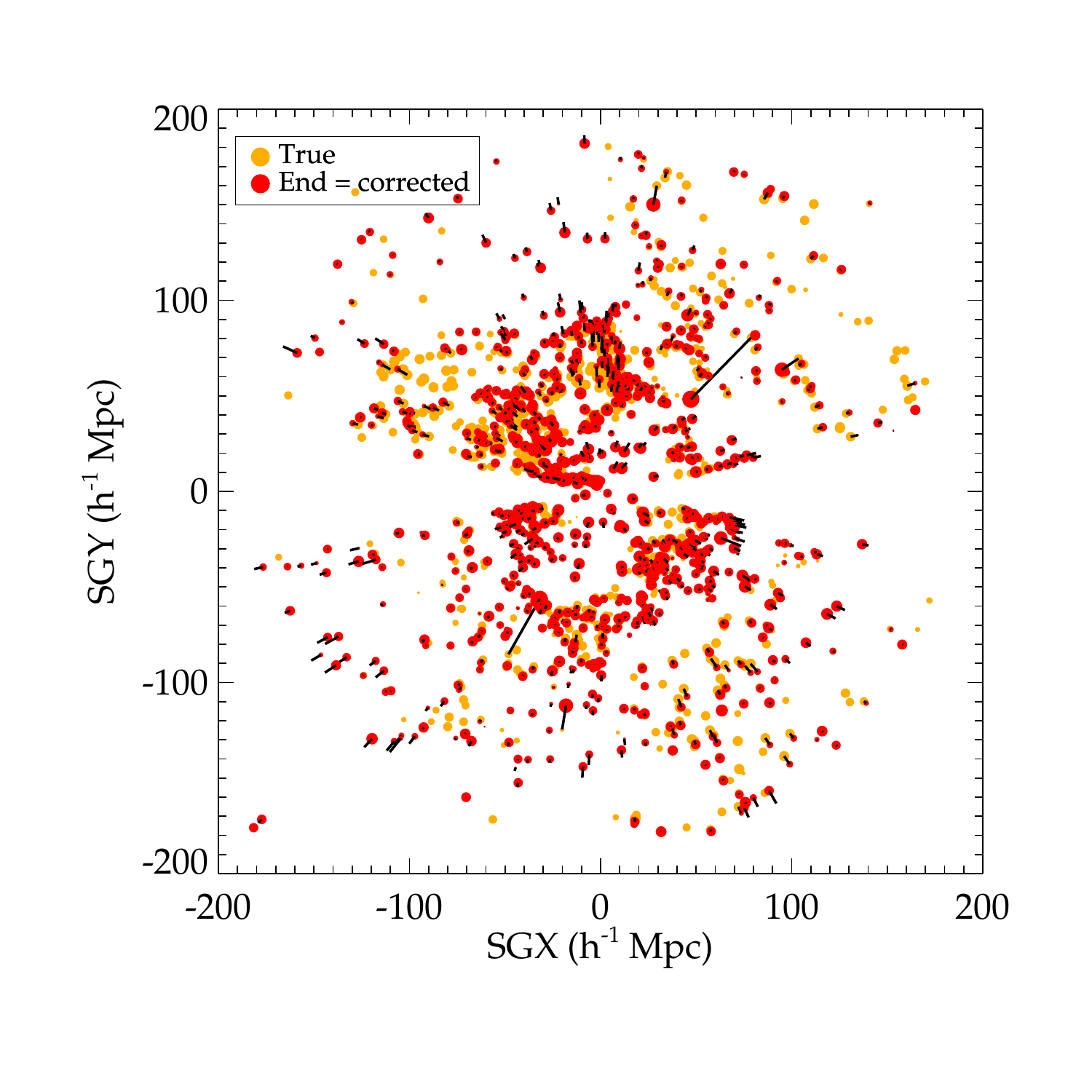}\hspace{2.5cm}
\includegraphics[width=0.36 \textwidth]{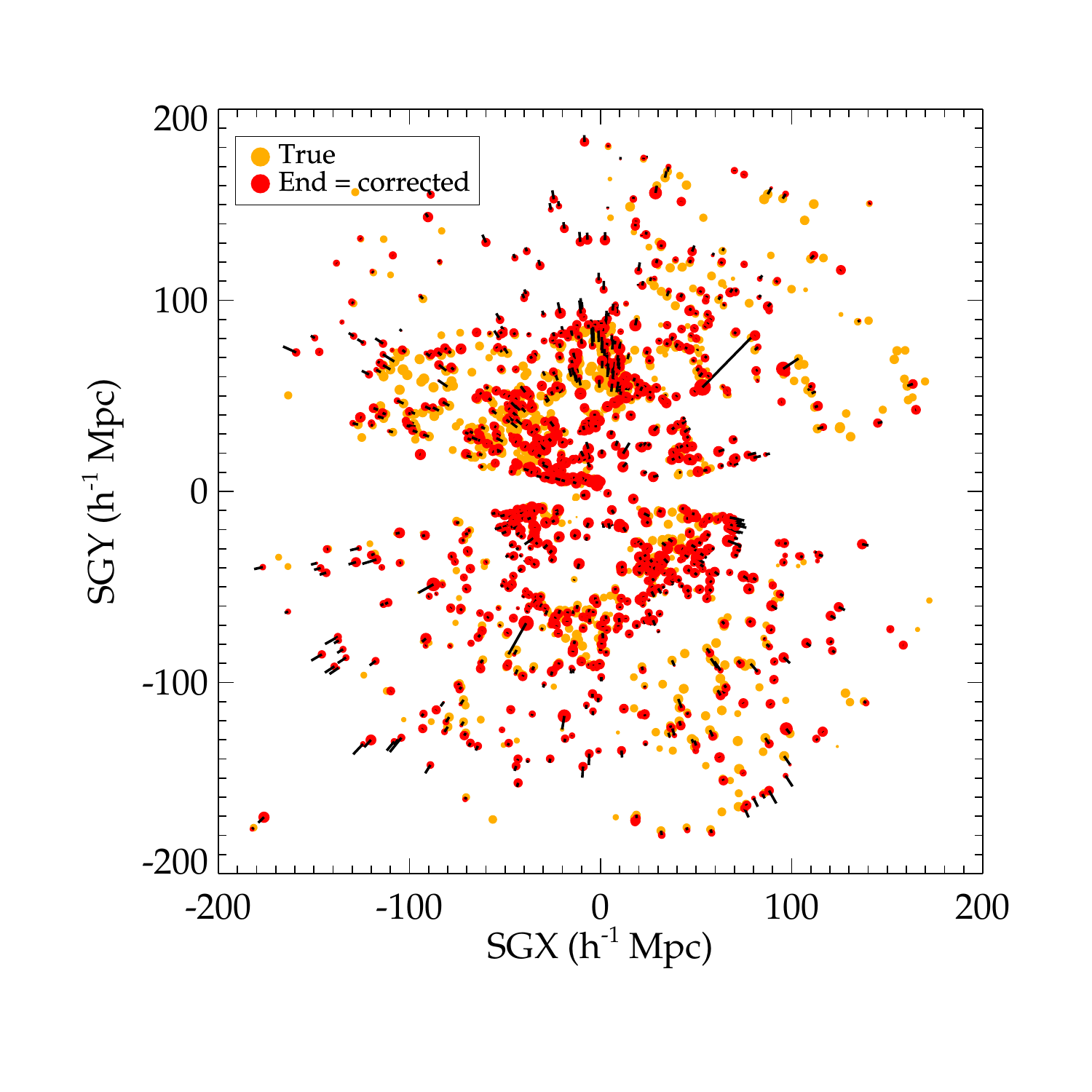}\\

\vspace{-1.5cm}
\hspace{-1.2cm}
\includegraphics[width=0.2 \textwidth]{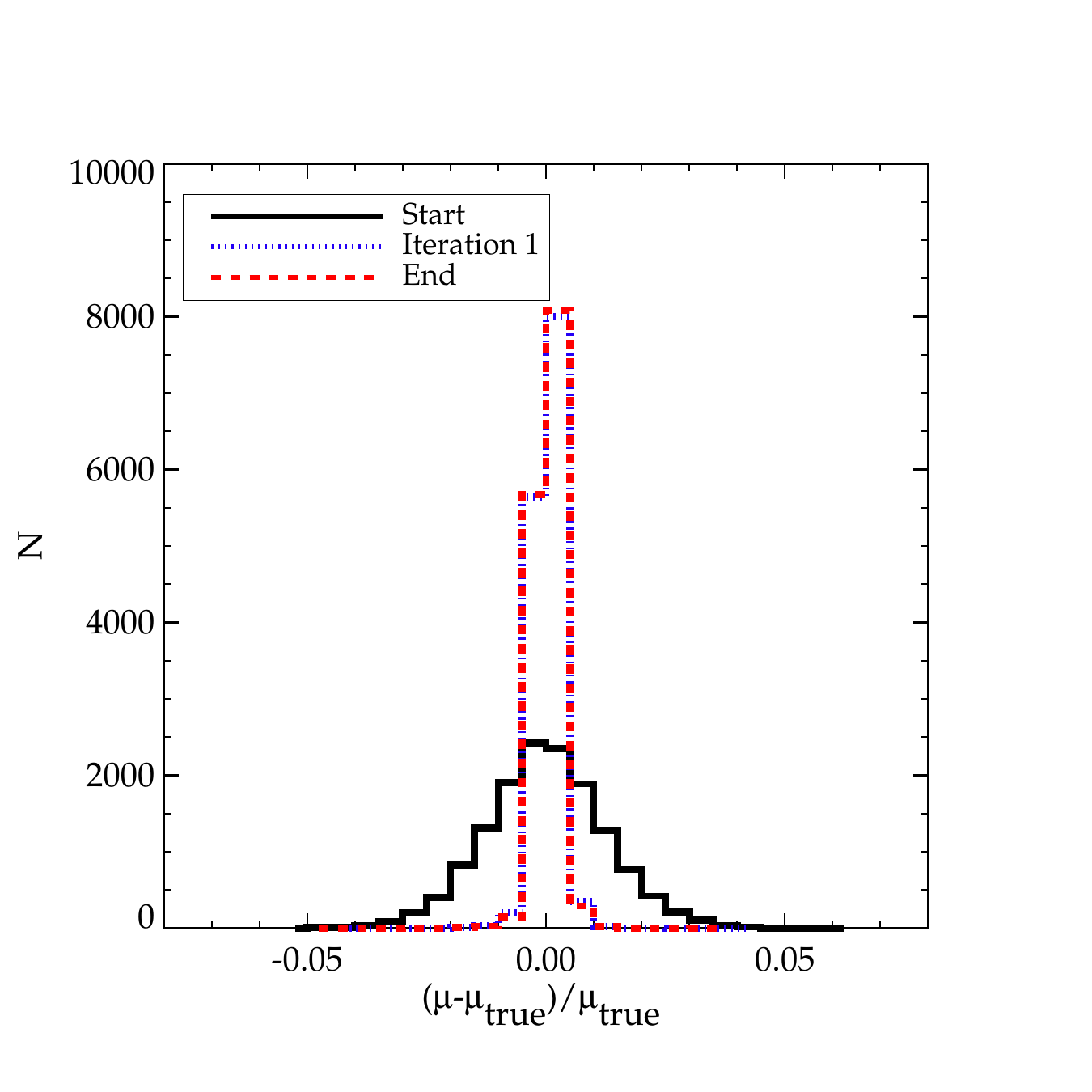}\hspace{-0.7cm}
\includegraphics[width=0.2 \textwidth]{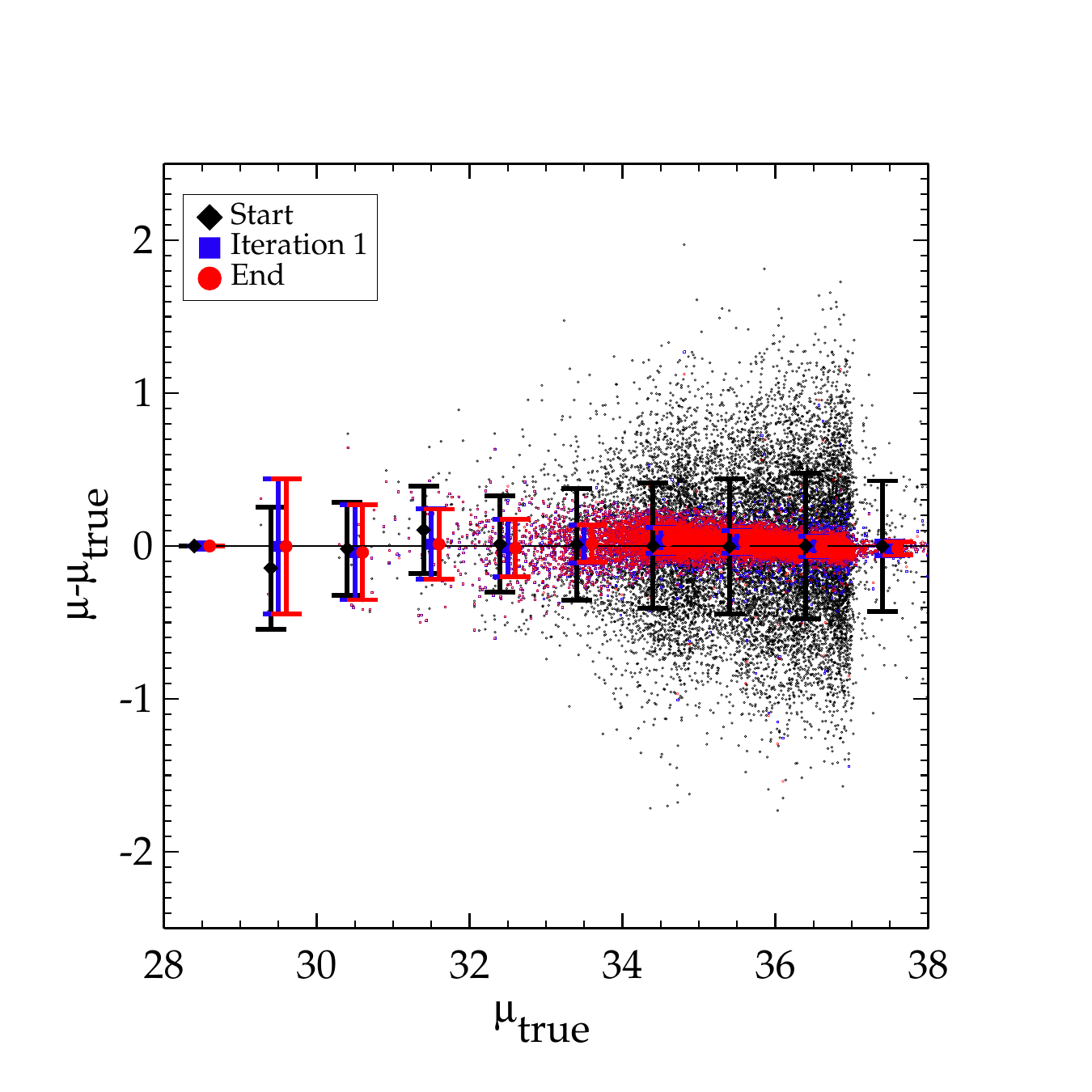}\hspace{-0.6cm}
\includegraphics[width=0.2 \textwidth]{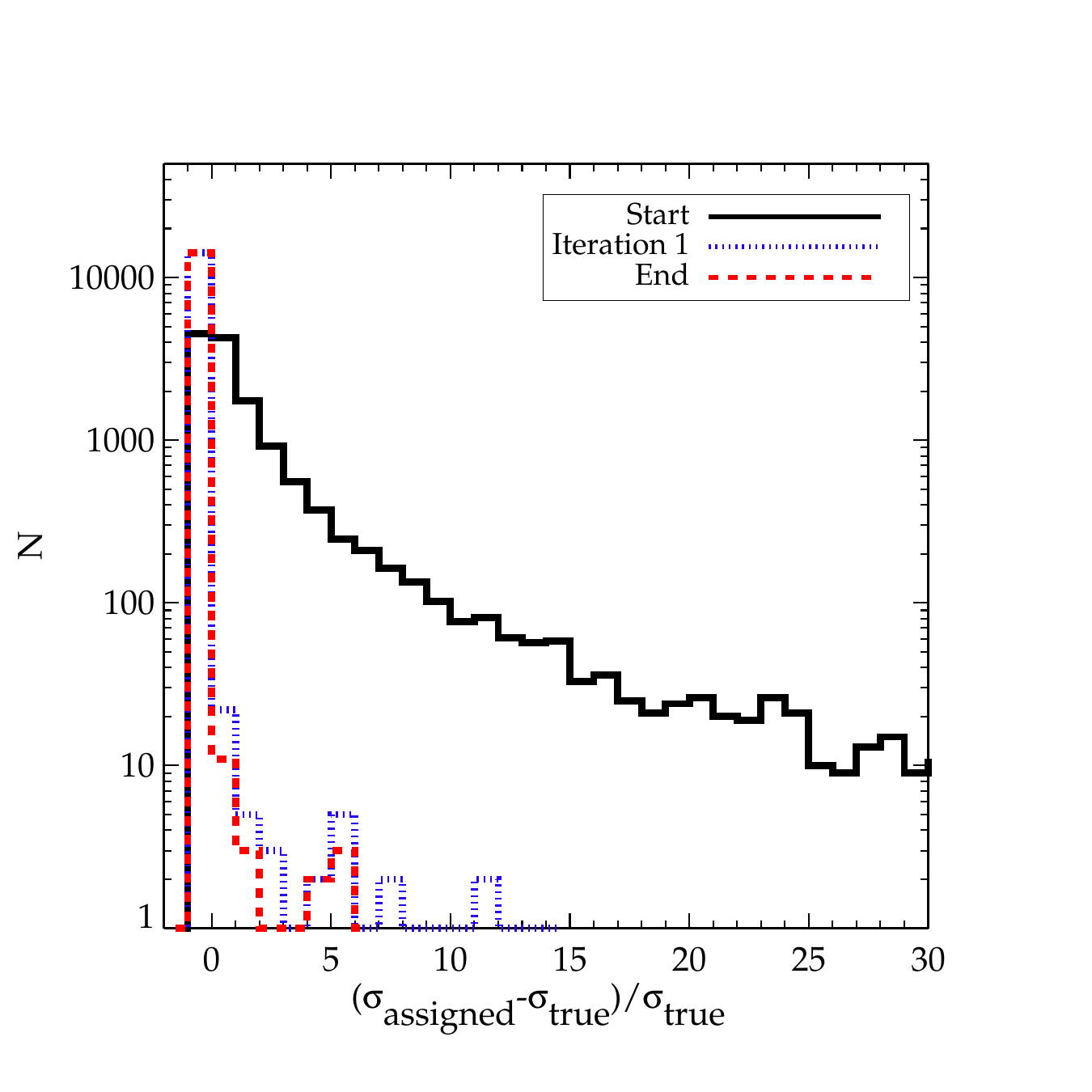}\hspace{-0.2cm}
\includegraphics[width=0.2 \textwidth]{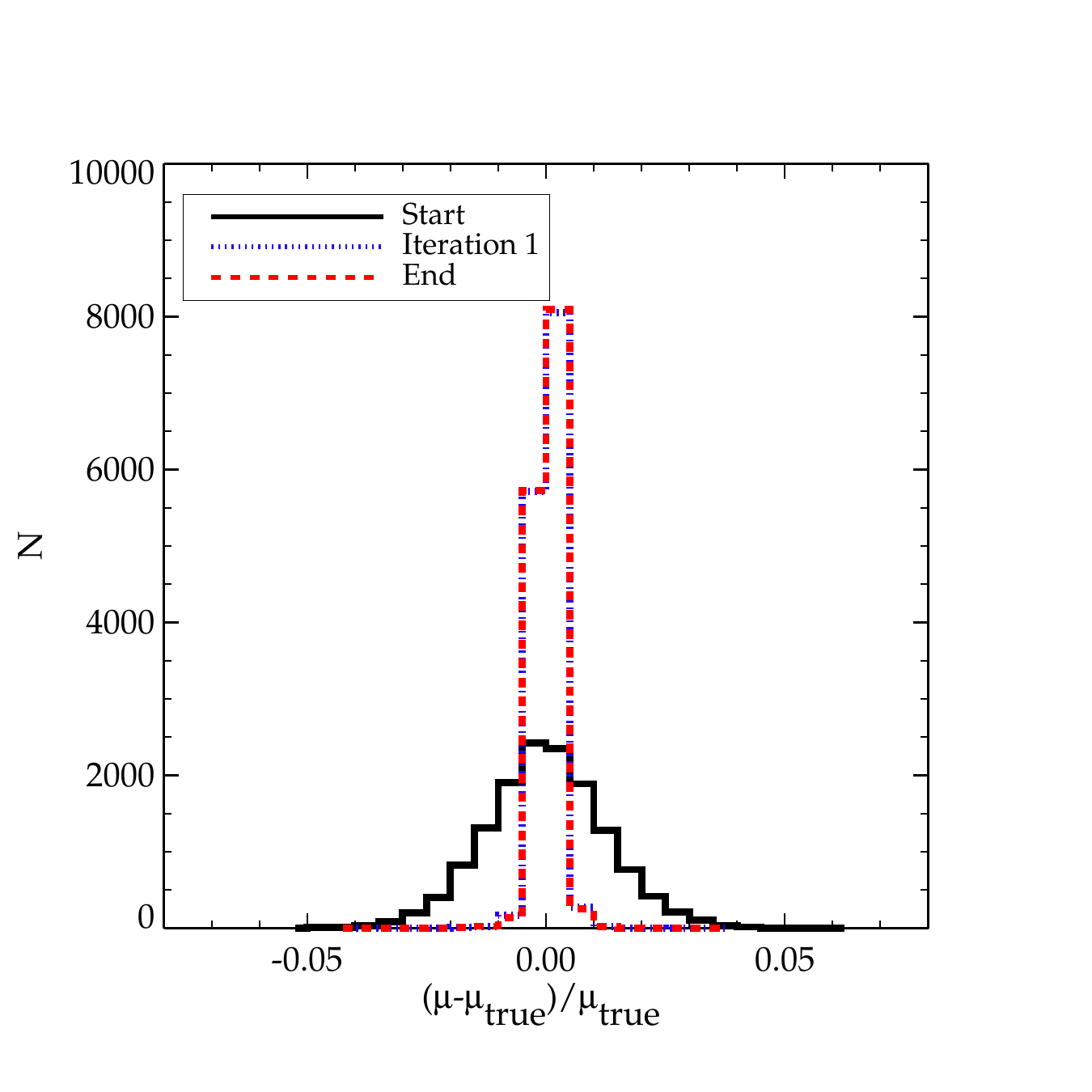}\hspace{-0.8cm}
\includegraphics[width=0.2 \textwidth]{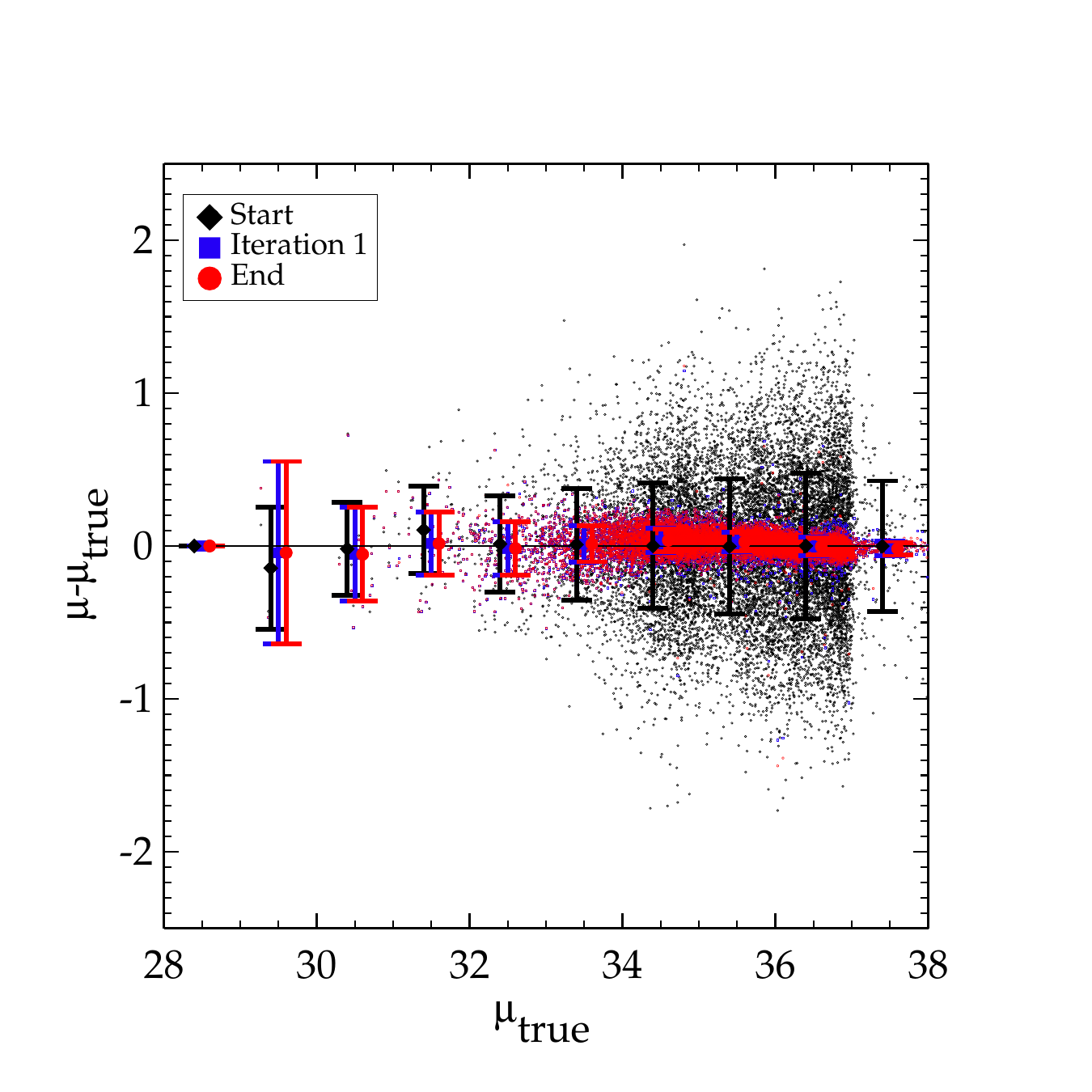}\hspace{-0.65cm}
\includegraphics[width=0.2 \textwidth]{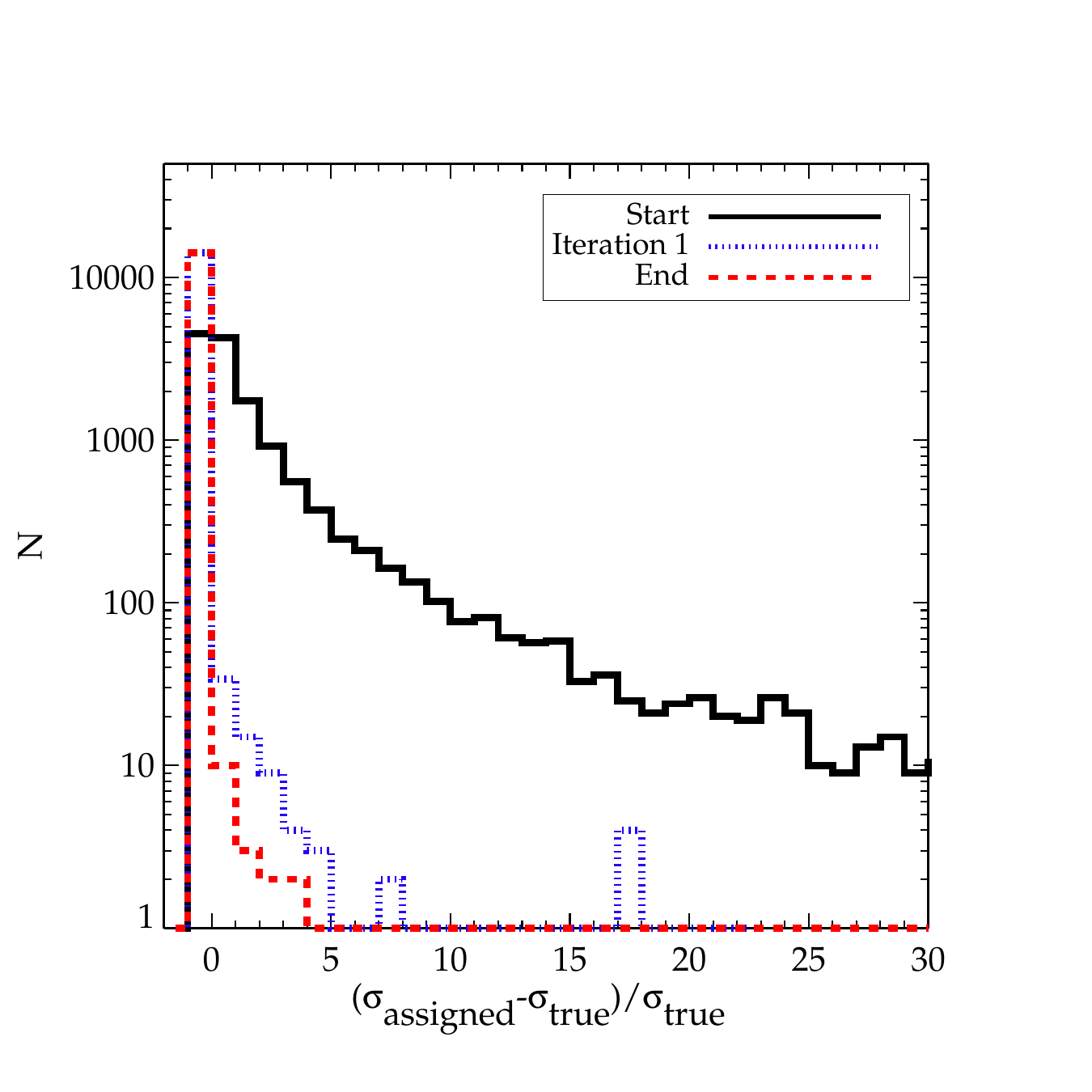}\\
\vspace{-0.8cm}

\hspace{-1.2cm}
\includegraphics[width=0.2 \textwidth]{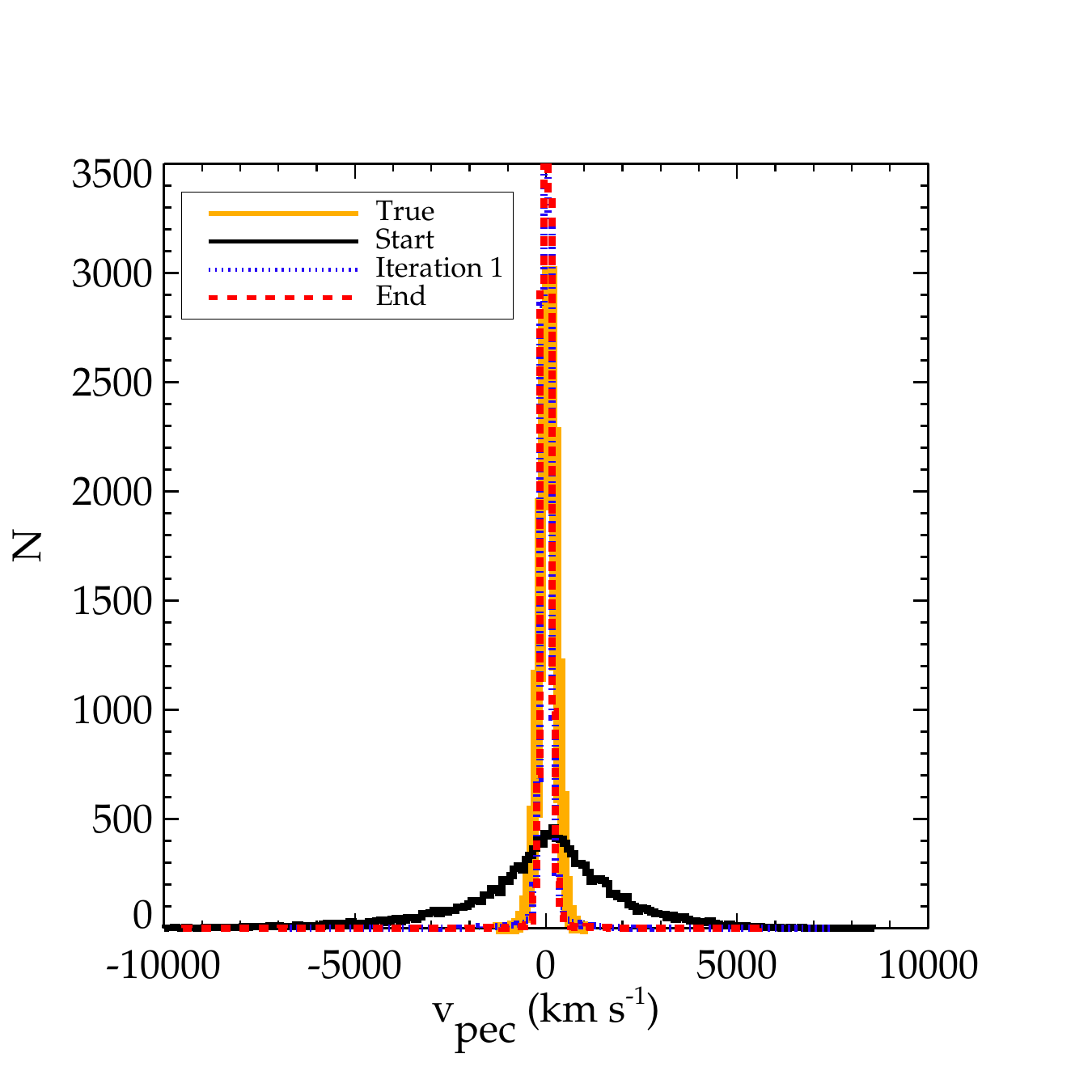}\hspace{-0.7cm}
\includegraphics[width=0.2 \textwidth]{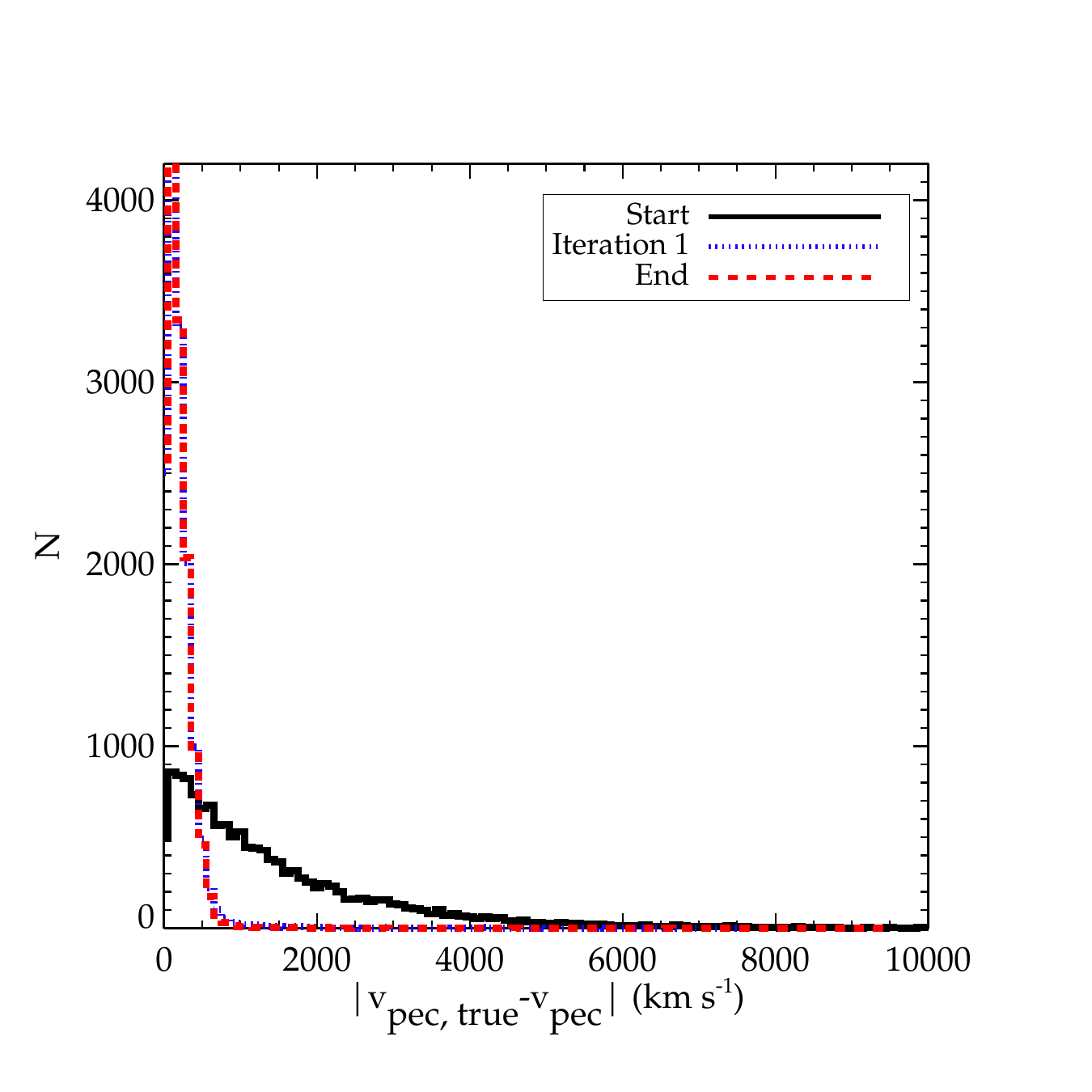}\hspace{-0.6cm}
 \includegraphics[width=0.2 \textwidth]{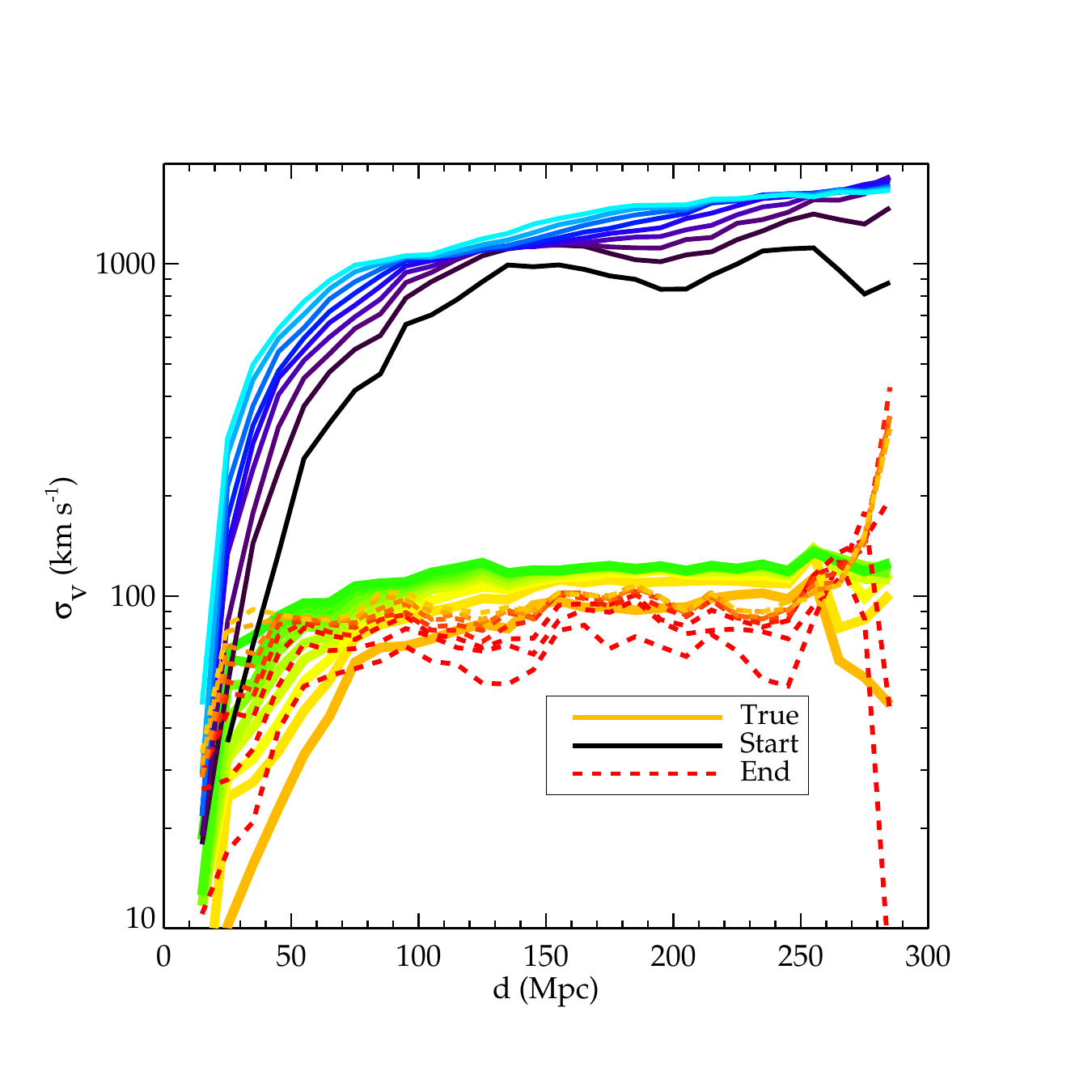}\hspace{-0.2cm}
\includegraphics[width=0.2 \textwidth]{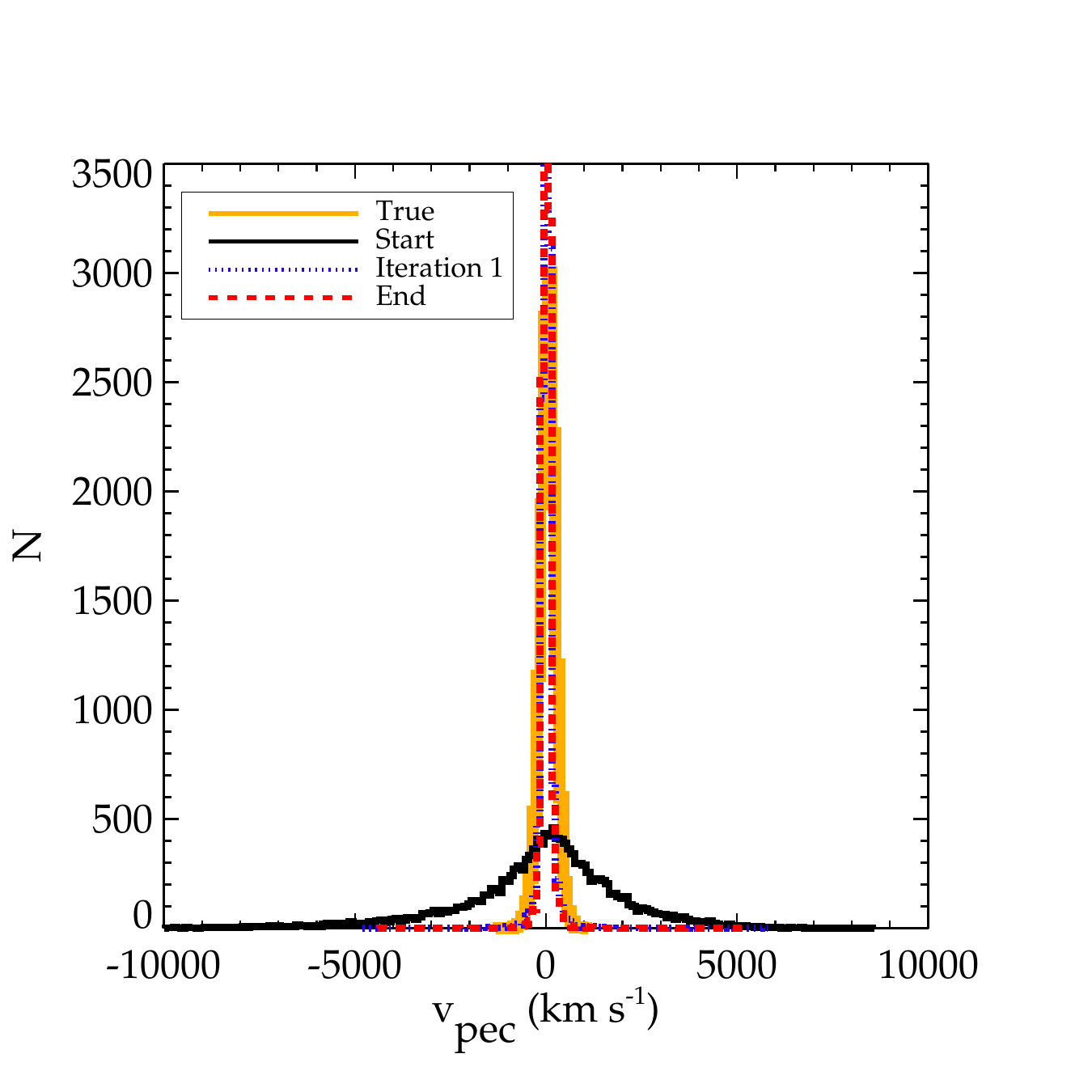}\hspace{-0.8cm}
\includegraphics[width=0.2 \textwidth]{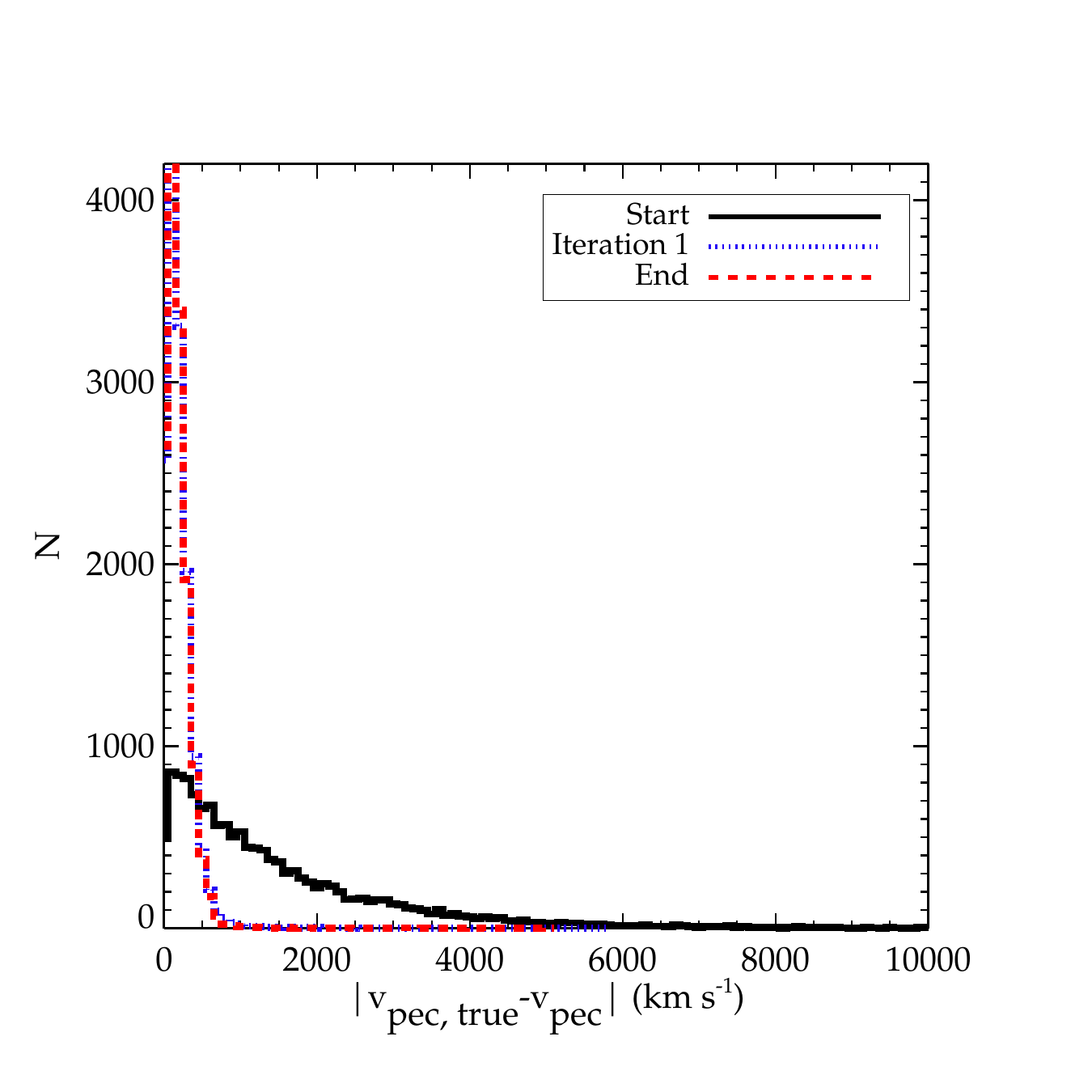}\hspace{-0.65cm}
 \includegraphics[width=0.2 \textwidth]{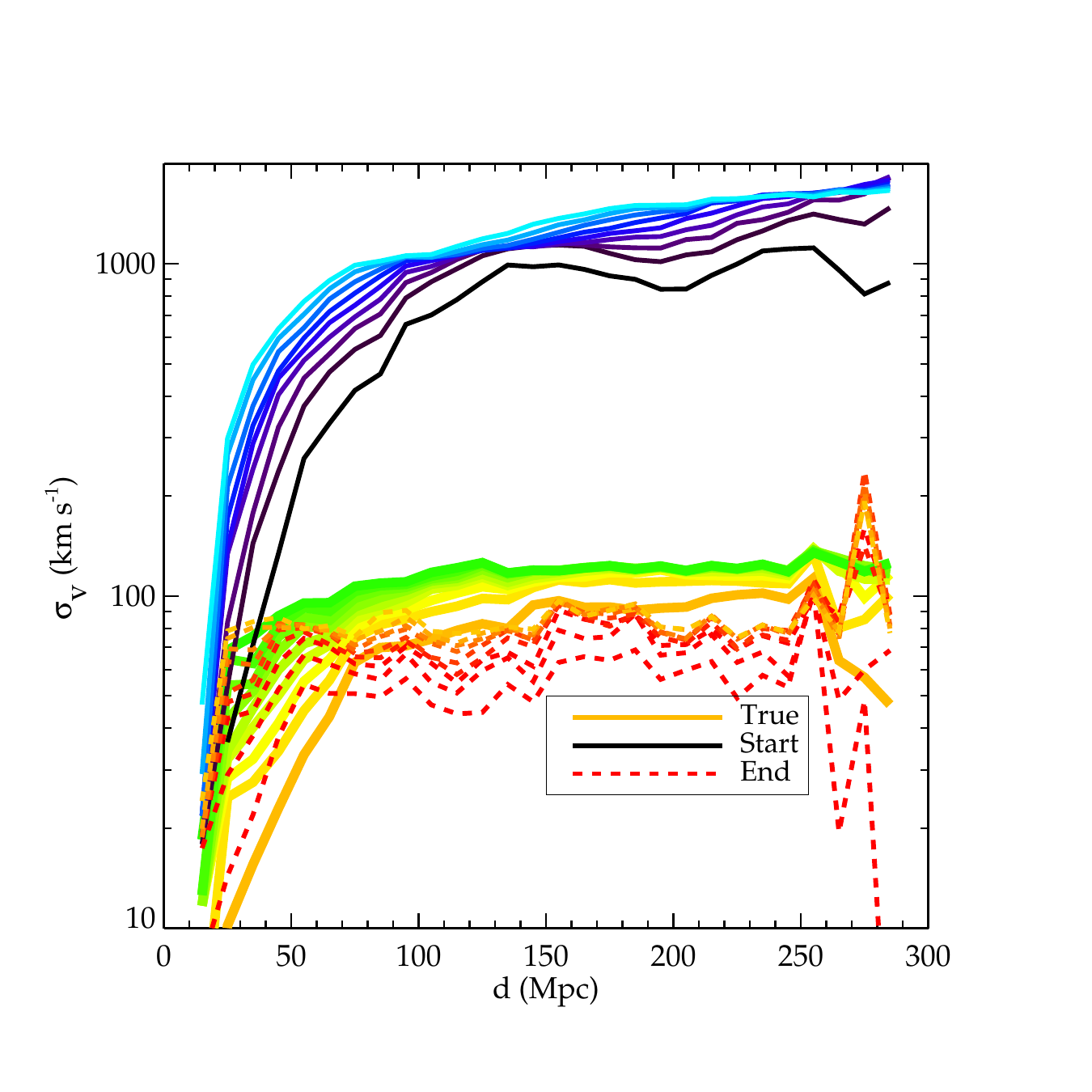}
\vspace{-0.cm}
\caption{Same as Fig. \ref{fig:befafter} right panel, Fig. \ref{fig:distmod} and \ref{fig:vpec} for five (left column) and ten (right column) stacked realizations of the corrected catalog obtained with the algorithm.}
\label{fig:stacked}
\end{figure*}

Finally, the two left panels of Figure \ref{fig:vpec} highlight that peculiar velocities (dotted and dashed blue and red vs. solid black lines), derived from statistically reduced-error distance moduli, have also on average reduced errors. In the left panel, the underlying Gaussian radial peculiar velocity distribution (thick solid yellow line) is recovered (dashed red line) like in \citet{2015MNRAS.450.2644S}. However, unlike in the latter case, it is not the main aim as efforts are focused on converging toward the most probable distance modulus distribution. It just happens that peculiar velocities derived from such a new distance modulus catalog have this property. Additionally, the expected {small-scale} velocity variance in different shape sizes (corresponding to different uncertainties and shown by a color gradient), at various distances from the center of the box, is almost recovered: warm color dashed lines (corrected) vs. warm color solid lines (true) with respect to cold color solid lines (biased) vs. warm color solid lines (true). \\

To confirm that the algorithm is not equivalent to a naive decrease of all the peculiar velocities that would also result in a reduced peculiar velocity variance (but fully intentional), we proceed with the following \textit{Gedankenexperiment}:
\begin{enumerate}
\item peculiar velocities rather than distance moduli constitute the starting point,
\item all the peculiar velocities are reduced by a constant factor chosen to find back the expected variance\footnote{Note that multiplying this factor by $\frac{1}{\sigma_{vpec,i}}$, to reduce less peculiar velocities that have smaller uncertainties, does not change the conclusions of the experiment.}.
\item new distance moduli are derived from these new peculiar velocities.
\end{enumerate}
We find that although the resulting peculiar velocity distribution presents the expected variance (by construction), close-by galaxies and groups have strongly biased distance moduli. Moreover, the correlation of velocities on small-scale velocity reaches extremely low values that are well below the expected values. In addition, bias b4 is fully present and more difficult to deal with. While it is possible to prevent from having it when starting from distance moduli to derive peculiar velocities, it is more difficult to extract non-b4-biased distance moduli from peculiar velocities as only one uncertainty is available (assumption of a symmetric distribution of the uncertainty). The consistency seen among all the panels of Figures \ref{fig:distmod} and \ref{fig:vpec} is another strong argument in favor of the algorithm capability.  \\

{Figure \ref{fig:biascheck} shows additional verifications of the algorithm results. In particular, it checks the reduction of  the effects of biases b2 and b3 (homogeneous and heterogeneous Malmquist biases). Bias b2 tends to scatter galaxies and groups closer to, rather than further away from, us. Although no data term enforces new distance moduli to be larger than starting ones, corrected distances should statistically be larger than biased ones. The left panel of the figure shows that indeed the median differences between after Iteration 1 (filled blue squares) and corrected (filled red circles) distances and biased distances (that have been used as a starting point) are statistically positive but at large distances. Errors on the median show that neither a positive nor a negative difference stands out at large distances. The effects of the catalog edges (e.g. sharp cut-off) indeed dominate over bias b2. Bias b3 reduces galaxy clustering by scattering objects from high to low density regions. Although no data term enforces datapoints to be close to each others\footnote{Indeed, the interaction term does not favor close-by datapoint configurations but if velocities are similar. Reversely, if they differ, it disfavors such close-by configurations.}, corrected distance moduli should increase back clustering. To derive an estimate of clustering, we build grids to split the Supergalactic coordinate space uniformly. We then fill in the grids with the catalog datapoints and proceed with a count-in-cell.  The top-right panel of Figure \ref{fig:biascheck} shows the resulting histograms of the number of datapoints per grid cells for the true (thick solid yellow line), biased (solid black line), after Iteration 1 (dotted blue line) and final or corrected (dashed red line) catalogs. The bottom panel shows the ratio of the different histograms to the true one (same color and linestyle codes). The thickness of the lines refers to the different grid-cell sizes. From the thickest to the thinner lines, the cell size ranges from $\sim$4.6 to 6.9 \hMpc. Because the catalogs are incomplete by construction, cells with zero datapoints are removed. On a few-megaparsecs scale, the biased catalog presents a strong excess (deficit) of cells with only one (several) datapoint(s) with respect to the true catalog. Conversely, the after Iteration 1 and corrected catalogs exhibit only a very small (if any) clustering difference with the true catalog.} \\

 \begin{figure*}
\vspace{0.cm}

\includegraphics[width=1 \textwidth]{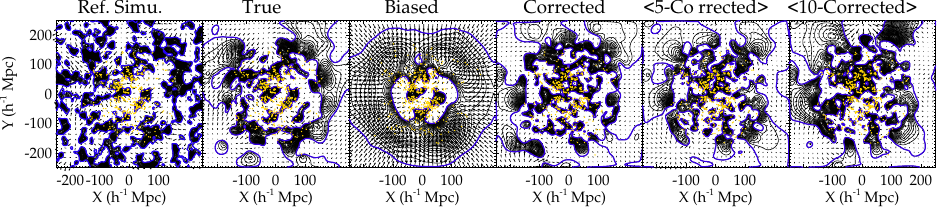}
\includegraphics[width=1 \textwidth]{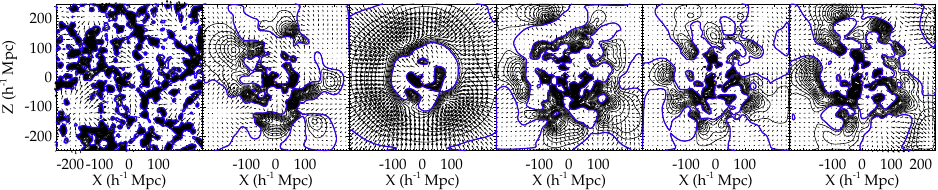}
\includegraphics[width=1 \textwidth]{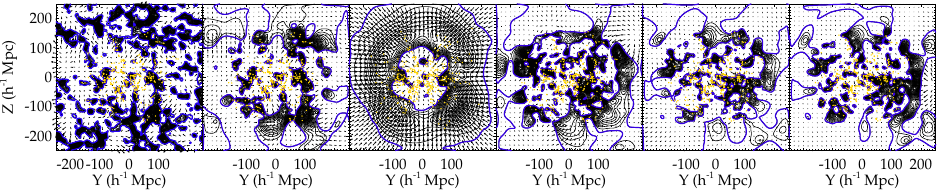}
\vspace{-0.4cm}

\caption{Supergalactic slices of the simulated (first column) and reconstructed (second to sixth columns) density and velocity fields. Contours show overdensities and arrows are velocities. The blue contours delimit the underdensities from the overdensities. Yellow points show halos constituting the different catalogs. The peculiar velocities of these halos, obtained from distance moduli, are actually used for the different reconstructions.}
\label{fig:mockWF}
\end{figure*}

{Figure \ref{fig:biascheck} confirms that the algorithm reduces  the effects of biases b2 and b3. Still, theoretically, the Metropolis-Hastings algorithm output is a sample distributed according to the probability distribution of interest. Similarly, the simulated annealing algorithm output is a sample distributed uniformly over the configuration subspace maximizing the probability distribution of interest. Under these circumstances, averaging realizations reduces the stochastic effects (variance) inherent to a single proposed solution.}

We stack several realizations of the corrected catalog obtained with the algorithm (i.e. slightly different realizations with a slightly different minimized energy term -- maximized probability density -- because of the \textit{a priori} non-concavity of the function). We treat each one of these realizations independently thus deriving a simple mean of their distance moduli and assigned uncertainties for each datapoint. We then derive the corresponding new velocities. Figure \ref{fig:stacked} shows the same plots as the right panel of Figure \ref{fig:befafter} and Figures \ref{fig:distmod} and \ref{fig:vpec} for five (left column) and ten (right column) stacked realizations. Small additional improvements are visible on a point-to-point basis. The mean length of the lines and its standard deviation are further decreased from $2.9\pm3.7$\hMpc\ to $2.2\pm2.7$\hMpc\ and $2.1~\pm~2.3$~\hMpc\ respectively. It is interesting to note that this is the characteristic average size of galaxy clusters. Overall, the major improvement is on the {small-scale} velocity variance (last panel of the last row in both columns). Note that stacking ten realizations rather than five does not seem to improve the {small-scale}  velocity variance anymore. It might even smooth the velocities a bit too much at least in the synthetic catalog case (see next section for the observational catalog). Stacking realizations permits also getting more realistic uncertainties on  the new distance moduli rather than globally converging towards zero ones. On the whole, the algorithm permits thus recovering statistically better distance moduli hence peculiar velocities for galaxies as per the peculiar velocity definition.\\

\begin{figure*}
\vspace{-0.5cm}
\centering
\includegraphics[width=0.45 \textwidth]{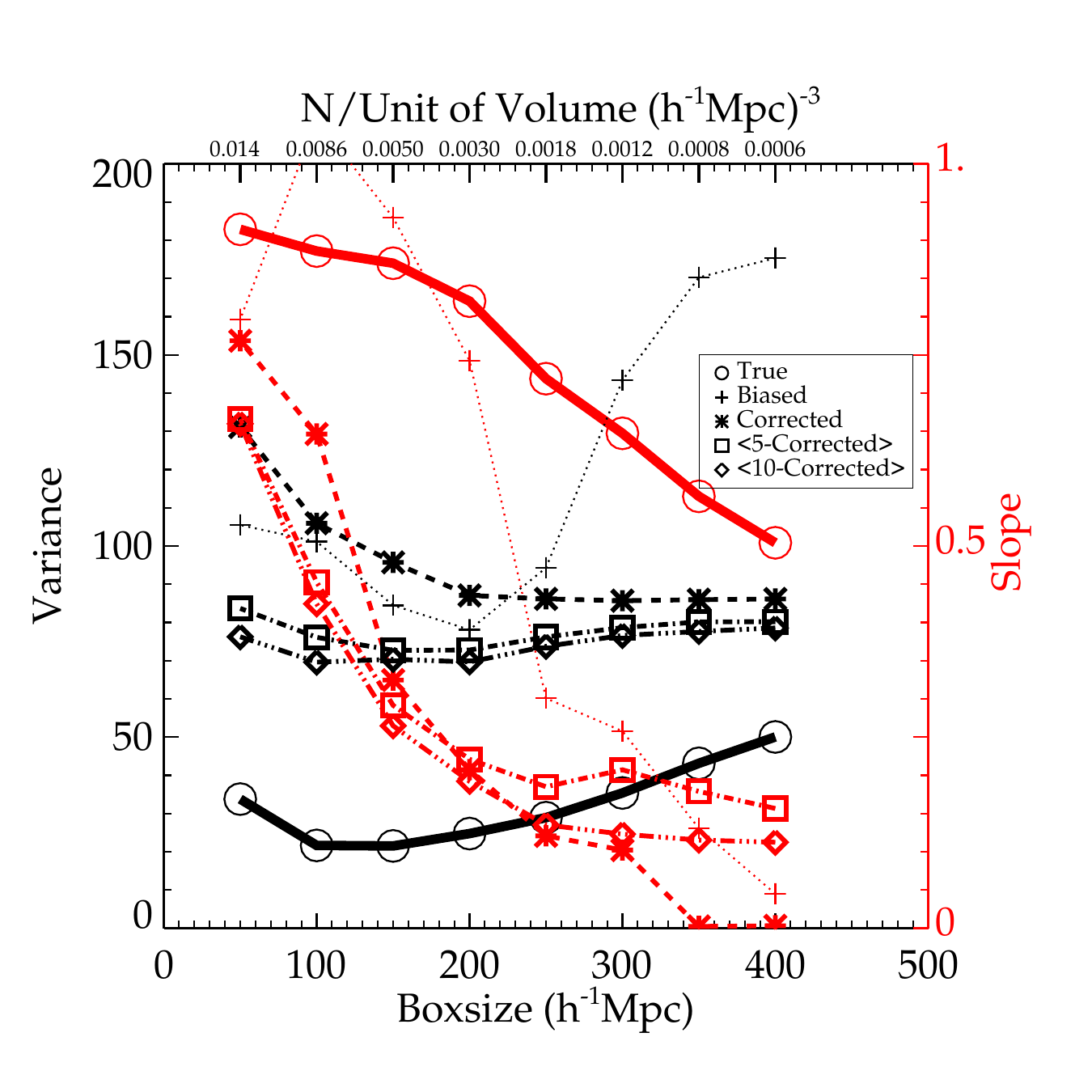}
 \includegraphics[width=0.45 \textwidth]{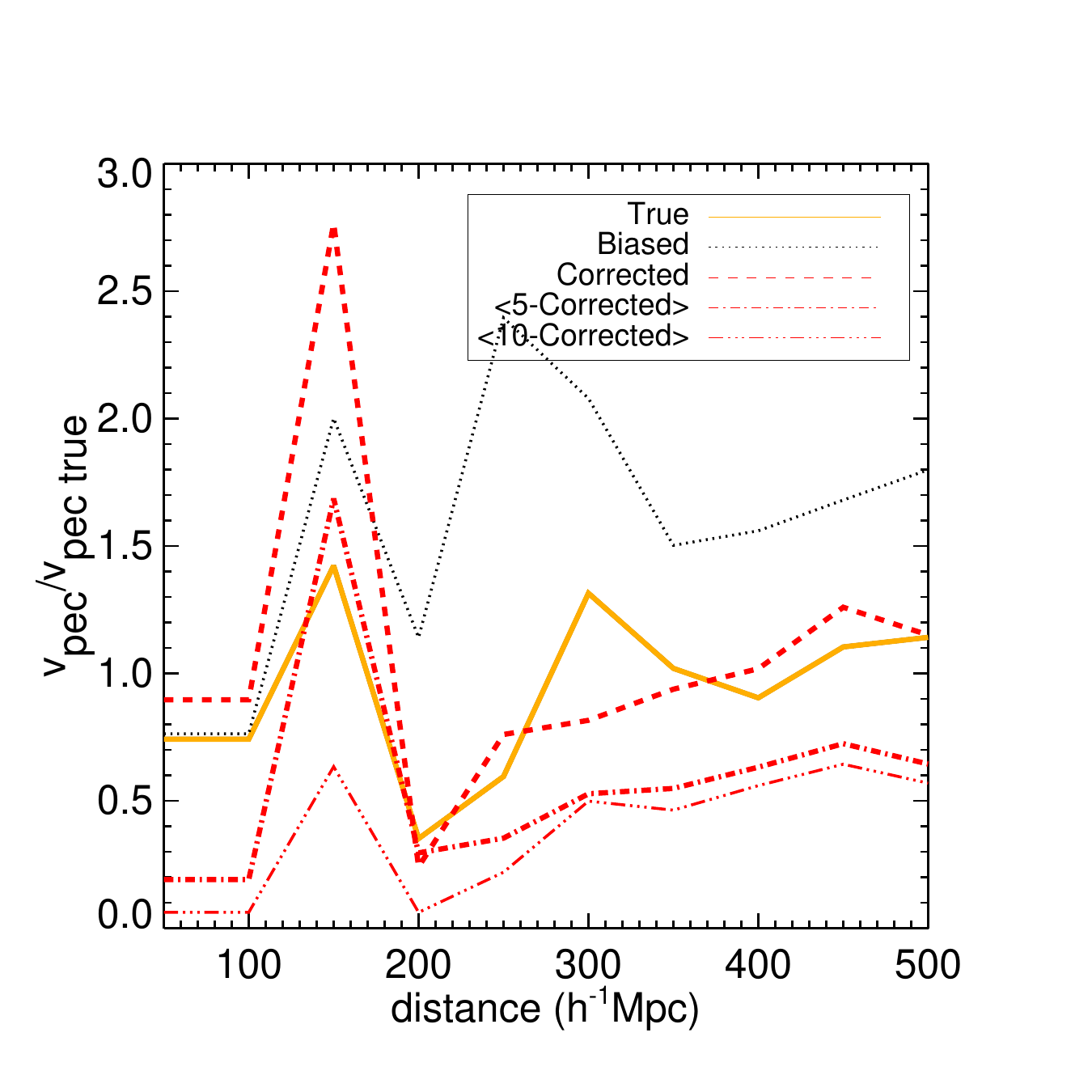}   \vspace{-0.4cm}
  
\caption{Comparisons of the velocity fields reconstructed from the different datasets. {\textit{Left}: Variances (black color) and slopes (red color) of linear fits to cell-to-cell simulated/reconstructed velocity field comparisons considering larger and larger sub-volumes of the total boxes. The variance, i.e. difference between reconstructed and simulated velocities, is the largest when using the biased catalog (black crosses) - especially when considering the largest sub-volumes - and the smallest when using the true catalog (black circles) to reconstruct the fields. It has intermediate values when the corrected catalogs (black stars, squares and diamonds) are used to reconstruct the fields. The slopes are smaller than 1 in all but one point for the biased-catalog-based reconstruction. It represents a well-known effect of the Wiener-filter technique that goes to the mean field in absence of data, implying reconstructed velocities with null values to be compared to simulated velocities in this case. \textit{Right}: Ratio between reconstructed and simulated velocities at the sole positions of the selected-for-the-mock-catalog halos as a function of the latter's distance. If the Wiener-filter tends to underestimate velocities when using corrected stacked-realization catalogs (dotted-dashed and three-dots-dashed red lines) for reconstruction, it does so quasi uniformly in the whole volume. It makes further corrections of Wiener-filter reconstructions easier than when using the biased catalog for the reconstructions. In the latter case (black dotted line), velocities are indeed alternatively under and overestimated with large fluctuations.}}  
\label{fig:compcell2cell}
\end{figure*}

\subsection{Mock field reconstructions}

Before applying the algorithm to the observational catalog, this subsection gives an example of how valuable the corrected catalog is as a whole. To that end, we reconstruct density and 3D velocity fields from the true, biased and corrected (both single and stacked) catalogs using the Wiener-filter technique \citep{1999ApJ...520..413Z}. {This technique is known for not being able to deal with the different biases.} Reconstructed fields are then compared to the initial simulation from which the synthetic catalog is built. Figure \ref{fig:mockWF} shows the three Supergalactic slices of the reconstructed local Universe obtained with the different catalogs of peculiar velocities. Black contours stand for the overdensity field. The thick blue solid lines delimit the overdensity from the underdensity. Arrows stand for the velocities. Yellow dots are datapoints constituting the synthetic catalogs. The biased catalog results in the worst reconstruction with round structures at the edge and a large infall onto the observer / center of the box. {Reconstructions based on corrected catalogs present more defined structures like for the reconstruction obtained with the true catalog.} In addition, the major infall onto the local Volume is suppressed. \\

{To quantify the similarity between the reconstructed fields and the simulation, cell-to-cell comparisons between simulated and reconstructed velocity fields are conducted. For each successive cell-to-cell comparison between two velocity fields, cells are selected in a larger and larger sub-volume of the reconstruction and simulation boxes. A linear fit to each one of the cell-to-cell comparison plots (reconstructed vs. simulated velocities) permits deriving the slope of the correlation between simulated and reconstructed x, y, z velocity components as well as its variance. Figure \ref{fig:compcell2cell} left shows the variances (black lines and symbols) and slopes (red lines and symbols) of the linear fits to the relations between reconstructed and simulated velocities. They are obtained by comparing cells in different sub-volumes of the boxes. As expected, the best (worst) reconstruction is obtained with the true (biased) catalog: the variance - difference between reconstructed and simulated velocity fields - is the smallest (largest) being below 50~km~s$^{-1}$ (about 100-170~km~s$^{-1}$) and the slope is the closest to 1 - i.e. almost a perfect match - (varying the most - i.e. no correlation)  as shown by the open black and red circles (crosses) respectively. Note that the slopes overall decrease with the increasing compared sub-volumes. As for reconstructions obtained with bias-minimized catalogs, those obtained with stacked realizations present linear fits to cell-to-cell comparison plots with smaller variances than that obtained with a single realization}: 80 against 100~km~s$^{-1}$ (black squares and diamonds with respect to stars). Slopes (same red symbols) present the same trend as for the true catalog. Namely, the larger the sub-boxsize considered for the cell-to-cell comparison is, the smaller the slope is. This is a Wiener-filter known effect as it goes to the mean field in absence of data. Indeed, the number of datapoints (N) per Unit of Volume drastically decreases with the distance to the center of the box (top axis of the figure).  \\

To confirm this effect, Figure \ref{fig:compcell2cell} right panel compares velocities only at selected-for-the-mock-catalog halo positions in the simulation and in the reconstructions. It highlights that indeed reconstructed velocities with the biased catalog are too large at large distances and too small close by with a large disparity (black dotted line). Reconstructed velocities obtained with the true and corrected catalogs are alternatively slightly too small, slightly too large (yellow and dashed red lines). Those obtained from the corrected stacked-realization catalogs are overall too small (dotted-dashed and three-dots-dashed red lines). Interestingly the ratio between reconstructed and simulated velocities has the smallest variance for the reconstructions obtained with stacked realizations making it easier to correct for the smoothing effect of the Wiener-filter.

 \begin{figure*}
 \vspace{-1cm}
\hspace{1.3cm}\includegraphics[width=0.3 \textwidth]{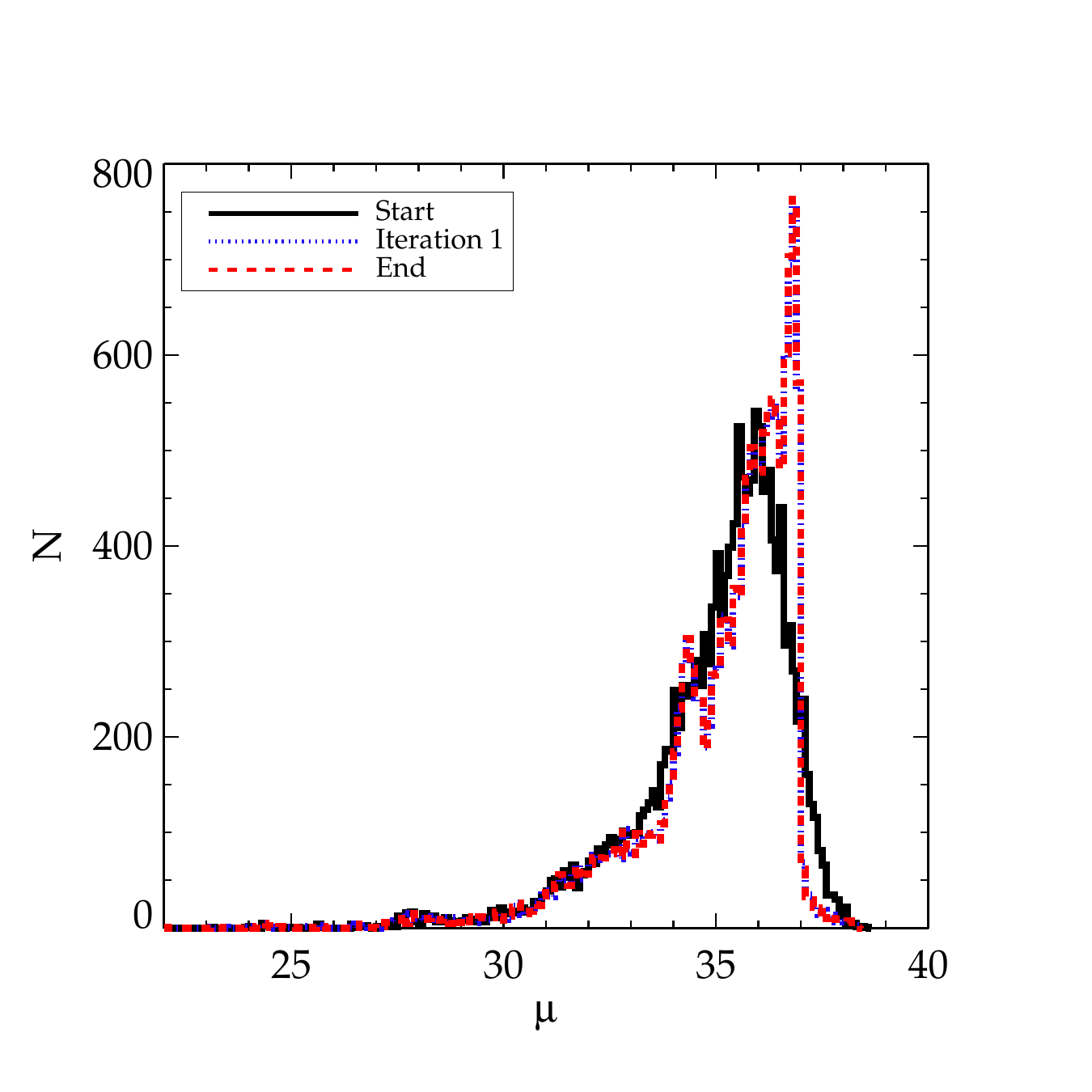}
\hspace{-0.8cm}\includegraphics[width=0.3 \textwidth]{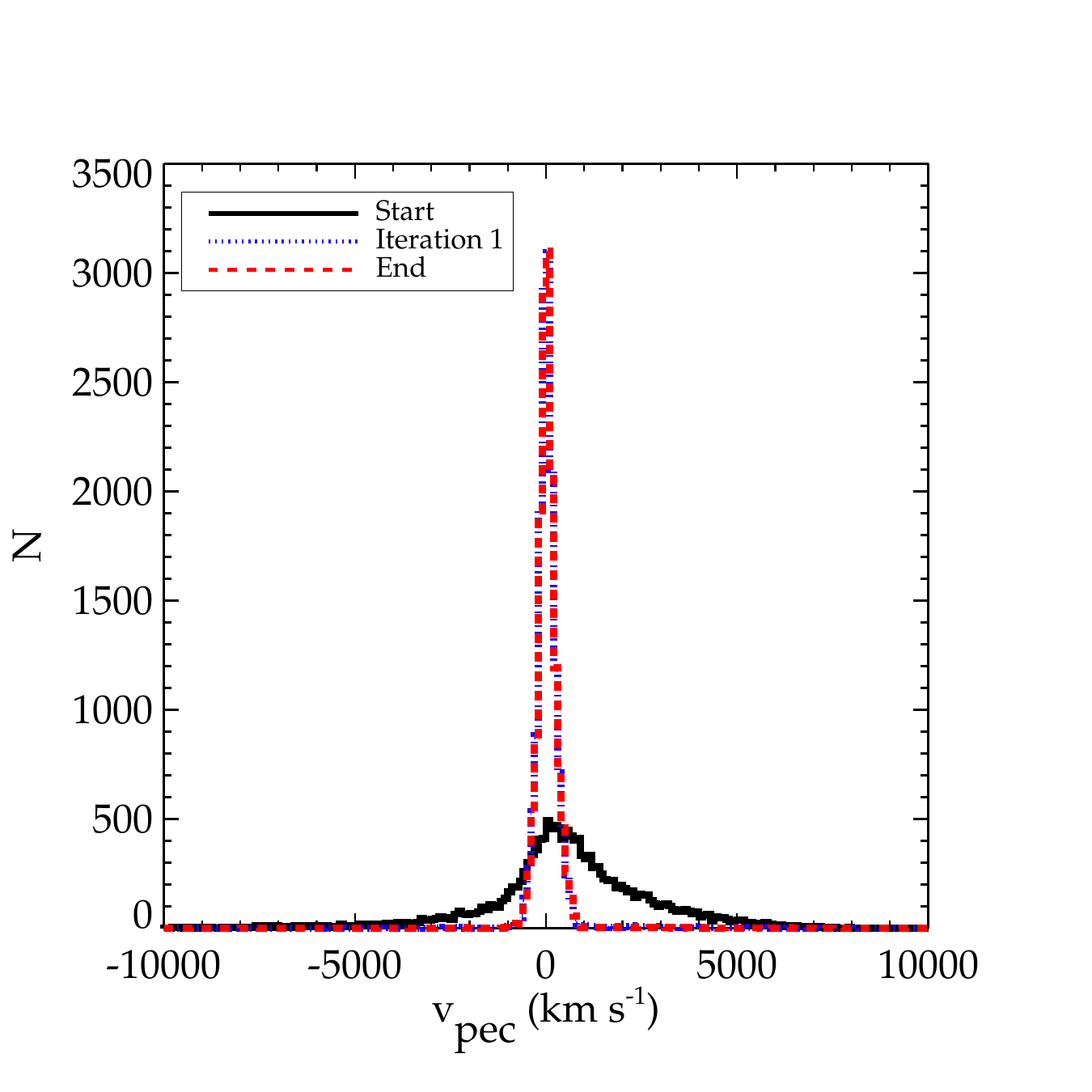}
\hspace{-0.8cm}\includegraphics[width=0.3 \textwidth]{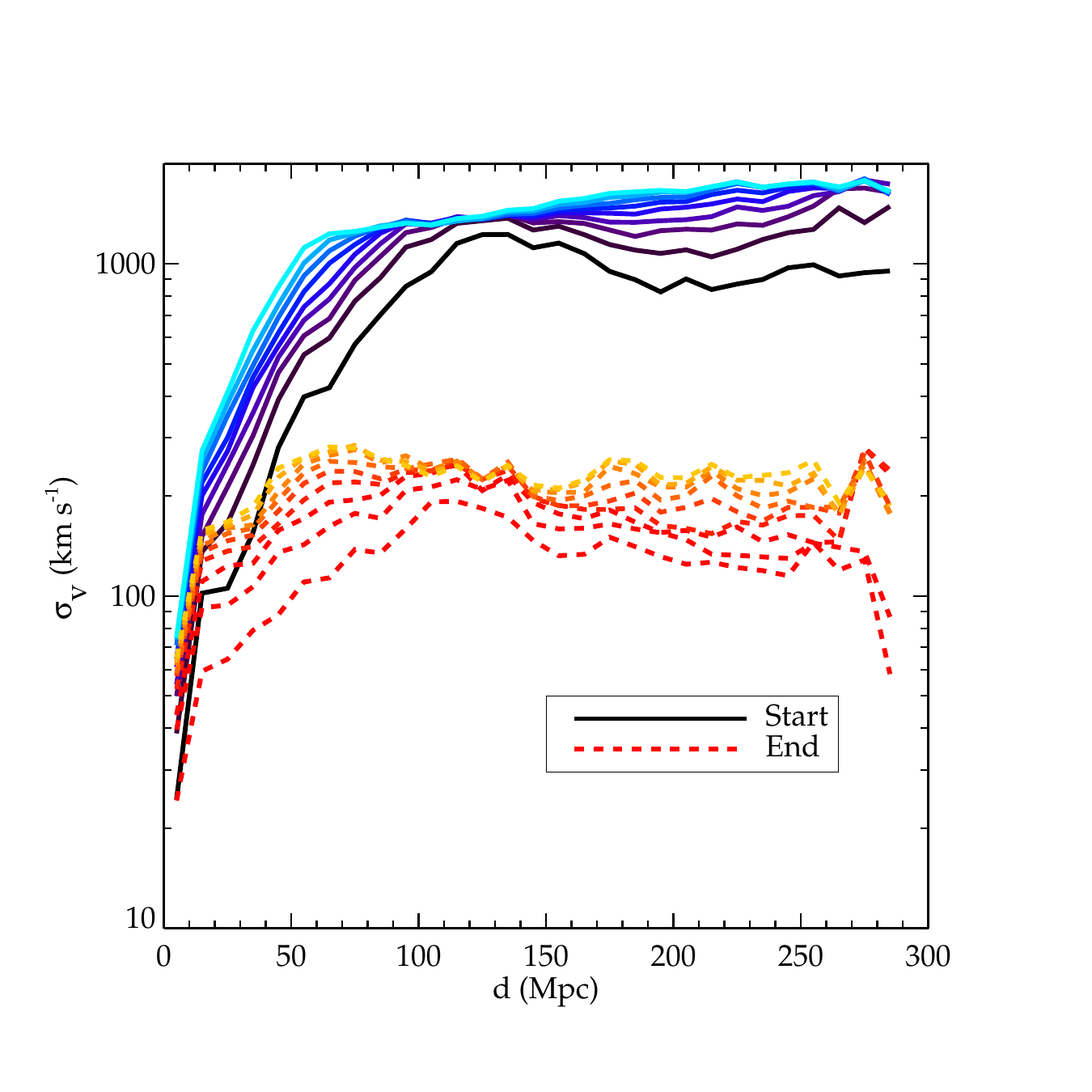}\\

\vspace{-0.8cm}
\hspace{-0.5cm}\includegraphics[width=0.2 \textwidth]{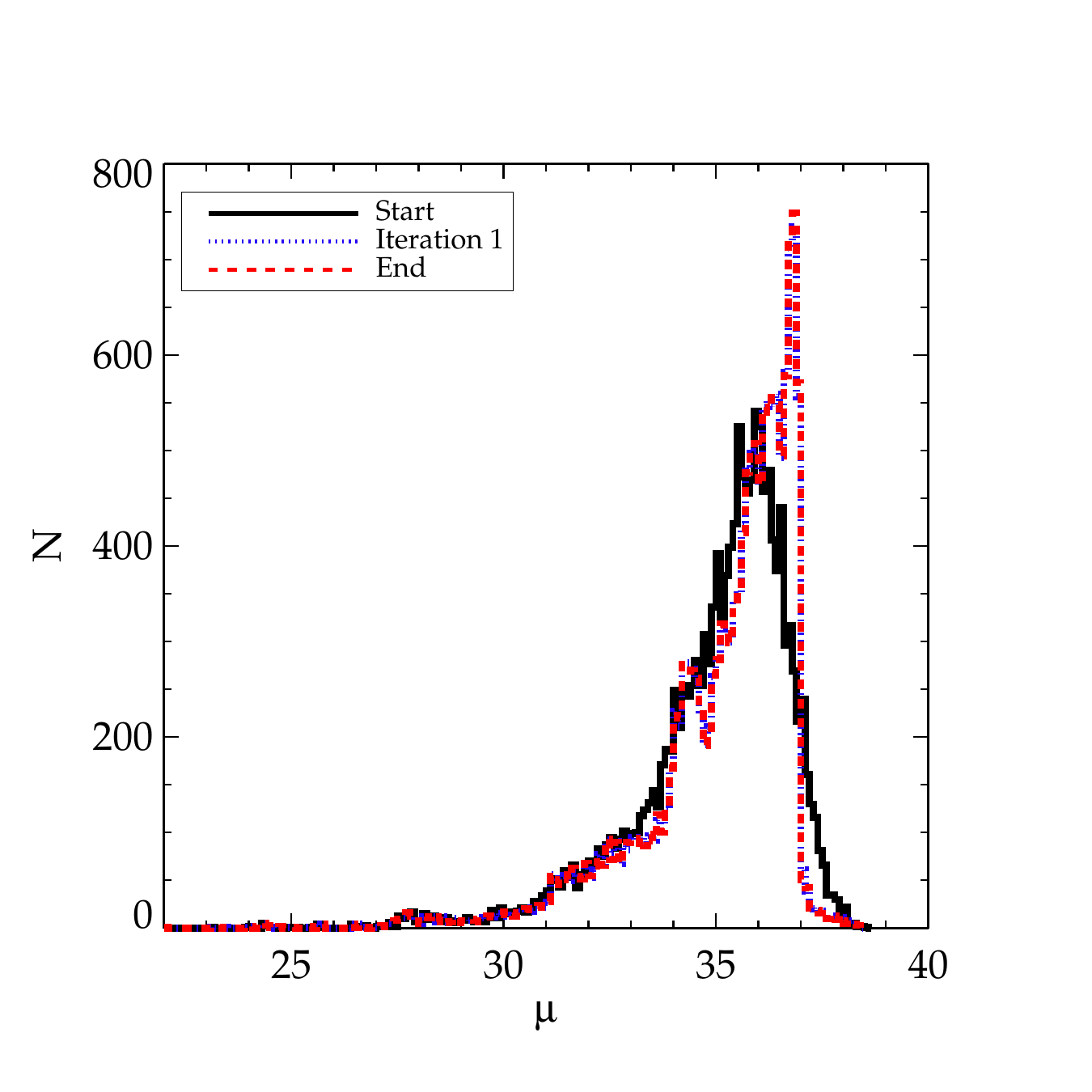}
\hspace{-0.75cm}\includegraphics[width=0.2 \textwidth]{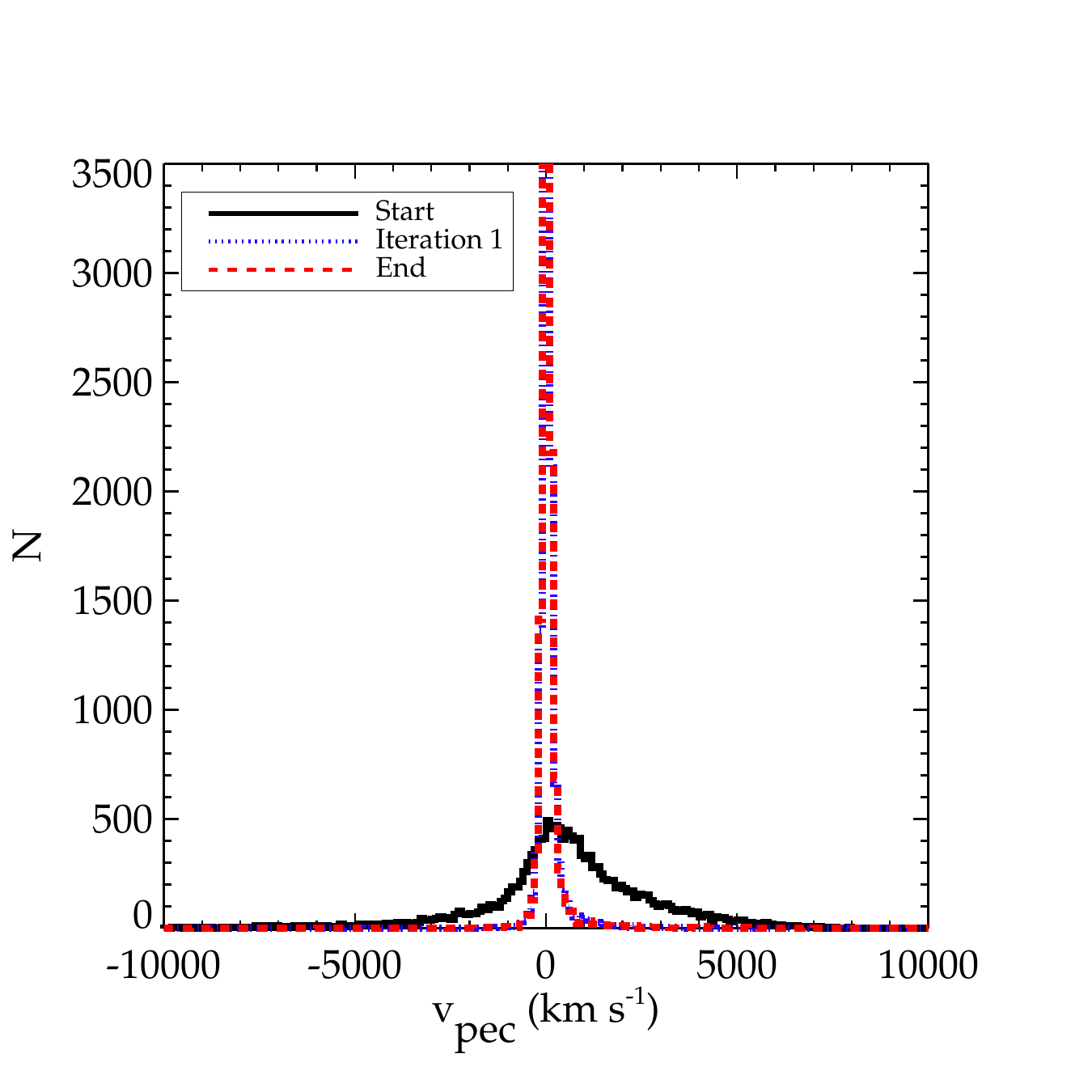}
\hspace{-0.75cm}\includegraphics[width=0.2 \textwidth]{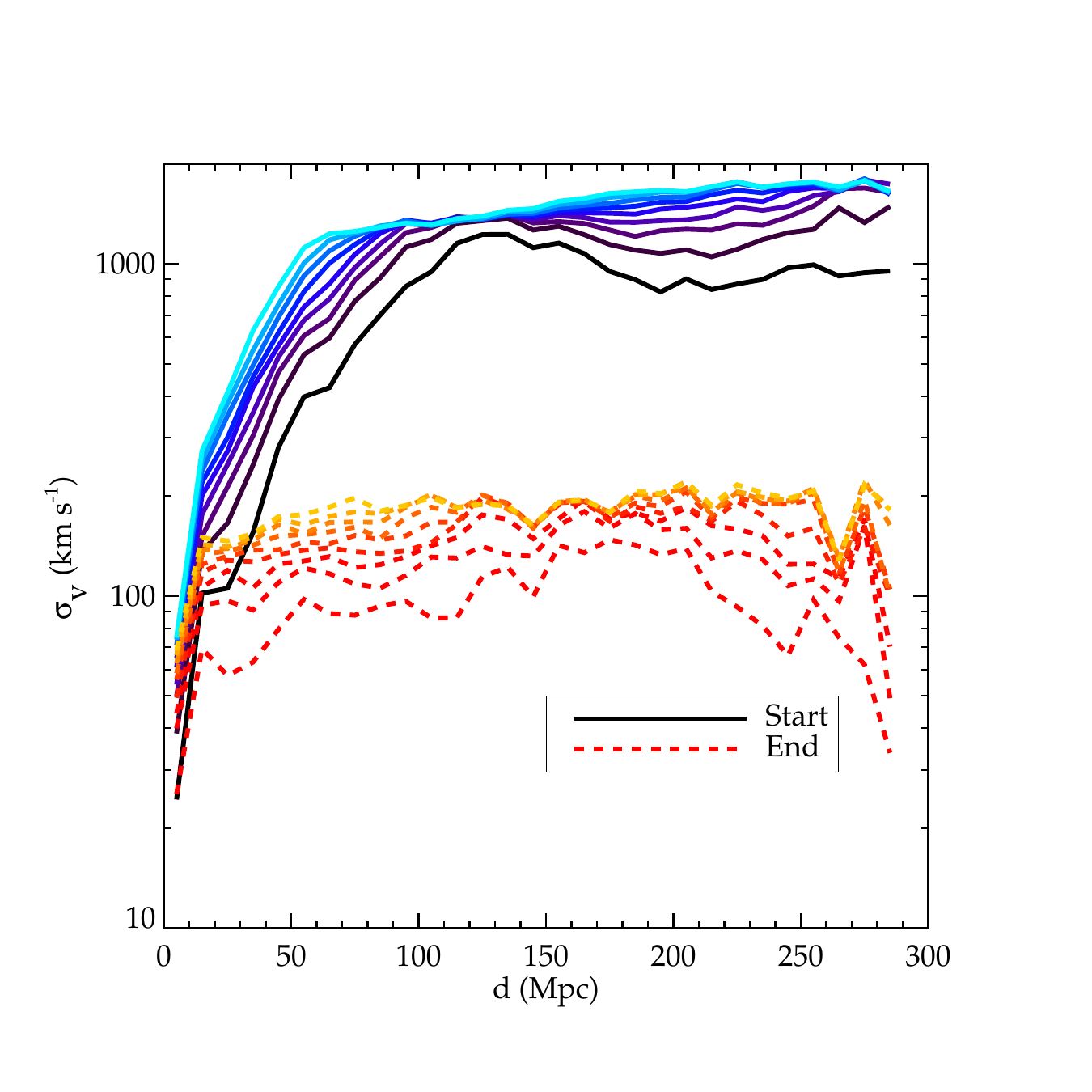}
\hspace{-0.6cm}\includegraphics[width=0.2 \textwidth]{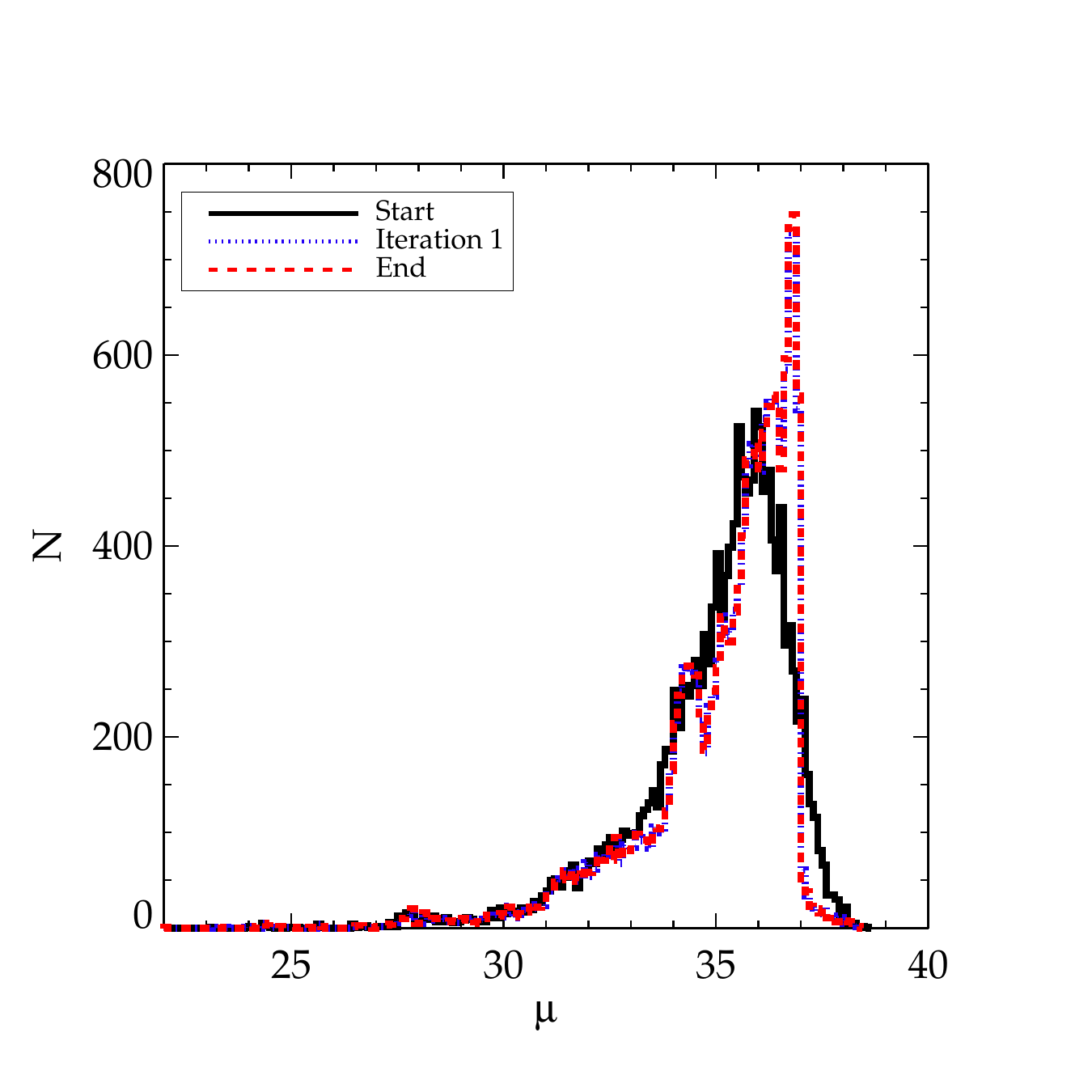}
\hspace{-0.75cm}\includegraphics[width=0.2 \textwidth]{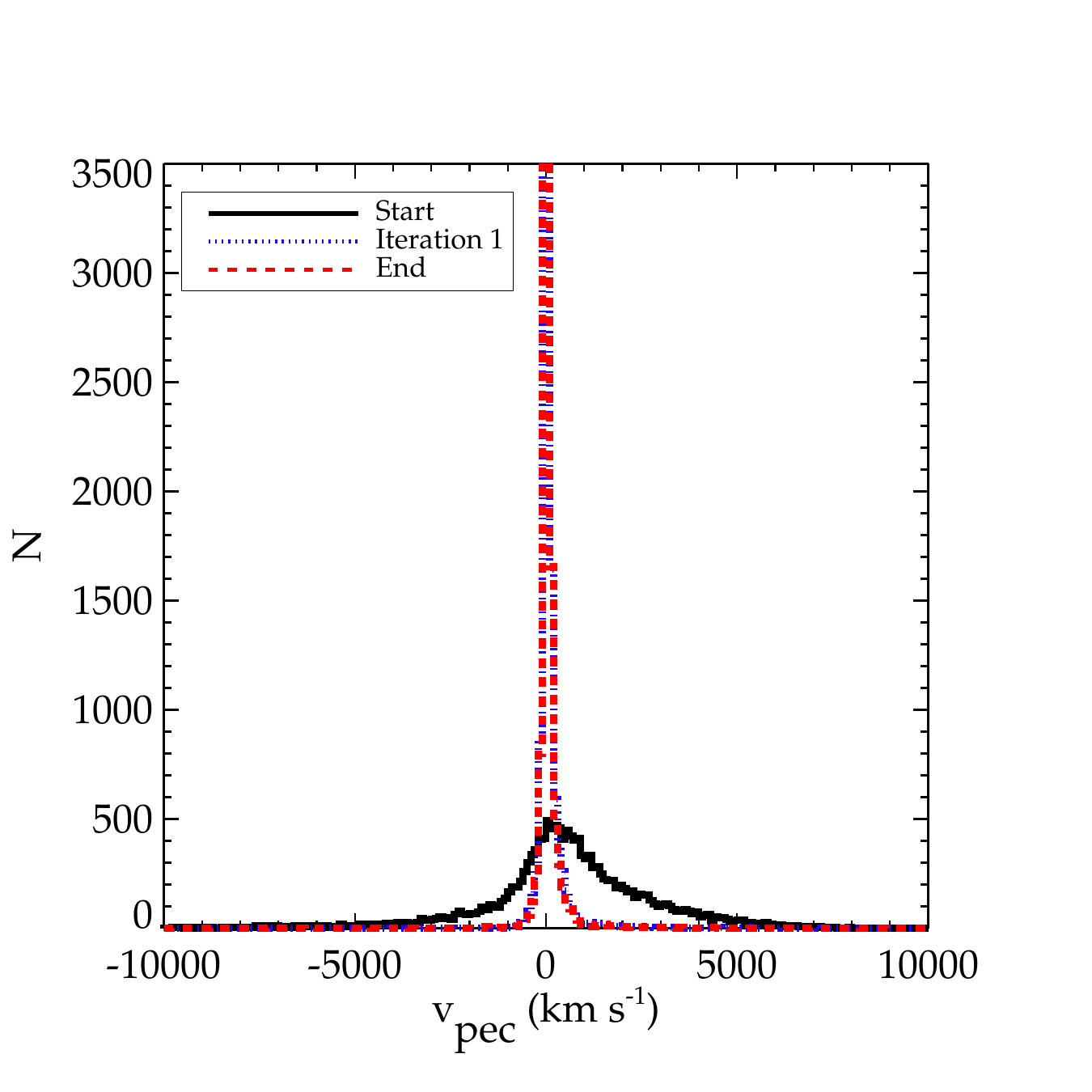}
\hspace{-0.75cm}\includegraphics[width=0.2 \textwidth]{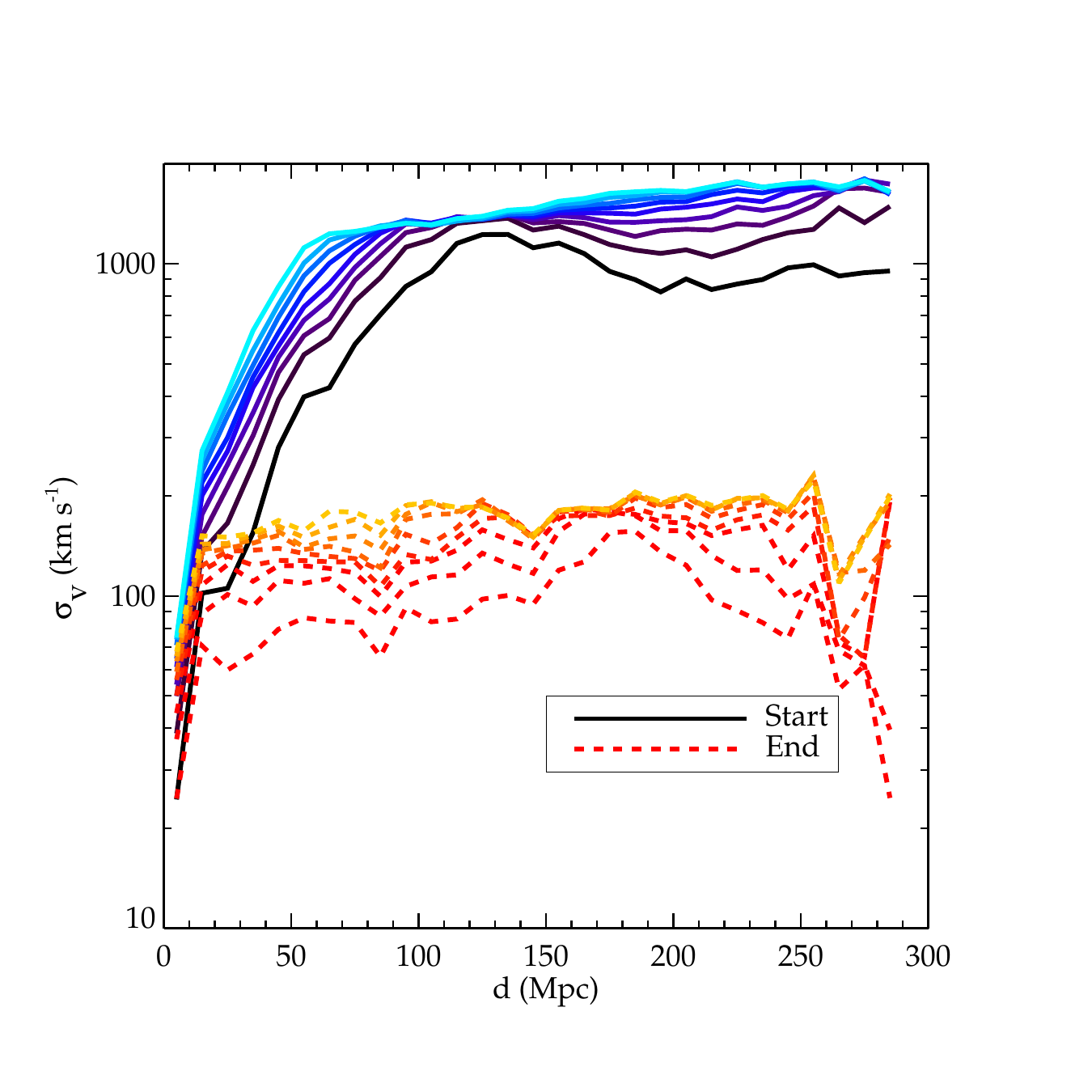}
\caption{Comparisons between spatial distribution and velocity properties of the observed catalog before and after bias-minimization. \textit{From left to right}: Distance modulus histograms, peculiar velocity distributions obtained from the raw  (black solid line) and corrected (blue and red dotted and dashed lines) observational catalogs as well as the {small-scale} velocity variance (solid cold color vs. dashed warm color lines respectively). \textit{Top to bottom left and right}: results for one distribution, for five and ten stacked realizations.}
\label{fig:cf3corr}
\end{figure*}

{Indeed, previously, to obtain better 3D reconstructed velocities, \citet{2015MNRAS.450.2644S} multiplied all the mock-catalog velocities by the average decrease caused by the Wiener-filter technique. Then, applying the Wiener-filter technique to this mock catalog, they got an average slope of 1 for the linear fits to the cell-to-cell comparisons between the simulated and reconstructed velocity fields. To obtain a slope value of 1 independently of the sub-boxsize used for the cell-to-cell comparisons, \citet{2018MNRAS.478.5199S} considered the tri-dimensional volume and applied an additional smoothing (uncertainties) inversely proportional to the number of points per sub-volume. We leave further comparisons as well as these additional steps with potential new improvements for the next paper of the series. The major improvement in the reconstructions obtained with the corrected catalogs is already visible: the difference between the reconstructed and simulated velocity fields or variance is drastically reduced whatever the subvolume considered is.}

 \begin{figure*}
 
 \vspace{-0.5cm}
 
\hspace{-1.1cm}\includegraphics[width=0.4 \textwidth]{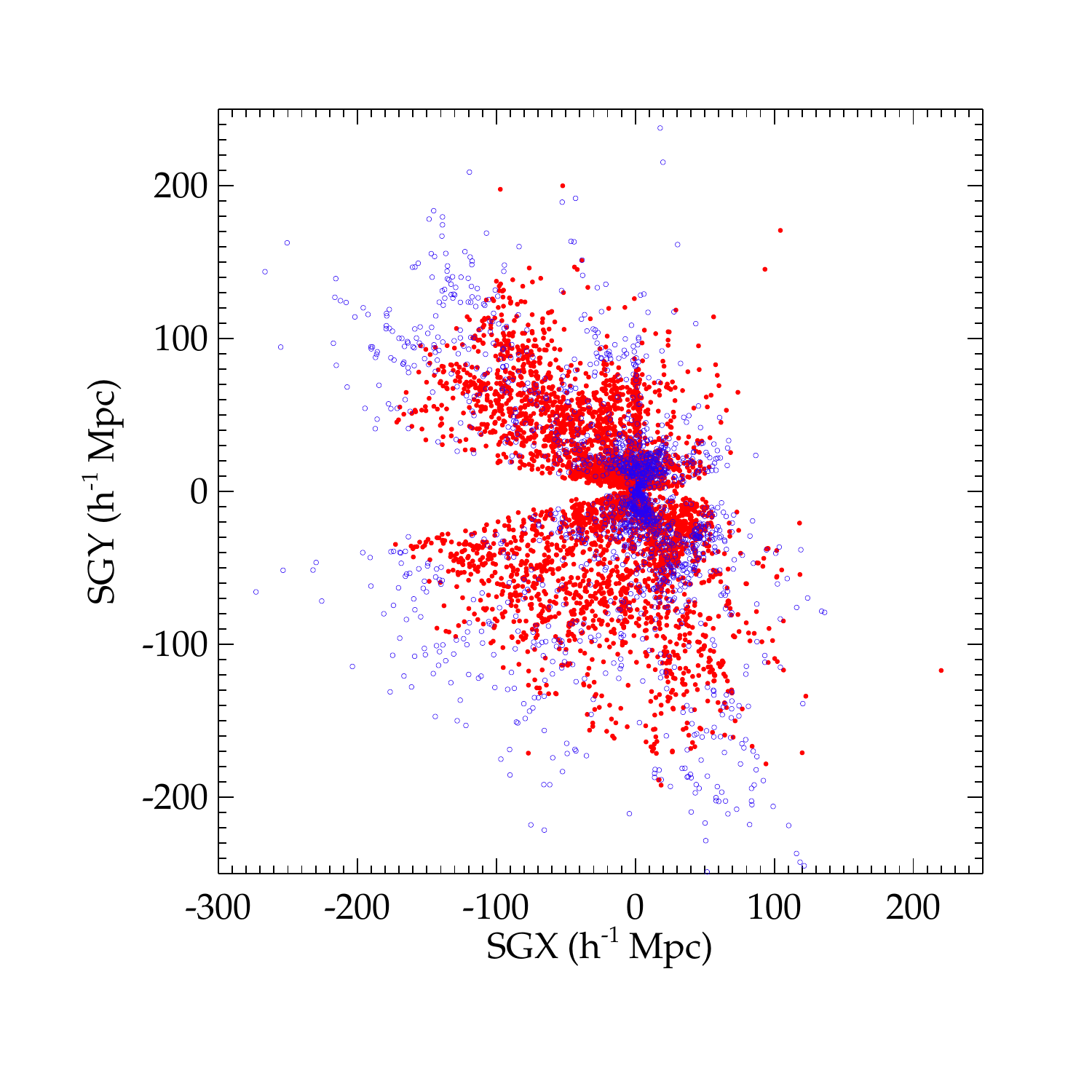}
\hspace{-0.72cm}\includegraphics[width=0.4 \textwidth]{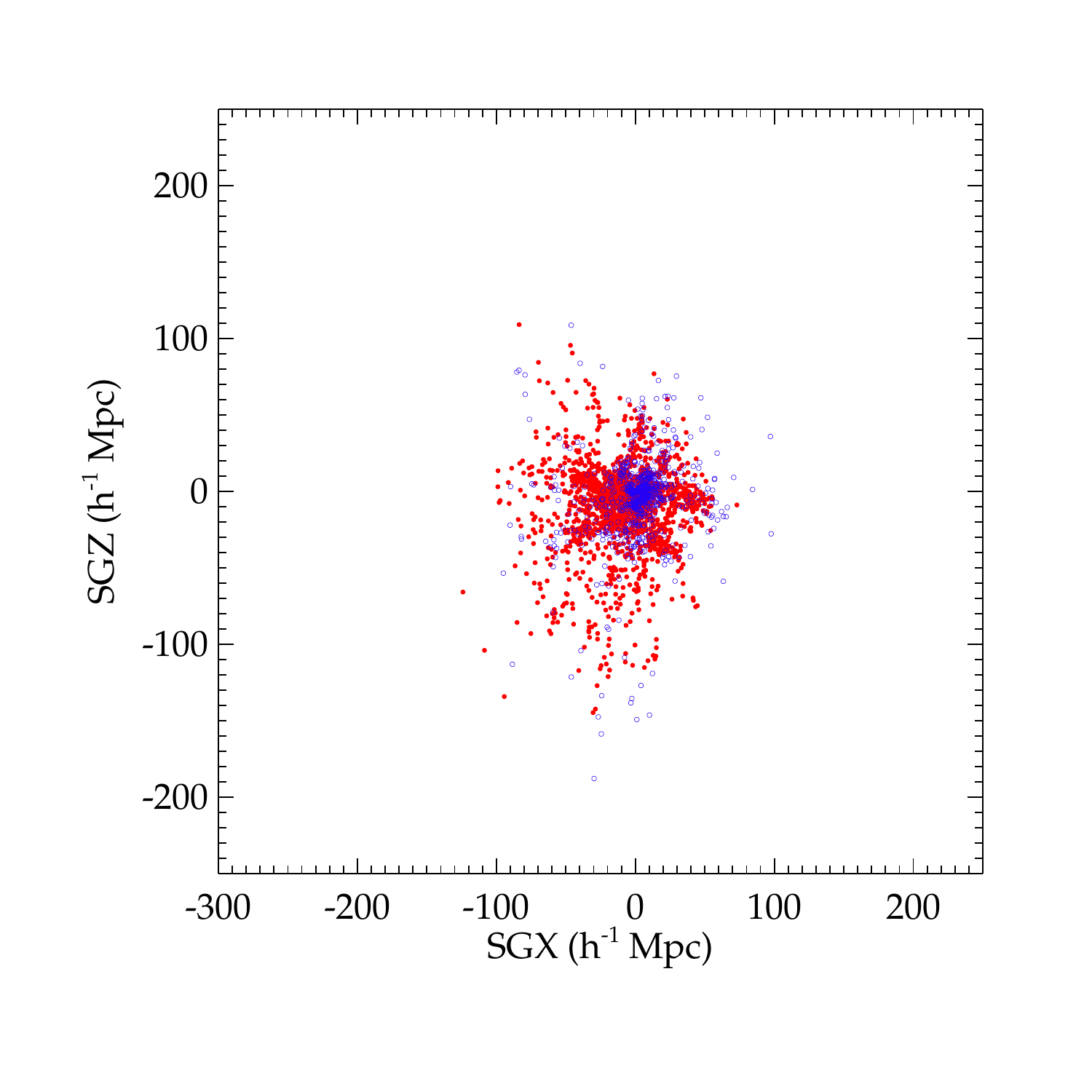}
\hspace{-0.72cm}\includegraphics[width=0.4 \textwidth]{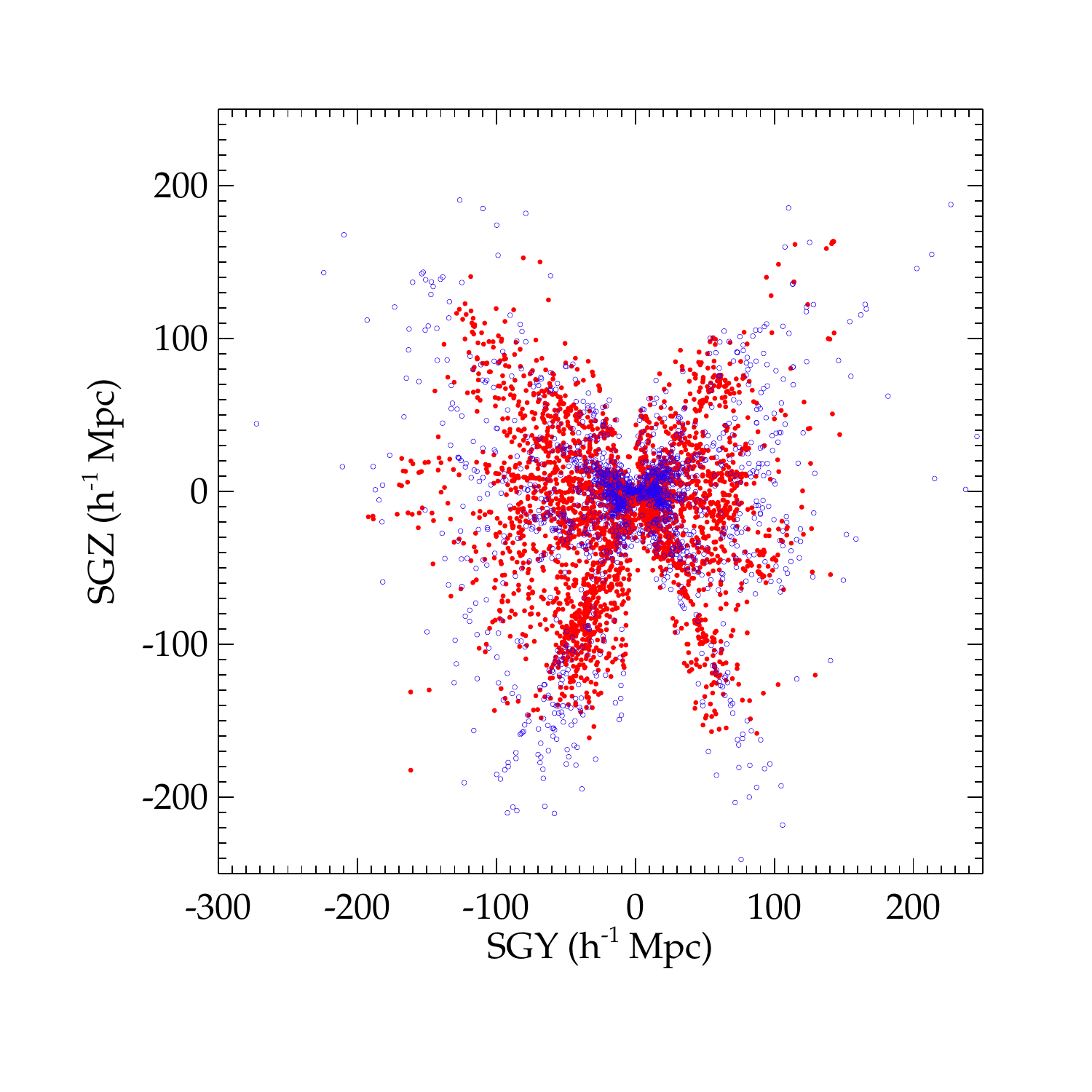}\\

\vspace{-1.6cm}

\hspace{-1.1cm}\includegraphics[width=0.4 \textwidth]{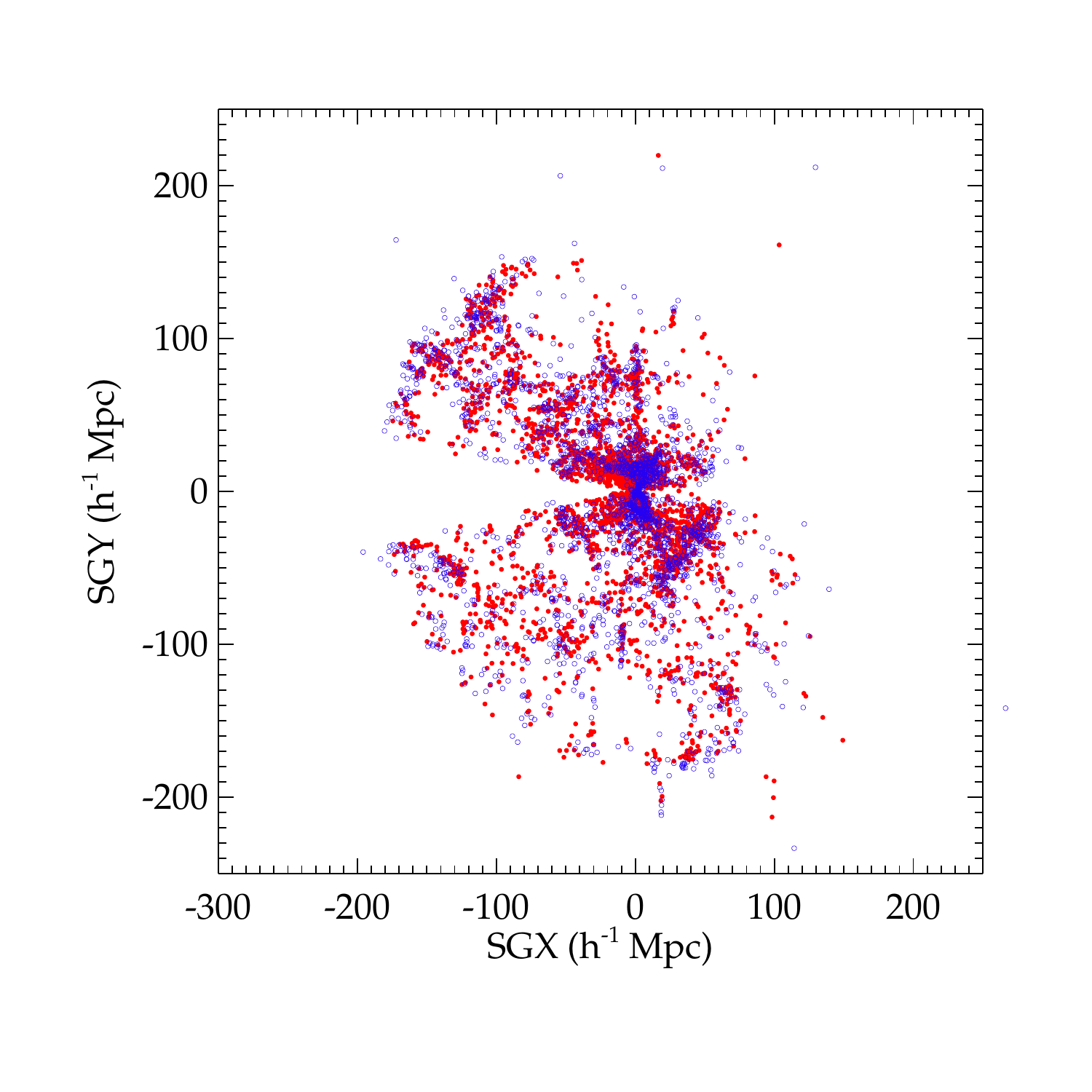}
\hspace{-0.72cm}\includegraphics[width=0.4 \textwidth]{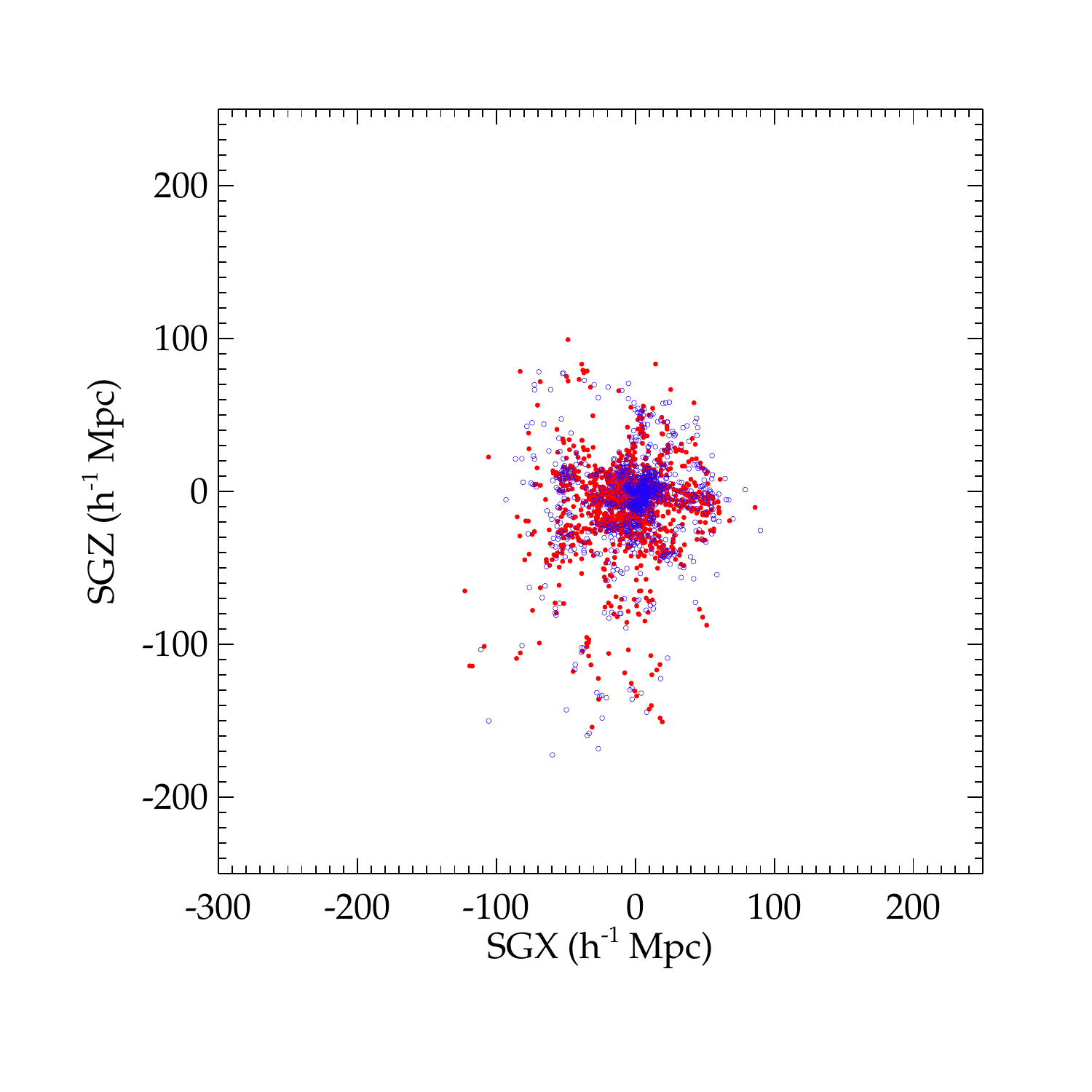}
\hspace{-0.72cm}\includegraphics[width=0.4 \textwidth]{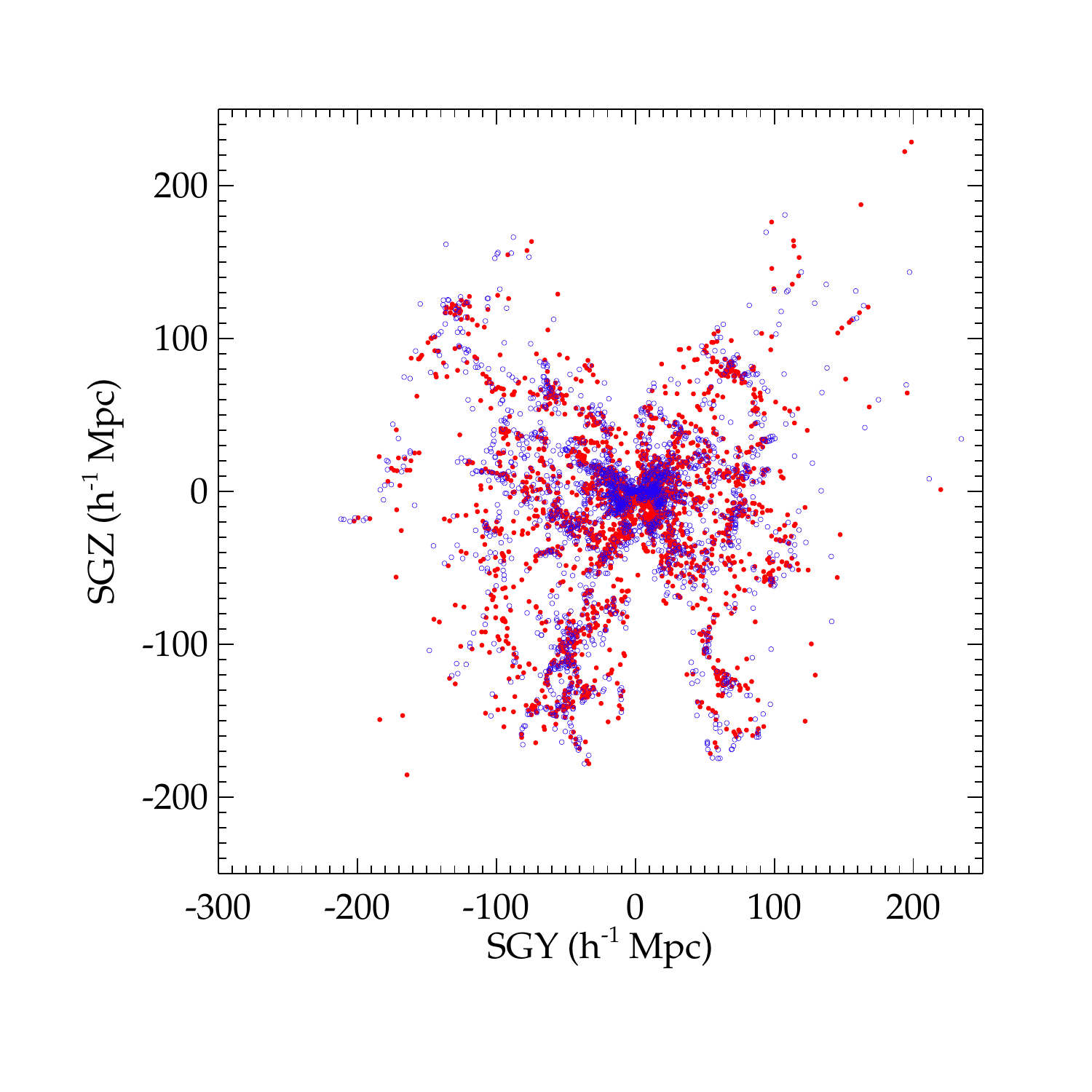}\\

\vspace{-1.6cm}
\hspace{-1.1cm}\includegraphics[width=0.4 \textwidth]{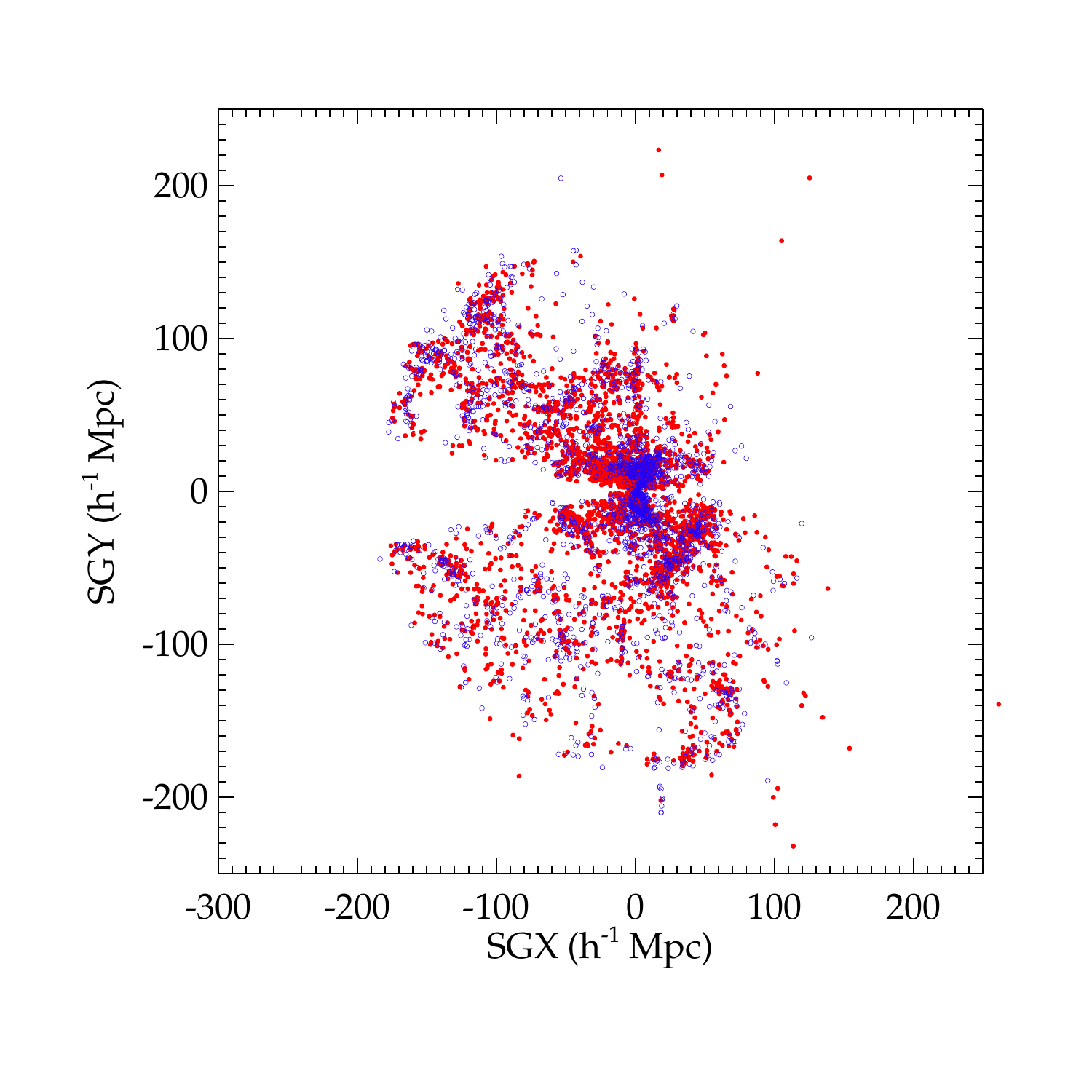}
\hspace{-0.72cm}\includegraphics[width=0.4 \textwidth]{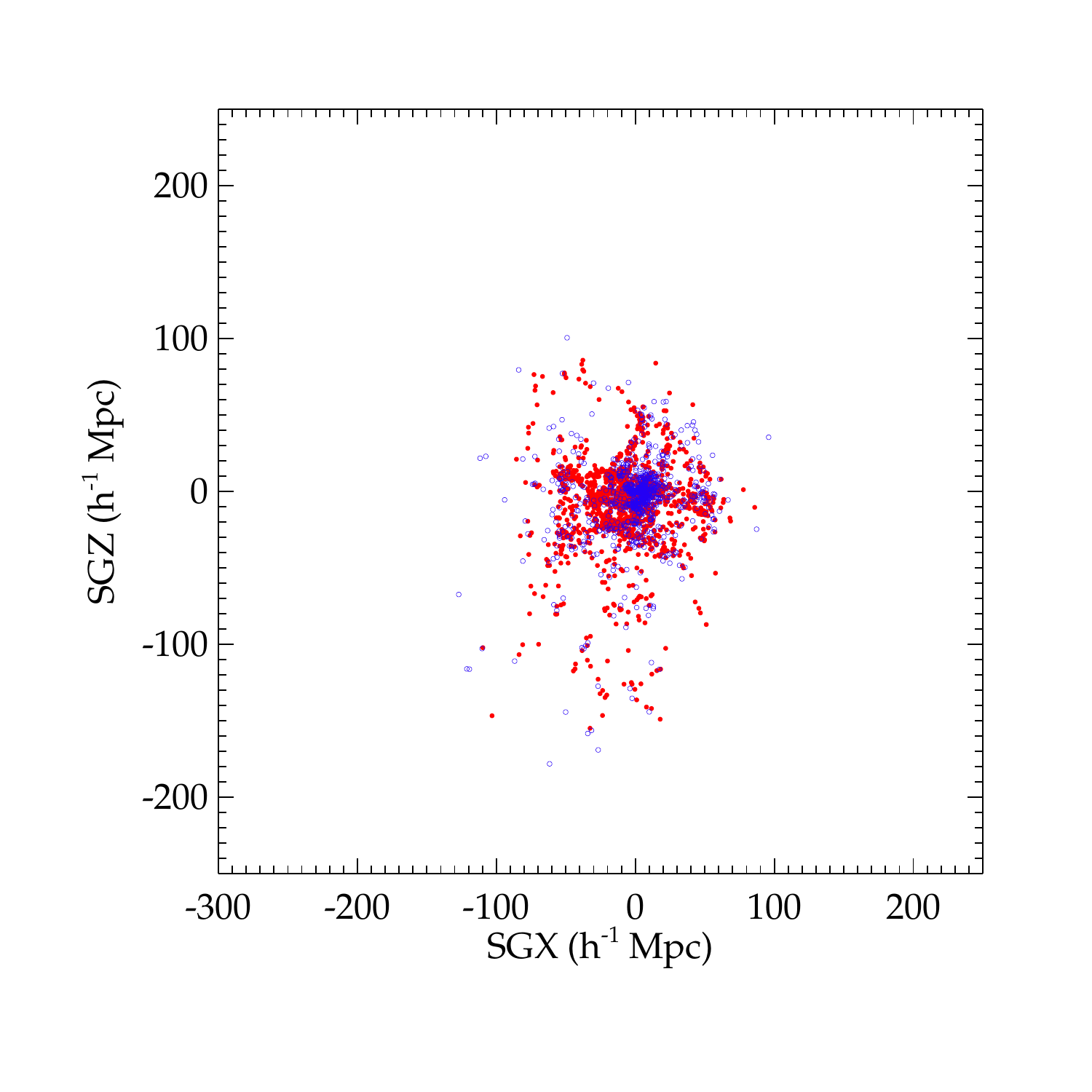}
\hspace{-0.72cm}\includegraphics[width=0.4 \textwidth]{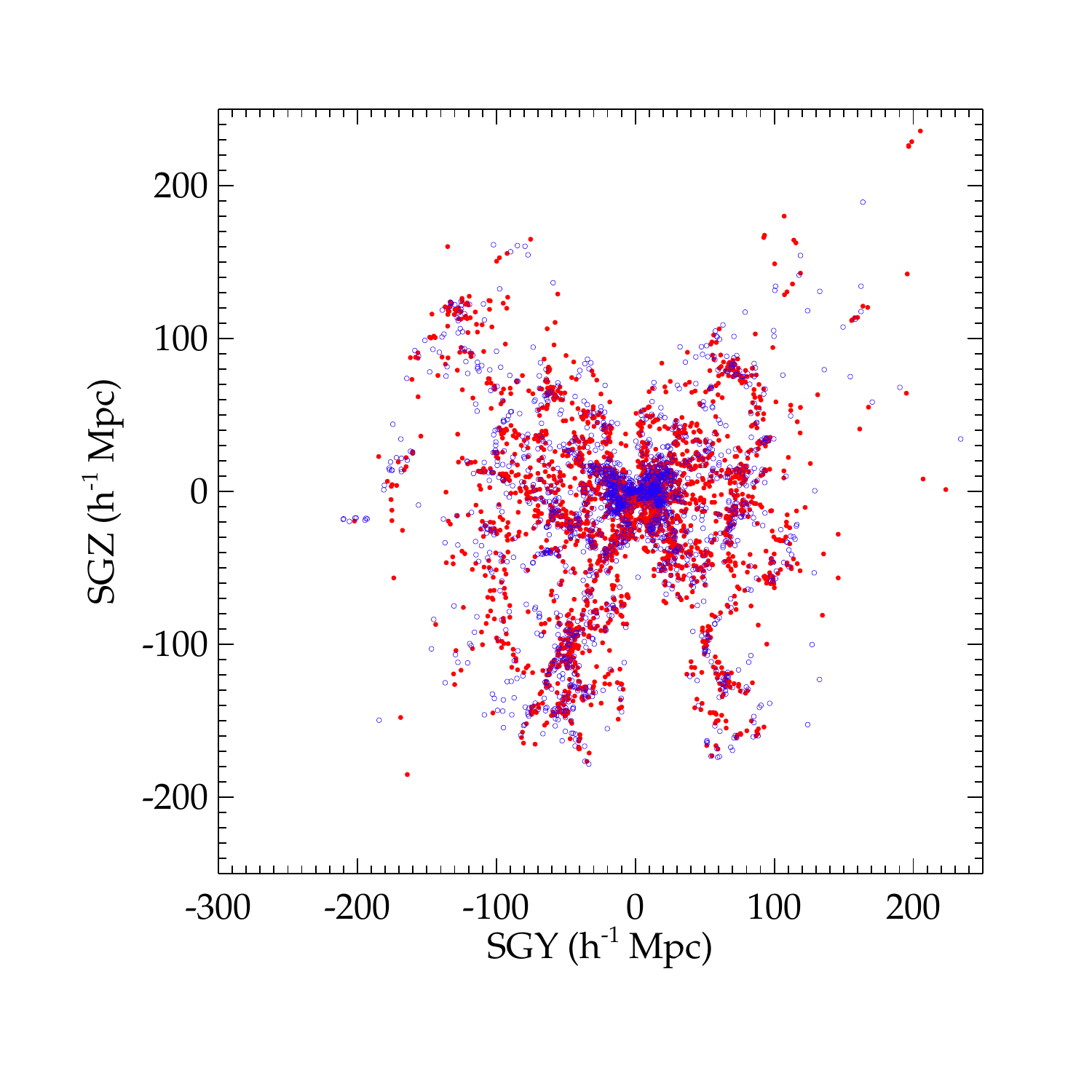}\\
 
 \vspace{-1cm}
 \caption{Distribution of galaxies. {\textit{From top to bottom}: in the raw, corrected, ten stacked realization third catalog of the Cosmicflows project in the three supergalactic 40~\hMpc\ thick slices (from left to right). Blue circles stand for galaxies with negative peculiar velocities. Red filled circles represent galaxies with positive peculiar velocities. The strong biases affecting the raw catalog (top) is clearly visible with negative velocities at large distances. The filamentary structure of the cosmic web is visible in the corrected catalogs}.}
\label{fig:distribvel}
\end{figure*}

\section{Application to the observational catalogs}

In this section, the algorithm is applied to the second and third catalogs of the Cosmicflows project. Again, for the sake of conciseness and results being identical for the two catalogs, results are shown only for the third catalog.

 \begin{figure*}
 \vspace{-0cm}\centering
\includegraphics[width=0.9\textwidth]{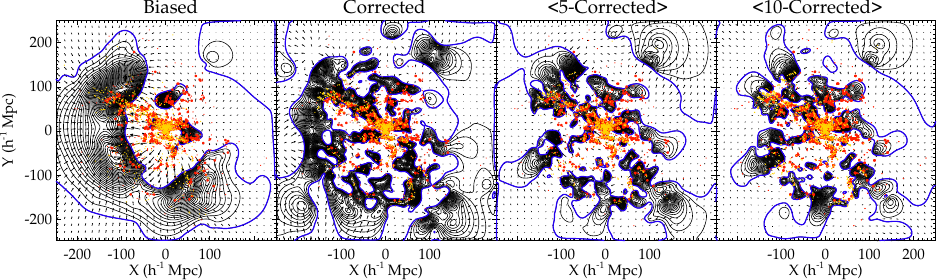}
\includegraphics[width=0.9\textwidth]{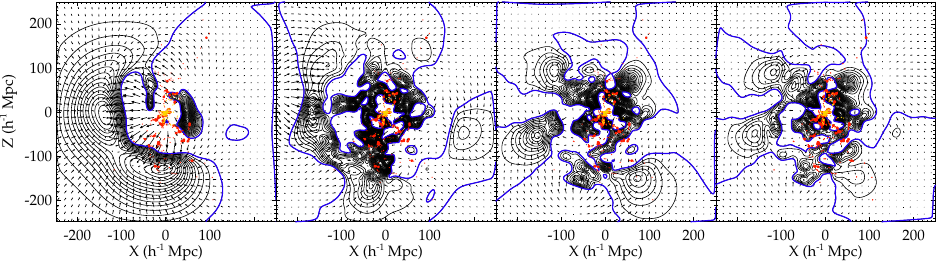}
\includegraphics[width=0.9\textwidth]{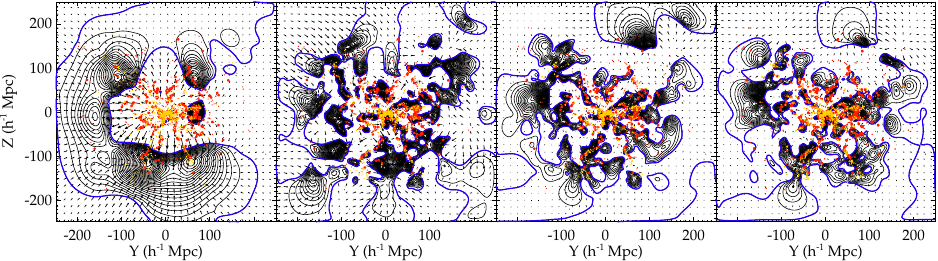}
\vspace{-0.1cm}

\caption{Supergalactic slices of reconstructed density (contours) and velocity (arrows) fields of the local Universe. The blue color delimits over- from under- densities. Red points are galaxies (small filled circles) and groups (larger filled circles) from the 2MRS Galaxy Redshift Catalog for comparison purposes only  (\citealp[2MRS,][]{2012ApJS..199...26H}; \citealp[groups from][]{2018A&A...618A..81T}). Yellow points show galaxies whose peculiar velocities obtained from distance moduli are actually used for the reconstructions. The bias effects are reduced in reconstructed fields obtained with corrected catalogs.}
\label{fig:cf3wf}
\end{figure*}

 \begin{figure*}
 \vspace{-0cm}\centering
\includegraphics[width=0.9\textwidth]{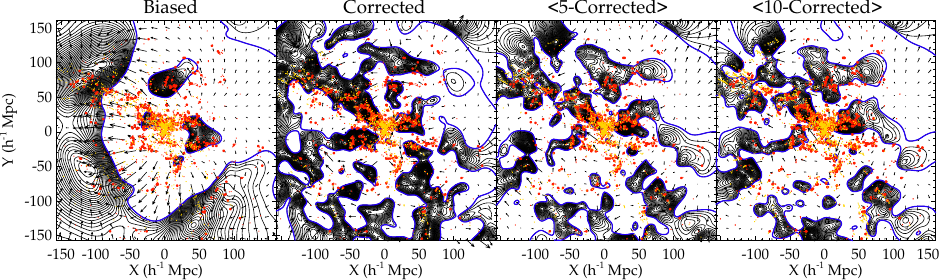}
\includegraphics[width=0.9\textwidth]{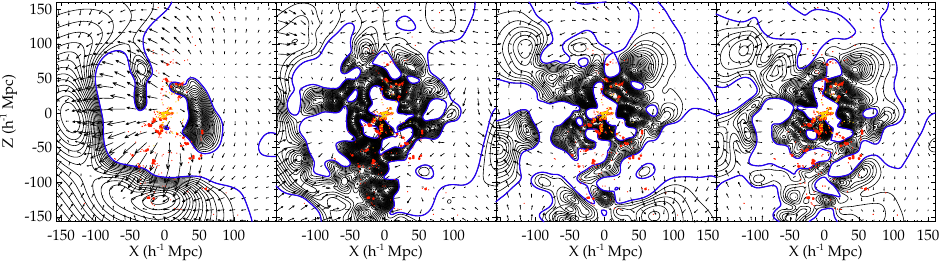}
\includegraphics[width=0.9\textwidth]{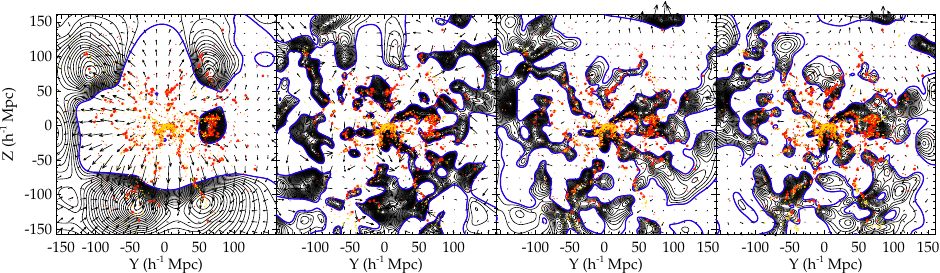}
\vspace{-0.1cm}

\caption{{Same as Fig. \ref{fig:cf3wf} but zoomed on the inner part of the box}.}
\label{fig:cf3wfzoom}
\end{figure*}

\subsection{The observational catalog}
We apply the algorithm to the third dataset of the Cosmicflows project \citep{2016AJ....152...50T}. This catalog contains 17,649 galaxy distance moduli. {Using several distance estimators, mostly from the Tully-Fisher \citep{1977A&A....54..661T} and Fundamental Plane \citep{2001MNRAS.321..277C} relations, this catalog allows probing distances as large as 500~Mpc. Still 50\% (90 and 99~\%) of the data are within 120~Mpc (225 and 330~Mpc).} The other distance indicators are Cepheids \citep{2001ApJ...553...47F}, Tip of the Red Giant Branch \citep{1993ApJ...417..553L}, Surface Brightness Fluctuation \citep{2001ApJ...546..681T}, supernovae of type Ia \citep{2007ApJ...659..122J} and other miscellaneous methods. We group the distance moduli into 15,050 galaxy/group distance moduli using the grouping technique of \citet{2016A&A...588A..14T} as described in \citet{2017MNRAS.469.2859S} using Planck cosmology in this paper. In a future paper of this series, we will study the impact of the cosmological parameter value choice in thorough details. {Still, since \citet{2016AJ....152...50T} estimate the third dataset of the Cosmicflows project to be compatible with H$_0$=75$\pm$2~km~s$^{-1}$~Mpc$^{-1}$, Appendix A shows the result obtained with WMAP7-like parameters: H$_0$=74~km~s$^{-1}$~Mpc$^{-1}$, $\Omega_m=0.27$ and $\Omega_\Lambda=0.73$. Additionally, section 6 quantifies the significance of the difference between results obtained with this second choice of cosmological parameter set values vs. the first one (Planck-like).}

\subsection{Results}

Figure \ref{fig:cf3corr} presents results obtained applying the algorithm to the observational catalog described above. The top row shows one realization of the catalog that maximizes the probability density of the point process model. The bottom row shows five (ten) stacked realizations.  In the first panel of each three-panel gathering, the distance modulus histograms reveal that indeed the Malmquist bias is overall corrected: objects those distances were underestimated (black solid line) are now further away (dashed and dotted blue and red lines). The sharp cut in distances due to the Fundamental plane-based 6-degree Field Galaxy Survey peculiar velocity sample \citep{2014MNRAS.445.2677S}, one of the main components of the cosmicflows-3 catalog, is also recovered. Note that unlike other bias-minimization techniques though, we do not use a prior on this sharp cut off.  In the second panel of each one of these three-panel gatherings, the 1D peculiar velocity distribution is also less flattened. The expected 1D Gaussian distribution is recovered. Finally, the last panel of each three-panel gathering shows that the {small-scale} velocity variance is reduced to reach values in better agreement with expectations. Note that contrary to realizations obtained with the mock catalog, stacking five or ten realizations does not further change evidently the  {small-scale} velocity variance.\\

In addition, Figure \ref{fig:distribvel} shows the distribution of galaxies in the three 40~\hMpc\ thick Supergalactic slices. Galaxies are represented as blue (filled red) circles when their associated velocities are negative (positive). While the top row shows galaxies in the raw third catalog of the Cosmicflows project, the second (third) row highlights the galaxy distribution in the corrected (ten stacked realization) catalog. No pattern emerges in the raw catalog distribution but for the biases: solely negative velocities at large distances. Conversely, in the corrected catalogs, the filamentary structure of the cosmic web starts to emerge: well-defined filaments with infalling galaxies on both side appear delimiting voids. {Note that while assuming redshift distances would also result in a filamentary like structure of the cosmic web, it would lack any information on peculiar velocities and thus on the true nature of the filamentary structures.}

\subsection{Reconstruction} 

Wiener-filter reconstructions of the density and velocity fields are built from the biased and corrected (both single and stacked realizations) observational catalogs. Supergalactic slices of these reconstructions are shown on Figures \ref{fig:cf3wf} {and \ref{fig:cf3wfzoom}. The latter being a zoom on the inner part of the box with respect to the former}. Galaxies and groups from the 2MASS Galaxy Redshift Catalog (\citealp[2MRS,][]{2012ApJS..199...26H}; \citealp[groups from][]{2018A&A...618A..81T}) are overplotted as red dots for comparison purposes solely. The yellow dots indicate galaxies those peculiar velocities have been used for the reconstructions. The large infall onto the center of the box / the observer and the rounded structures observed in the reconstruction obtained with the biased velocity catalog are suppressed in reconstructions obtained with corrected catalogs. {These infalls and rounded structures are the result of biases that the Wiener-filter cannot not take into account by itself. In reconstructions obtained with bias-minimized catalogs, structures are more sharply defined. The velocity field presents several islands of convergence/divergence in agreement with the clustering of galaxies as given by the redshift survey.}  \\

Note that the central outflow is more pronounced than the overall infall in the reconstruction obtained with the biased observational catalog than in that obtained with the biased synthetic catalog. This is because we use in both cases the same Planck Hubble constant value, while the cosmicflows-3 catalog zeropoint is set to a higher local Hubble constant value. The detailed effect of the Hubble constant choice and a possible estimation of the best fit to the data will be thoroughly investigated in a subsequent paper. \citet{2017MNRAS.469.2859S} already showed the impact of the Hubble constant value and the capabilities of bias minimization techniques in standardizing the results to suppress at several levels and first order this dependence. {Appendix A seems to further comfort this capability while section 6 quantifies the significance of the differences between results obtained assuming different H$_0$ values.} In any case, one can notice that both outflow and infall are drastically reduced in reconstructions obtained with bias-minimized catalogs. 

 \section{On the choice of H$_0$}
 
{ This section aims at quantifying the impact of H$_0$ value on the results shown in this paper. It starts with quantifying differences  and estimating their significance using synthetic catalogs before propagating the study to the observational catalogs.}

  \subsection{Synthetic catalogs}
  
  \begin{table}
\begin{center} 
\begin{tabular}{llrl}
\hline
\hline
Catalog$_1$ & Catalog$_2$ & Type & $\sigma_\mu$ \\
\hline
\hline
C67 & C67 & V & 0.06 $\pm$ 0.06\\
C74 & C74 & V & 0.06 $\pm$ 0.06\\
C74with67 & C74with67 & V & 0.06 $\pm$ 0.06\\
C67 & C67 & VoR & 0.09 $\pm$ 0.08\\
C74 & C74 & VoR  & 0.09 $\pm$ 0.08\\
C74with67 & C74with67 & VoR & 0.09 $\pm$ 0.09\\
C67 & C74with67 & VoR & 0.09 $\pm$ 0.08\\
\hline
\hline
\end{tabular}
\end{center}
\vspace{-0.25cm}
\caption{{\textit{Three first lines}: Average variance between distance moduli (V) in C67, C74 and C74with67 obtained applying the algorithm, initiated with ten different seeds, on B67, B74 and B74with67 respectively. \textit{Four last lines}: Average variance of distance modulus residuals (VoR) between pairs of bias-minimized catalogs (Catalog$_1$ vs. Catalog$_2$).}}
\label{tbl:2}
\end{table}

{ To conduct this study, we build two synthetic catalogs following the procedure we described hereabove (in subsection 4.1). One of this catalog uses  H$_0$=67.77 km~s$^{-1}$~Mpc$^{-1}$, the other one uses H$_0$=74 km~s$^{-1}$~Mpc$^{-1}$. This constitutes two \textbf{true} catalogs containing ($sgl$,$sgb$,$z_{obs}$,$\mu$)  denoted respectively T67 and T74. Our building procedure implies that selected halo-points are not identical in both catalogs. From these two catalogs, we build three \textbf{biased} catalogs. Two of them are obtained adding uncertainties as detailed in subsection 4.1 to the two \textbf{true} catalogs. The third one is obtained from the T67-catalog but assuming H$_0$=74~km~s$^{-1}$~Mpc$^{-1}$ and propagating uncertainties. This configuration permits having the same selected halo-points in both biased catalogs. These \textbf{biased} catalogs are respectively called B67, B74 and B74with67. Only B67 and B74with67 share the same ground truth in terms of catalogs. However there is only one ground truth in terms of the reference simulation. Note that we could have built instead B67with74 without affecting the following conclusions. Since H$_0$=67.77 km~s$^{-1}$~Mpc$^{-1}$ is not favored though for the observational catalogs, presenting the results for B74with67 is more relevant.} \\
 
 {We apply the algorithm to the three biased catalogs to get the (stacked) bias-minimized catalogs called C`H$_0$\_N' where the `H$_0$' string is 67 or 74 and `\_N' is used only for the stacked realizations with N=5 or 10.}  \\
 
 {Table \ref{tbl:2} compiles the average variance (V) of the distance moduli between the ten different realizations used for the stacked version of the bias-minimized catalogs. Namely, the algorithm is applied ten times with a different initial seed on B67 (or B74 or B74with67). The variance between the distance moduli obtained for a halo-point in ten resulting bias-minimized catalogs, C67 (or C74 or C74with67) is then derived. The average of the variances for all halo-points is then reported in Table \ref{tbl:2}. Additionally, the variance of the residuals between the distance moduli is computed for pairs of realizations. The average variance of the residuals (VoR) is then reported in Table \ref{tbl:2}. The same can be done for pairs made of one C67 and one C74with67 as they share exactly the same halo-points. Note that the mean of these residuals is zero in all the cases but for C67 vs. C74with67. Final distance moduli are on average slightly smaller ($\sim$0.18$\pm$0.05) for C74with67. This change of zeropoint is related to the larger assumed value of H$_0$ but is still smaller than the initial uncertainties on average ($\sim$0.5 against $\sim$0.18 mag).}
 
{It is notable that the average variance (V) between the different realizations is almost an order of magnitude smaller than initial uncertainties on average ($\sim$0.5 against $\sim$0.06 mag). Although slightly higher, the average variance of the residuals (VoR) is also about the same order of magnitude ($\sim$0.09 mag). The most interesting fact is that this average variance of the residuals is not larger when using pairs of realizations obtained with the algorithm assuming the same H$_0$ value with respect to using pairs based on different H$_0$ values. Statistically, the differences between distance moduli have the same variance ($\sim$0.09 mag).}  \\

 \begin{figure}[!ht]
 \vspace{-1cm}
\includegraphics[width=0.5 \textwidth]{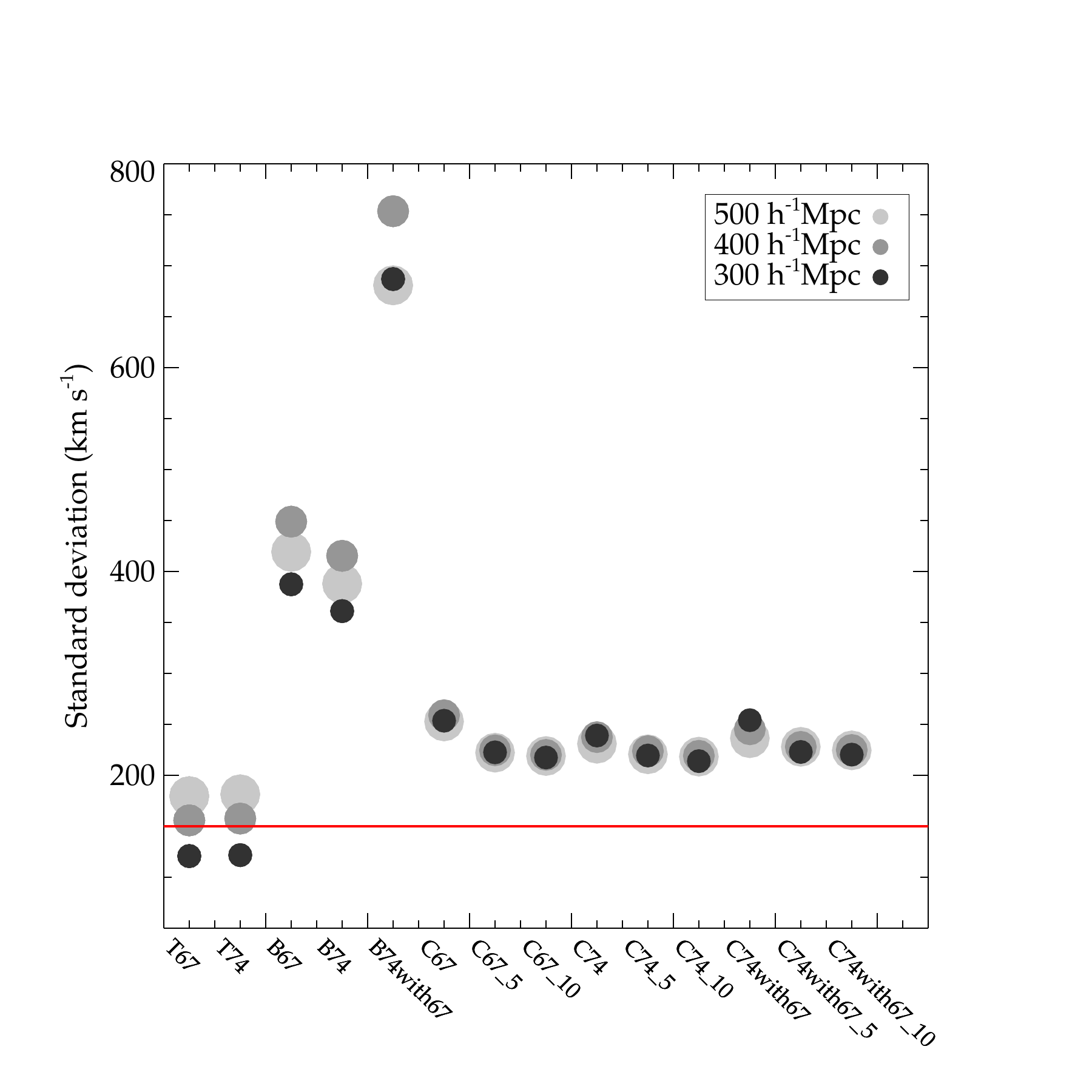}
\vspace{-2cm}

\includegraphics[width=0.5 \textwidth]{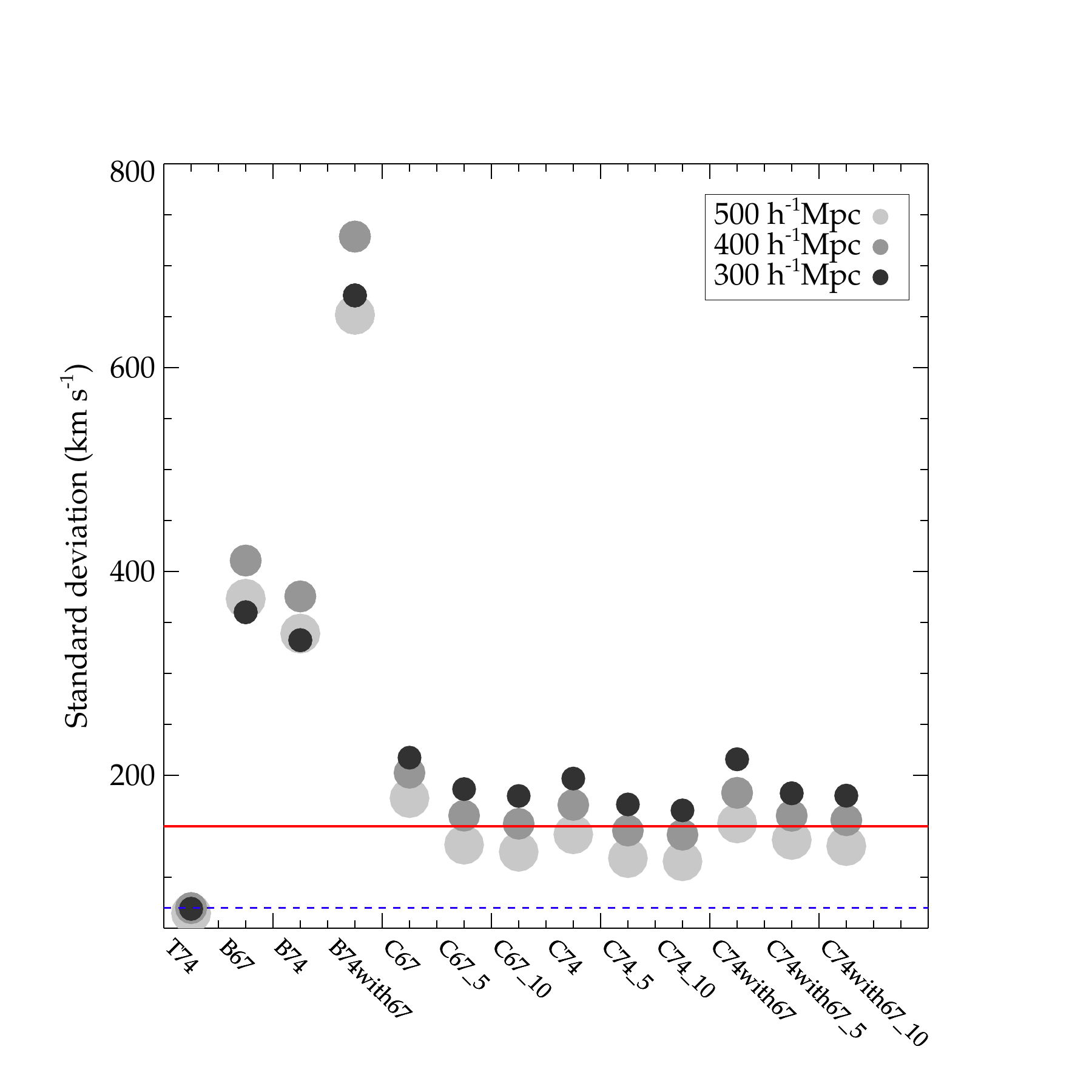}
\caption{Comparisons between reconstructed velocity fields. {\textit{Top}: Variance (filled circles) between the velocity field of the reference simulation and those reconstructed from the different catalogs. \textit{Bottom}: Variance (filled circles) between the reconstructed velocity field obtained with the true catalog T67 and those derived from the other catalogs. See the text for a detailed explanation of the different catalogs. Names are given as follows: the letter indicates the type - T for True, B for biased and C for bias-minimized ; the number gives the Hubble constant value - 67 for 67.77~km~s$^{-1}$~Mpc$^{-1}$, 74 for 74~km~s$^{-1}$~Mpc$^{-1}$ and 74with67 when assuming 74~km~s$^{-1}$~Mpc$^{-1}$ but for a 67.77~km~s$^{-1}$~Mpc$^{-1}$ based true catalog ; any additional suffix means that several realizations have been stacked (either 5 or 10). The size and color of the filled circles stand for the sub-box size within which fields are compared. The solid red line stands for the average variance between the reference simulated velocity field and that reconstructed from T67. The dashed blue line shows the average variance between the reconstructed fields obtained from both true catalogs with different H$_0$ values (T67 and T74).  Fields reconstructed from the bias-minimized catalogs differ from those obtained with the true catalog by the same amount as the latter differs from the reference field.}}
\label{fig:H0groundtruth}
\end{figure}

{To understand better the impact of these slight differences, comparisons can be extended to the Wiener-filter reconstructed velocity fields obtained with the different catalogs. They can be compared between themselves or with the reference simulation velocity field smoothed at the same scale. Indeed, in order to determine whether the difference between counterparts (i.e. fields obtained with the same type - true, biased, bias-minimized - of catalogs but different H$_0$ values) is significant or not, we need first to estimate by how much the reconstructed field obtained with the true catalog differs from the simulated field. }

{Figure \ref{fig:H0groundtruth} top shows the variance (filled circle) between the reference simulation velocity field and that reconstructed from the different synthetic catalogs i.e. from the true ones to the bias-minimized ones through the biased ones. The solid red line highlights the average variance between the simulated velocity field and a reconstruction obtained with a true synthetic catalog. The velocity fields are compared in the full box as well as in different sub-boxes (gradient of gray).Velocity fields obtained from both true synthetic catalogs differ from the simulated velocity field at the same level. Fields reconstructed with the biased catalogs differ the most from the simulated one. Variances derived from the bias-minimized catalogs are all of the same order and are intermediate between those derived using the true catalogs and those obtained with the biased ones.}

{Figure \ref{fig:H0groundtruth} bottom goes further by showing the variance (filled circle) between reconstructed velocity fields obtained from T67 and from all the other catalogs. The same color and linestyle code as the top panel applies. The additional dashed blue line highlights the variance between reconstructed velocity fields obtained with both true catalogs built out of different H$_0$ values. Notably, velocity fields reconstructed with the bias-minimized catalogs (especially the stacked ones) differ from that obtained with the true catalog (T67) by the same amount that the latter differs from the reference simulated field (solid red line). }\\

\begin{figure}[ht!]
\vspace{-1cm}
\includegraphics[width=0.5 \textwidth]{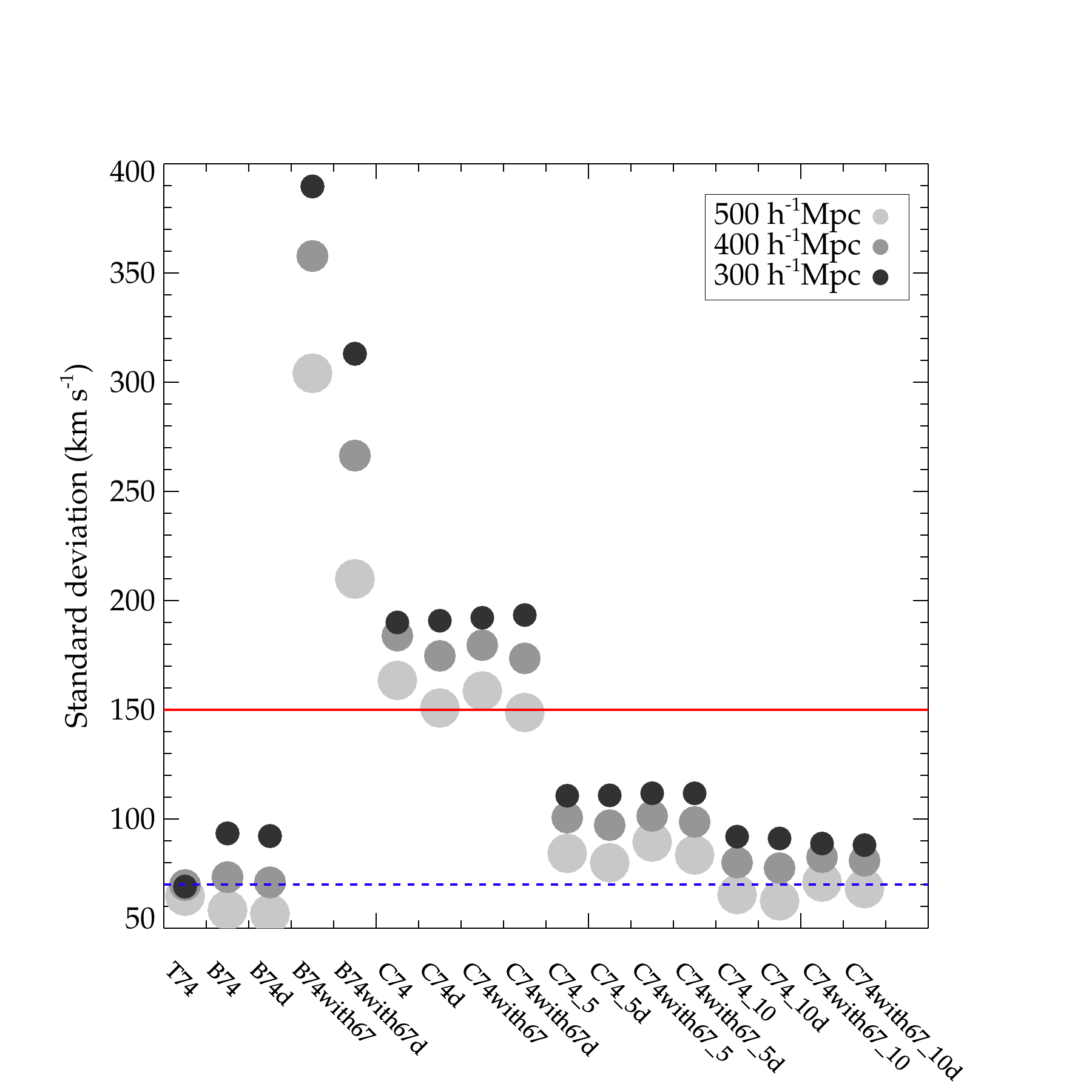}

\caption{{Same as Figure \ref{fig:H0groundtruth} but comparing reconstructed velocity fields obtained from the same type - true, biased, bias-minimized - of catalogs but different H$_0$ values. The additional `d' suffix stands for divergent to be opposed to full velocity field. Fields reconstructed from the ten stacked bias-minimized realizations but different H$_0$ values differ on average by the same order of magnitude as those obtained from the true catalogs but different H$_0$ values.}}
\label{fig:H0counterpart}
\end{figure}

{Finally, Figure \ref{fig:H0counterpart} compares pairs of velocity fields obtained with the same type of catalogs but different H$_0$ values. It looks not only at the total velocity fields but also only at its divergent part denoted by the additional `d' letter at the end of the name for `divergent'.   Fields reconstructed with the ten stacked bias-minimized realizations but different H$_0$ values differ no more than those obtained from the true catalogs but different H$_0$ values (dashed blue line). In any case, they differ less than the reconstructed velocity field obtained from the true catalog differs from the reference field (solid red line) or than the reconstructed fields derived from the bias-minimized catalogs differ from that reconstructed from the true catalog. Differences can then be considered insignificant at this first order. }

 \subsection{Observational catalogs} 
 
{In this section, the study is repeated on the observational catalog with the exception that the ground truth is unknown. Neither the true field nor the true catalog are available. The letter `o' is added to the names of the different catalogs for `observed'. Table \ref{tbl:3} reports the average variance between distance moduli (V) and the average variance of the residuals (VoR).  As for the synthetic catalogs, the mean of the residuals is zero but when comparing C67o with C74o. Final distance moduli are on average slightly smaller ($\sim$0.16$\pm$0.09) for C74o. All the values are very similar to those obtained with the synthetic catalogs and in that respect the same conclusions can be drawn.} \\

\begin{figure}
\vspace{-1cm}
\includegraphics[width=0.5 \textwidth]{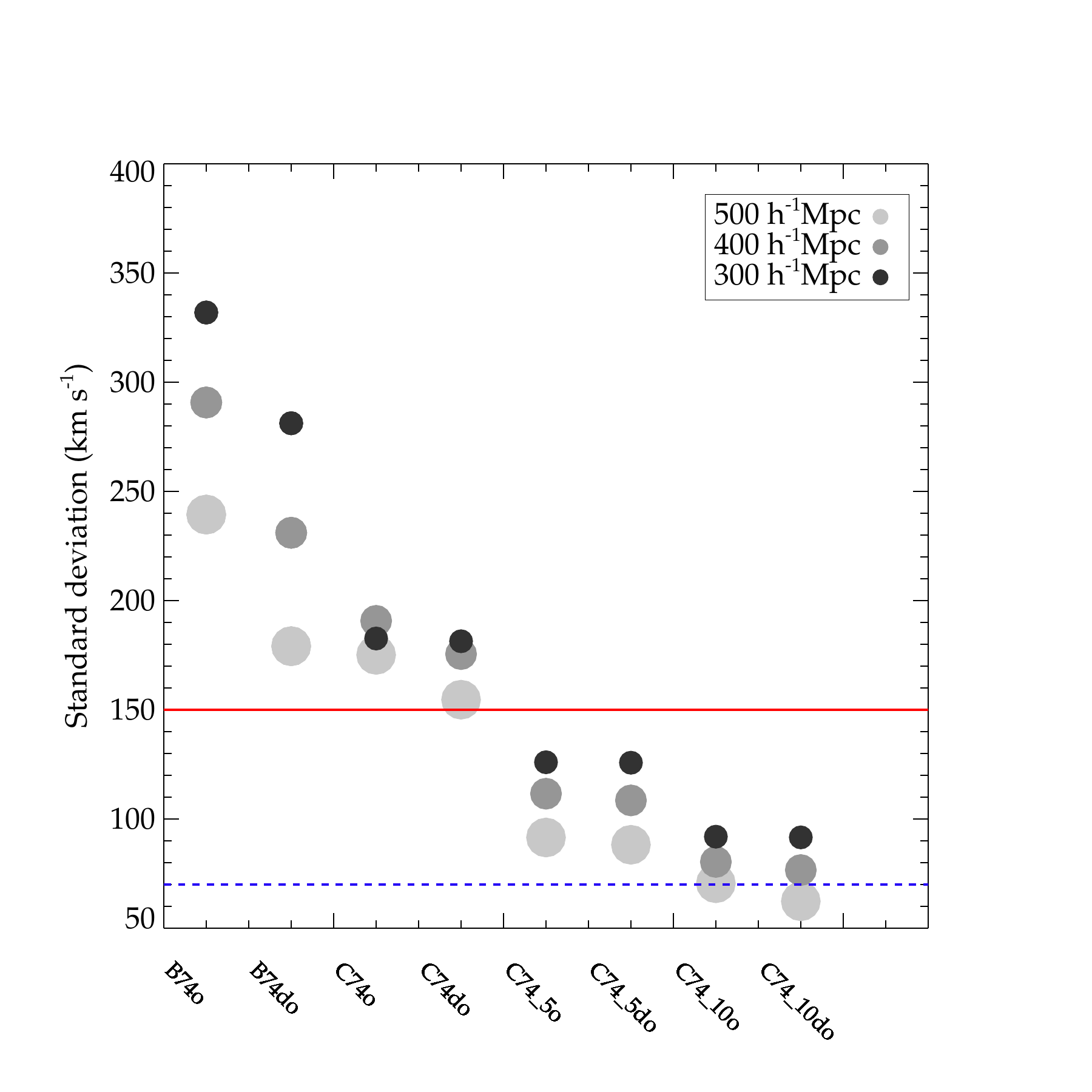}
\vspace{-1cm}

\caption{{Same as Figure \ref{fig:H0counterpart} but for the observational catalog (hence the additional letter `o'). Fields reconstructed from the ten stacked bias-minimized realizations but different H$_0$ values differ on average by the same order of magnitude than those obtained from the true synthetic catalogs but different H$_0$ values. Stacked bias-minimized catalogs permit obtaining reconstructions that differ at most by the same amount as the reconstructed field from the true synthetic catalog differs from the reference simulated field.}}
\label{fig:H0obs}
\end{figure}

{Wiener-filter reconstructed velocity fields are also compared. Variances are shown on Figure \ref{fig:H0obs} in the same fashion as Figure \ref{fig:H0counterpart}. The solid red line highlights the average variance found when comparing the velocity field reconstructed with the true synthetic catalog and the reference simulated field. The dashed blue line shows the average variance between the reconstructed velocity fields obtained with the true synthetic catalogs but different H$_0$ values. Conclusions are similar to those drawn with the synthetic catalogs. Differences between reconstructed velocity fields derived from the stacked bias-minimized realizations obtained with different H$_0$ values are insignificant at this first order.} \\

 \begin{table}
\begin{center} 
\begin{tabular}{llrl}
\hline
\hline
Catalog$_1$ & Catalog$_2$ & Type & $\sigma_\mu$ \\
\hline
\hline
C67o & C67o & V & 0.09 $\pm$ 0.10\\
C74o & C74o & V & 0.08 $\pm$ 0.09\\
C67o & C67o & VoR & 0.13 $\pm$ 0.15\\
C74o & C74o & VoR & 0.12 $\pm$ 0.13\\
C67o & C74o & VoR & 0.13 $\pm$ 0.13\\
\hline
\hline
\end{tabular}
\end{center}
\vspace{-0.25cm}
\caption{{\textit{Two first lines}: Average variance between distance moduli (V) in C67o and C74o obtained applying the algorithm, initiated with ten different seeds, on B67o and B74o respectively. \textit{Three last lines}: Average variance of distance modulus residuals (VoR) between pairs of bias-minimized catalogs (C67o vs. C74o)} }
\label{tbl:3}
\end{table}

\begin{figure*}[!ht]
   \centering
   \vspace{-0.2cm}
   
\includegraphics[width=1 \textwidth]{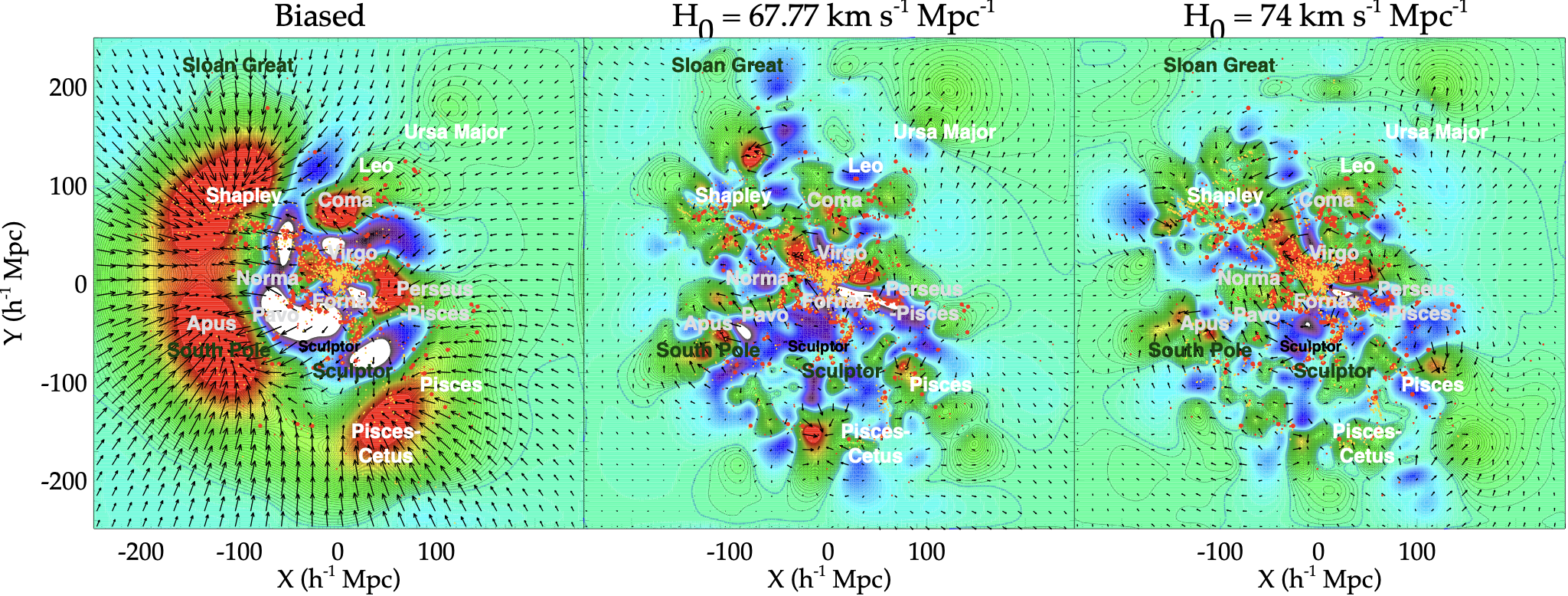}
\includegraphics[width=1 \textwidth]{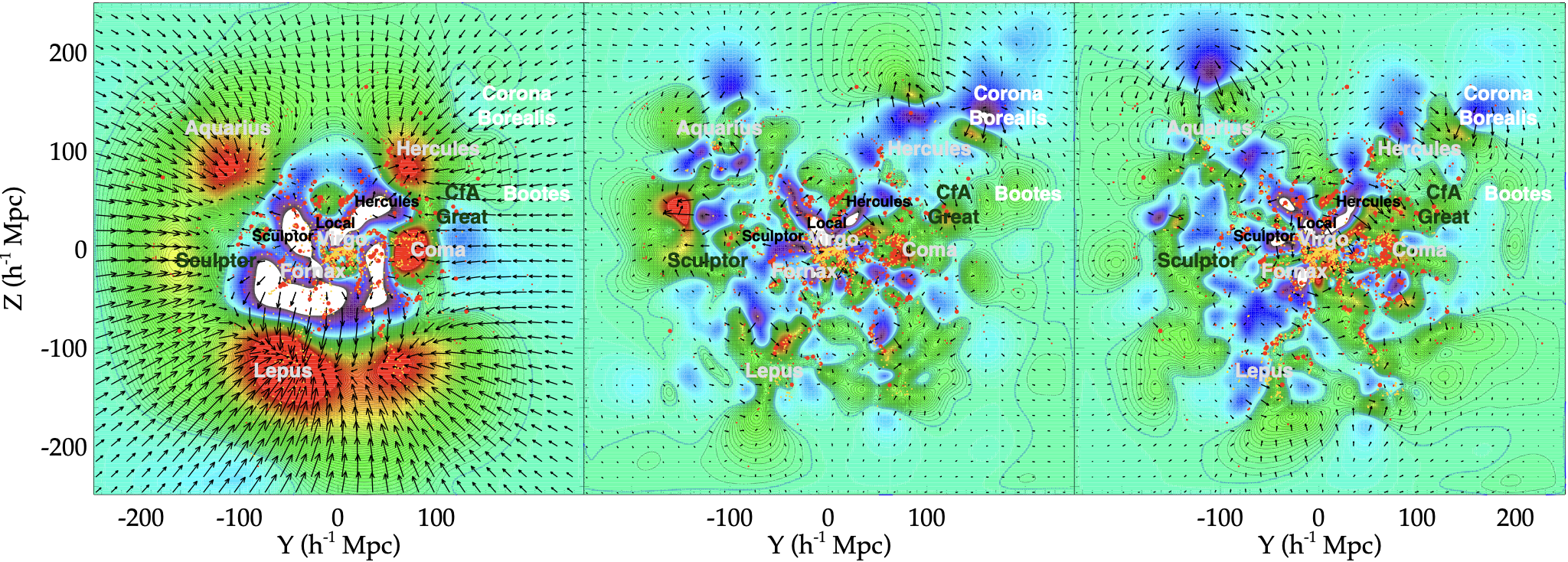}
\includegraphics[width=1 \textwidth]{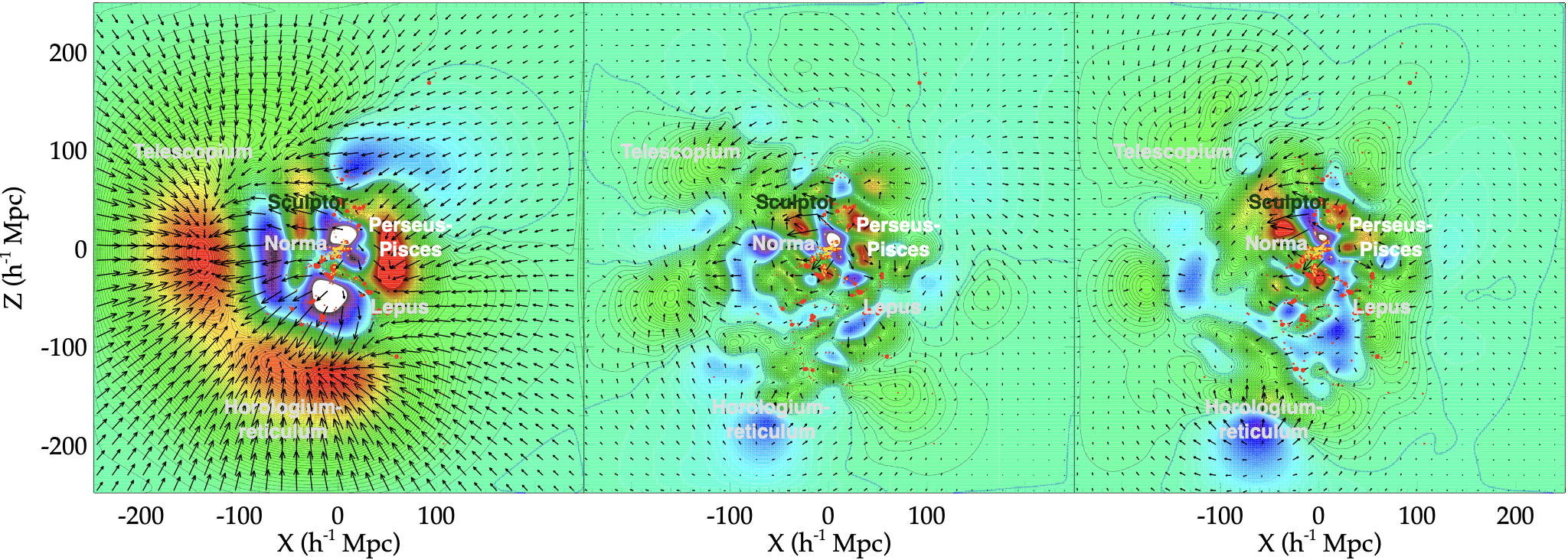}
\vspace{-0.6cm}

\caption{{Supergalactic slices of reconstructed density (filled contours) and velocity (arrows) fields of the local Universe. Hot to cold colors go from over- to under- densities. Red points are galaxies (small filled circles) and groups (larger filled circles) from the 2MRS Galaxy Redshift Catalog for comparison purposes only  (\citealp[2MRS,][]{2012ApJS..199...26H}; \citealp[groups from][]{2018A&A...618A..81T}). Yellow points show galaxies whose peculiar velocities obtained from distance moduli are actually used for the reconstructions. Names indicate superclusters (white), clusters (grey), walls (dark green) and voids (smaller size characters in black). The bias effects are reduced in reconstructed fields obtained with corrected catalogs using either H$_0$=67.77~km~s$^{-1}$~Mpc$^{-1}$ or H$_0$=74~km~s$^{-1}$~Mpc$^{-1}$.}}
 \label{fig:colorlabel}
   \end{figure*}

{Finally, Figure \ref{fig:colorlabel} shows the three supergalactic slices of the overdensity and velocity fields reconstructed from the biased, bias-minimized assuming H$_0$=67.77~km~s$^{-1}$~Mpc$^{-1}$ and   H$_0$=74~km~s$^{-1}$~Mpc$^{-1}$ observational catalogs. The color gradient from red to violet-white highlights high over- to under-densities.  Names of structures are visible. White is used for superclusters, grey for clusters, dark green for walls and black with smaller character sizes for voids. Red points are still galaxies (small filled circles) and groups (larger filled circles) from the 2MRS Galaxy Redshift Catalog for comparison purposes only  (\citealp[2MRS,][]{2012ApJS..199...26H}; \citealp[groups from][]{2018A&A...618A..81T}). Yellow points still show galaxies whose peculiar velocities obtained from distance moduli are actually used for the reconstructions.}

{Clearly the reconstruction obtained with the original catalog is biased although high density regions and structures are not completely off in the sense that there are over- and under- densities where expected although too pronounced. The velocity field though is clearly wrong. Reconstructions obtained with the ten-stacked bias-minimized catalogs present, on the contrary, several zones of velocity convergence and divergence. The agreement with known structures is good with both values of H$_0$. The overall fields are very similar in agreement with the variance obtained when comparing the velocity fields. Over- and under- densities are at similar locations with tiny fluctuations that may hint at a better reconstruction alternatively using one or the other value of H$_0$ with some difficulties in concluding in absence of ground truth. Future constrained simulations obtained with these catalogs will however also provide us with the mass of the clusters that can be compared with the observational mass estimates. Because the major differences between the reconstructions seems to be at the level of the intensities of the over- and under-densities, simulations seem like an excellent alternative to push further this study. We will certainly do so in a subsequent paper.}\\

{This section can be concluded comparing reconstructed structures from this paper to those obtained for instance by \citet{2019MNRAS.488.5438G}. Structures are named like in their figure 8 to ease the comparisons. Similar structures are found back. Additionally, the residual spherical imprint of biases in the structures centered on the observer seems more dissipated. Several radial structures make their apparition. This similarity between reconstructions is to be pointed out especially because:
\begin{itemize}
\item their grouping is different from ours and we showed in \citet{2017MNRAS.469.2859S} that it may lead to drastic changes in the reconstruction.
\item their reconstruction technique is different (Wiener-filter vs. Hamiltonian Monte-Carlo - HMC),
\item leading to maximum a posteriori fields vs. mean of HMC sampling fields.
\end{itemize}}

\section{Conclusion} 

As a response to the full underlying gravitational field, galaxy peculiar velocities can be extraordinary cosmological probes provided that biases inherent to their catalog construction are controlled. To minimize the effect of the different biases, this paper proposes a new technique based on a point process model whose density probability is maximized with Metropolis-Hastings samplings embedded into a simulated annealing scheme.  The algorithm determines realizations maximizing the density probability. They correspond to sets of galaxy distance moduli and uncertainties with the highest probabilities given their corresponding radial peculiar velocities.\\

This new algorithm builds on our 2015 work \citep{2015MNRAS.450.2644S} and improves it by determining the most probable position of a galaxy and its associated peculiar velocity given not only the 1D peculiar velocity probability distribution but also the 3D {small-scale} velocity correlation. This concept is at the core of the algorithm proposed.\\

{Moreover, the model, tailored for this purpose, relies neither on very detailed prior knowledge of the catalog nor on prior hypotheses specific to the catalog:
\begin{itemize}
\item it offers a great flexibility,
\item it does not require as many priors as usual Bayesian techniques used in the field,
\item these priors are independent on the datasets unless the data sampling in space varies from one extreme to another, 
\item the algorithm thus makes it possible to easily switch from one dataset to another,
\item the cosmological model and its parameters can also be modified in the most convenient way\footnote{{As a matter of fact, Appendix A gives the results for the observational catalogs using a set of cosmological parameter values consistent with WMAP7 while Planck values are used in the core of the paper.}}. \\
\end{itemize}}

The proposed method diminishes the effects of the biases. The obtained results and the conducted statistical tests show the reduction of the effects of the biases  in two situations. Applied to synthetic catalogs\footnote{N.B.: one run requires between  700 and 1,500 cpu hours for a catalog of cosmicflows-3 size, i.e. 15,000+ datapoints. Typically, one stacked realization of ten realizations requires then 7,000 to 15,000 cpu hours.},  when comparisons with both the true and biased data are possible, the algorithm results in statistically corrected datasets i.e. distance moduli (thus distances and peculiar velocities) are in agreement at better than an average 1\%. This step permits also setting the parameters of the algorithm. \\

Subsequently, applied to observational catalogs of the Cosmicflows project, the algorithm gives datasets that are input into the Wiener-filter technique to reconstruct the local Universe. The Wiener-filter technique, a classical restoration method, is specifically chosen for its inefficiency in taking into account the biases. Resulting reconstructed density fields are a great match to the 2MASS Galaxy Redshift Catalog (2MRS) and the velocity field does not present any significant outflow/infall, signs of biases, out of/onto the local Volume. Within this context, the new proposed method improves the quality of the peculiar velocity catalogs as a whole.\\

The newly derived version of the peculiar velocity datasets will be used for future constrained simulations of the local Universe. {Nevertheless, reducing biases in cosmological data and  those appearing while exploiting the data is still an open and challenging problem. It leads to new questions. We will tackle them in future work.} For instance, following \citet{2018MNRAS.478.5199S}, the extra-smoothing of the velocity field by the Wiener-filter technique, due to the fading of the number of datapoints with the distance from the observer, will be taken care of. \\

Meanwhile, studies will focus on using different H$_0$ values {to probe on the possibility to extract an estimate of the H$_0$ value directly from the data}. {A first quantitative study based on the fields reconstructed from the bias-minimized catalogs obtained with different H$_0$ values hints at insignificant changes at first order that could become significant with deeper studies like going to constrained simulations and looking at the resulting cluster mass function.}\\

Later on, the cosmological model itself could be relaxed. In addition, the observational redshift, z$_{obs}$, assumed to have negligible errors with respect to distance moduli, might also be unfixed in future developments. 

\section*{Acknowledgements}
The authors would like to thank the referee for their careful reading of the manuscript and their sensible comments that helped improved its quality. The authors acknowledge the Gauss Centre for Supercomputing e.V. (www.gauss-centre.eu) and GENCI (https://www.genci.fr/) for funding this project by providing computing time on the GCS Supercomputer SuperMUC-NG at Leibniz Supercomputing Centre (www.lrz.de) and Joliot-Curie at TGCC (http://www-hpc.cea.fr). JS acknowledges support from the ANR LOCALIZATION project, grant ANR-21-CE31-0019 of the French Agence Nationale de la Recherche. ET acknowledges support by ETAg grant PRG1006 and by the EU through the ERDF CoE grant TK133. This work was supported by the Programme National Cosmologie et Galaxies (PNCG) of CNRS/INSU with INP and IN2P3, co-funded by CEA and CNES.

\bibliographystyle{aa}

\bibliography{biblicompletenew}

\newpage
\section*{Appendix A: H$_0$=74~km~s$^{-1}$~Mpc$^{-1}$}

\begin{figure*}
 
 \vspace{-0.5cm}
 
\hspace{-1.1cm}\includegraphics[width=0.4 \textwidth]{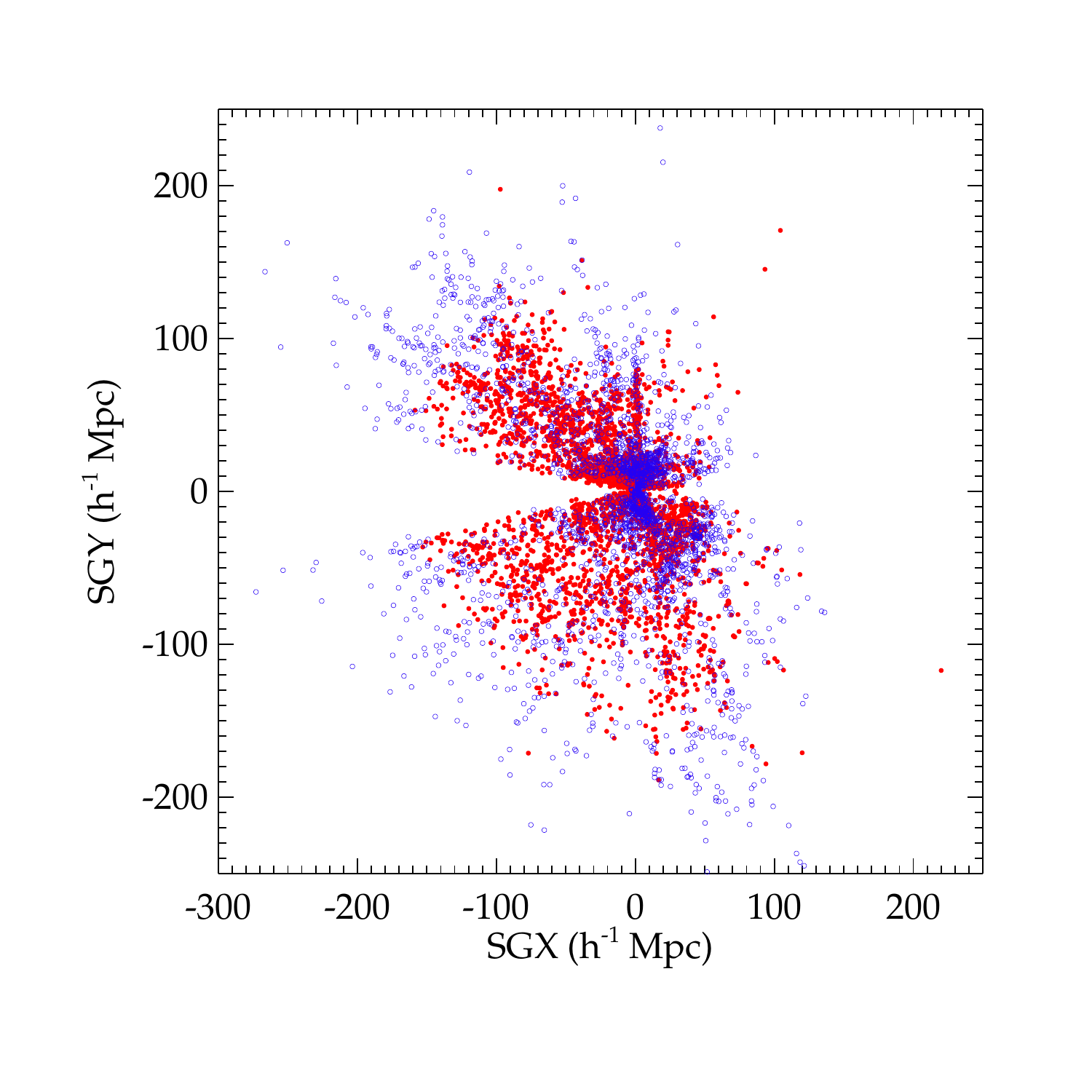}
\hspace{-0.72cm}\includegraphics[width=0.4 \textwidth]{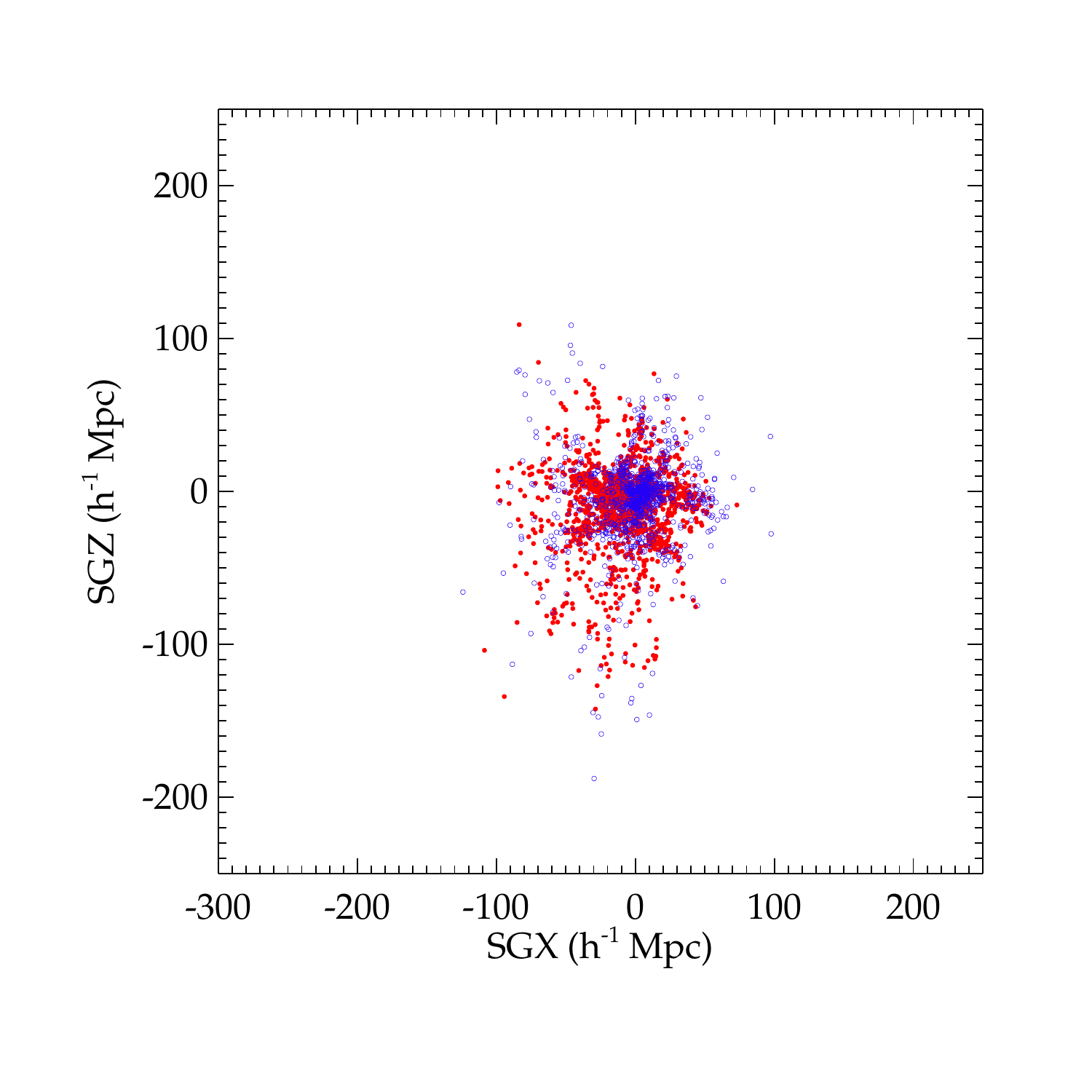}
\hspace{-0.72cm}\includegraphics[width=0.4 \textwidth]{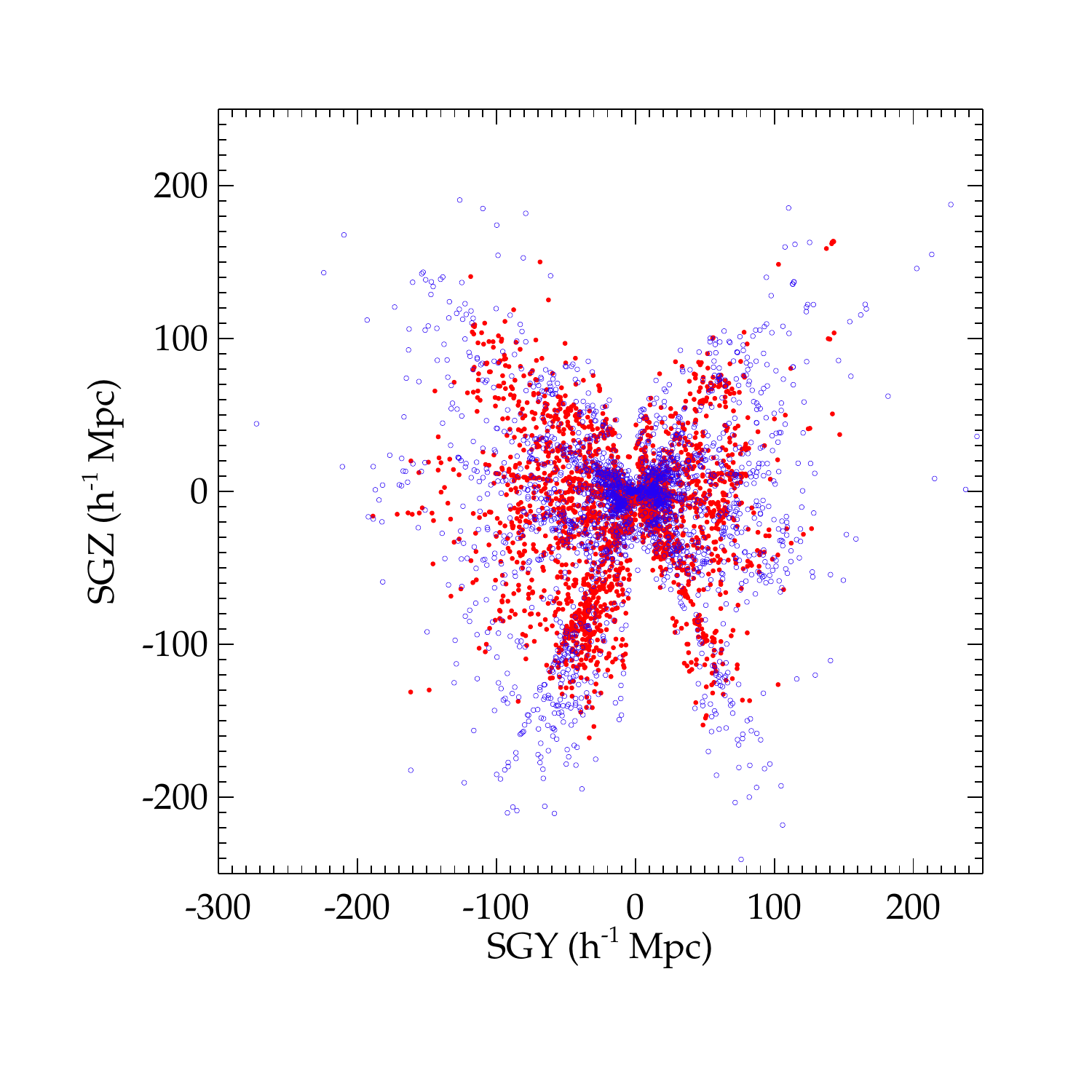}\\

\vspace{-1.6cm}

\hspace{-1.1cm}\includegraphics[width=0.4 \textwidth]{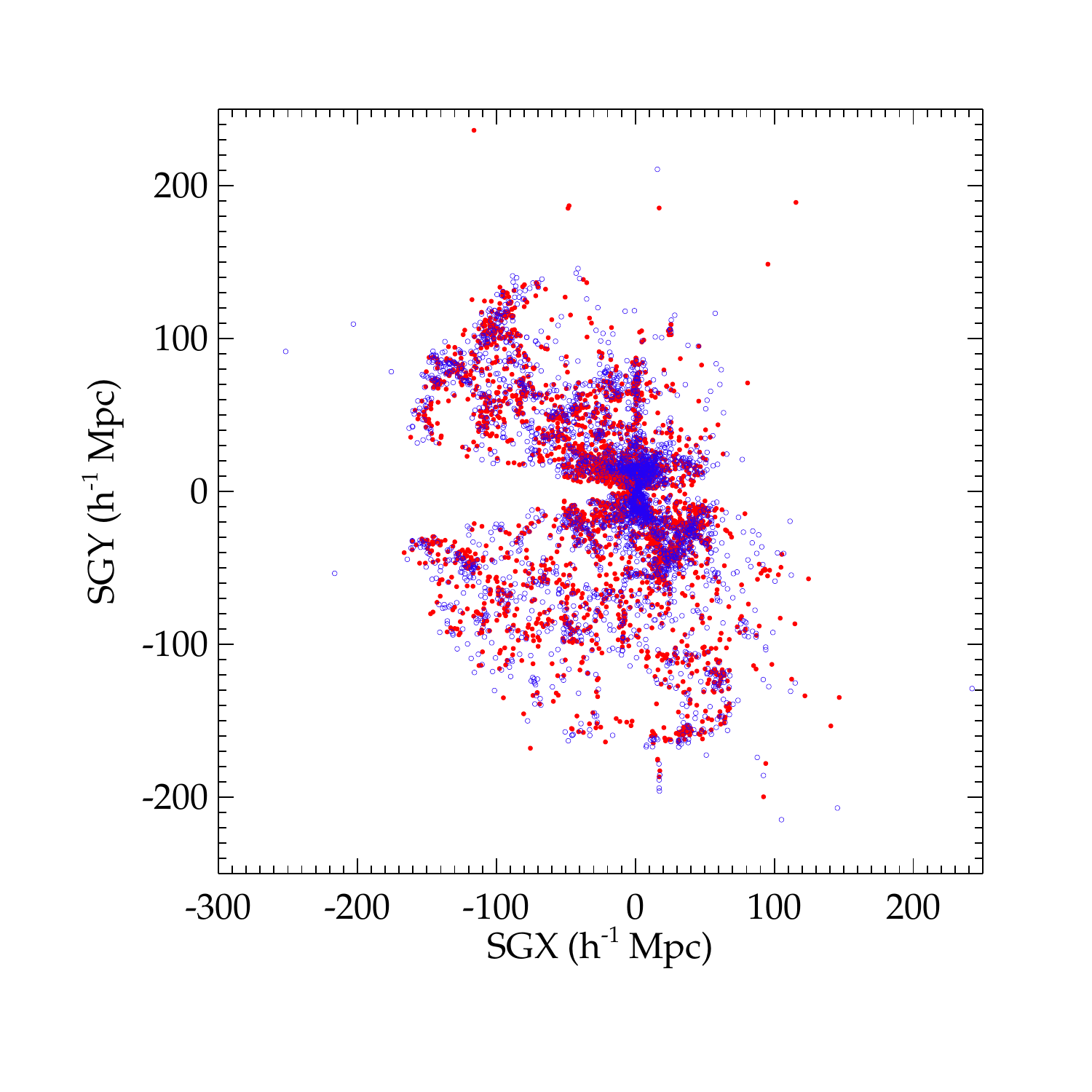}
\hspace{-0.72cm}\includegraphics[width=0.4 \textwidth]{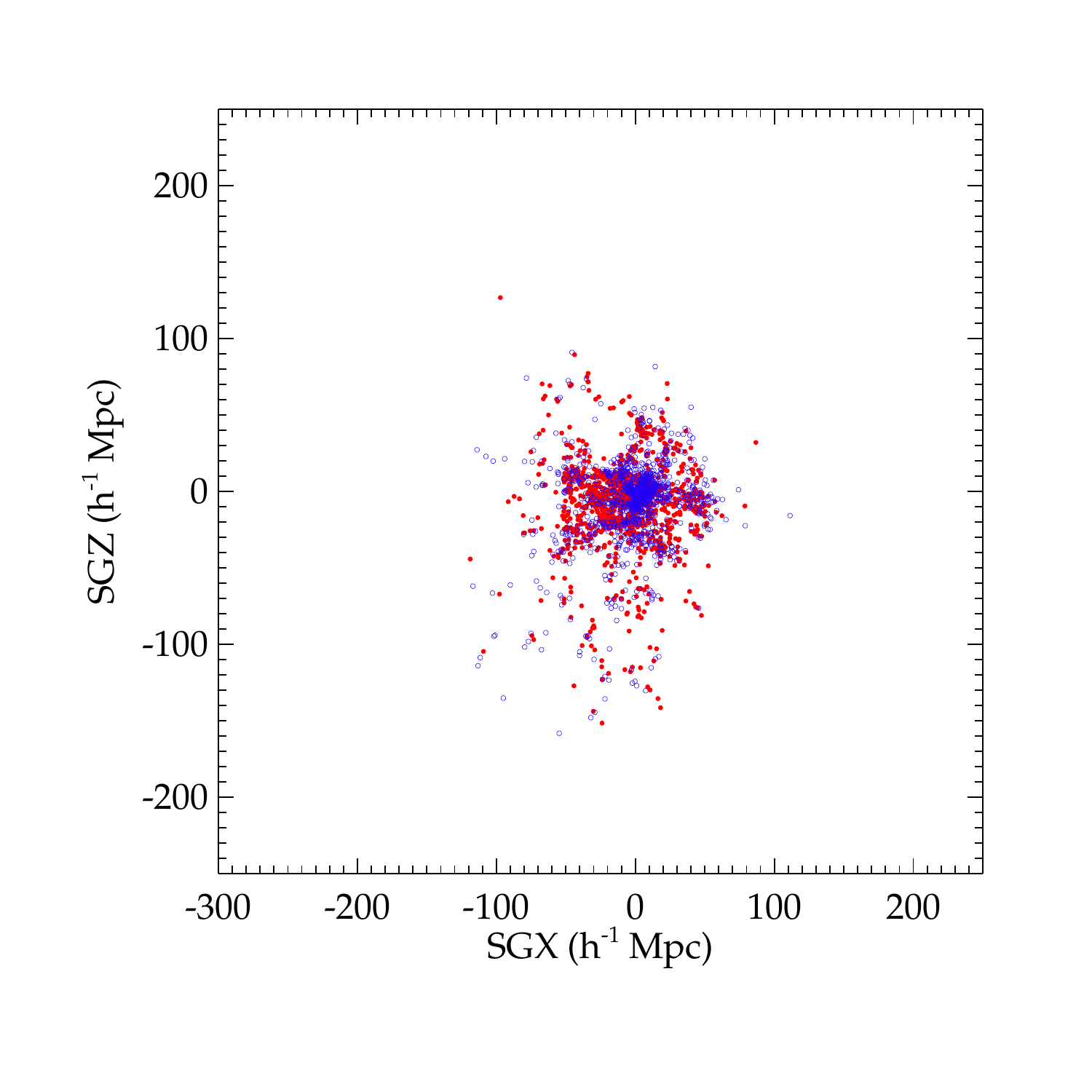}
\hspace{-0.72cm}\includegraphics[width=0.4 \textwidth]{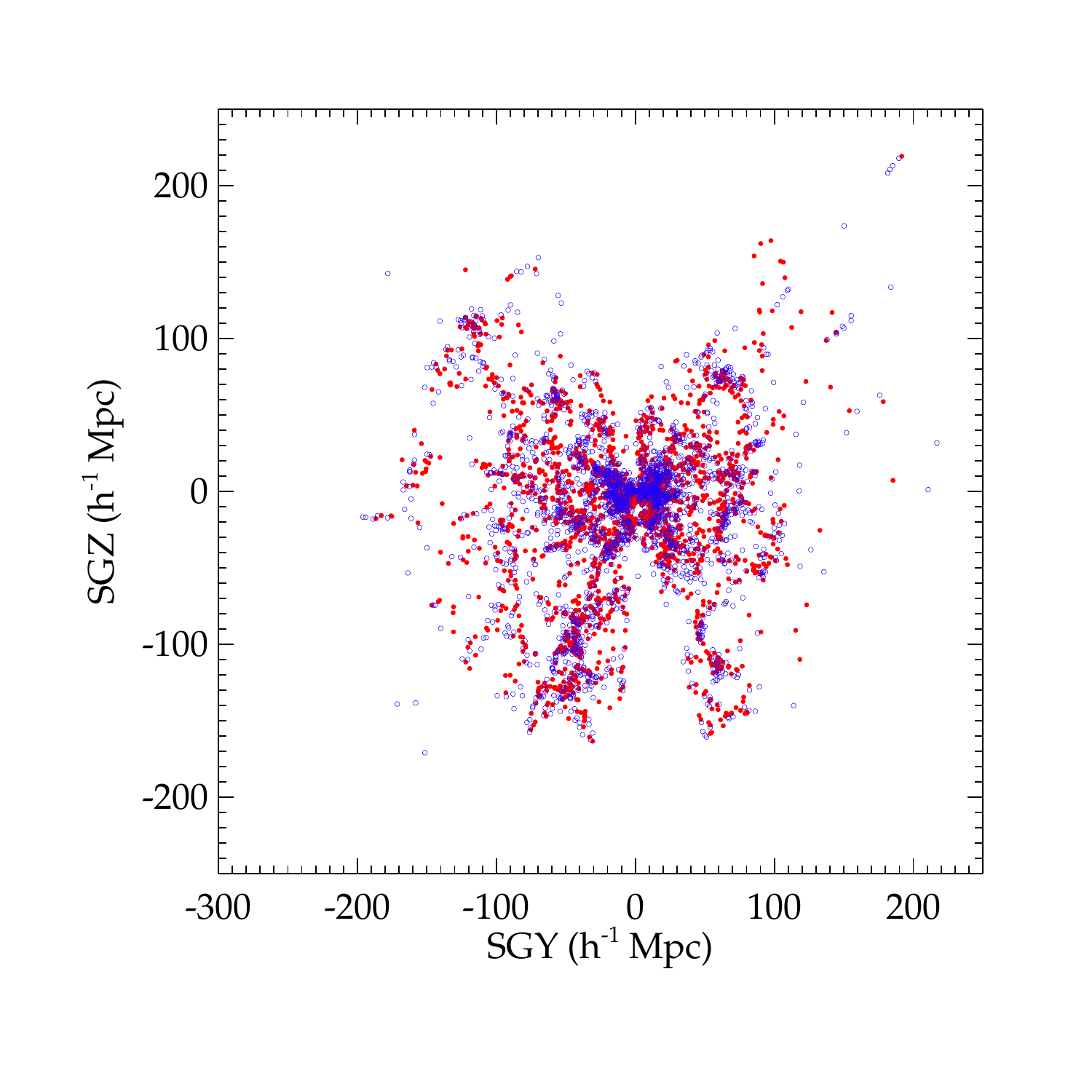}\\

\vspace{-1.6cm}
\hspace{-1.1cm}\includegraphics[width=0.4 \textwidth]{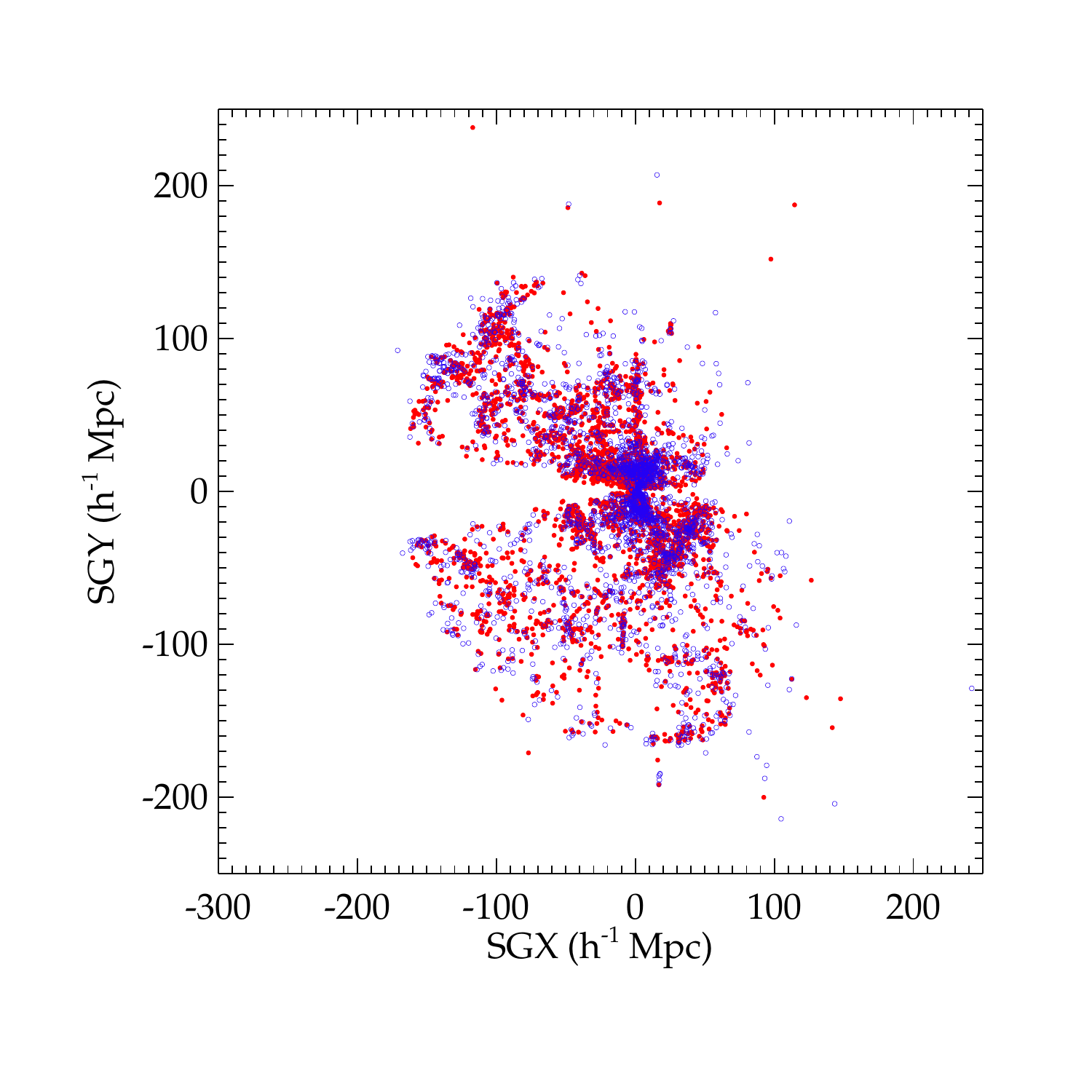}
\hspace{-0.72cm}\includegraphics[width=0.4 \textwidth]{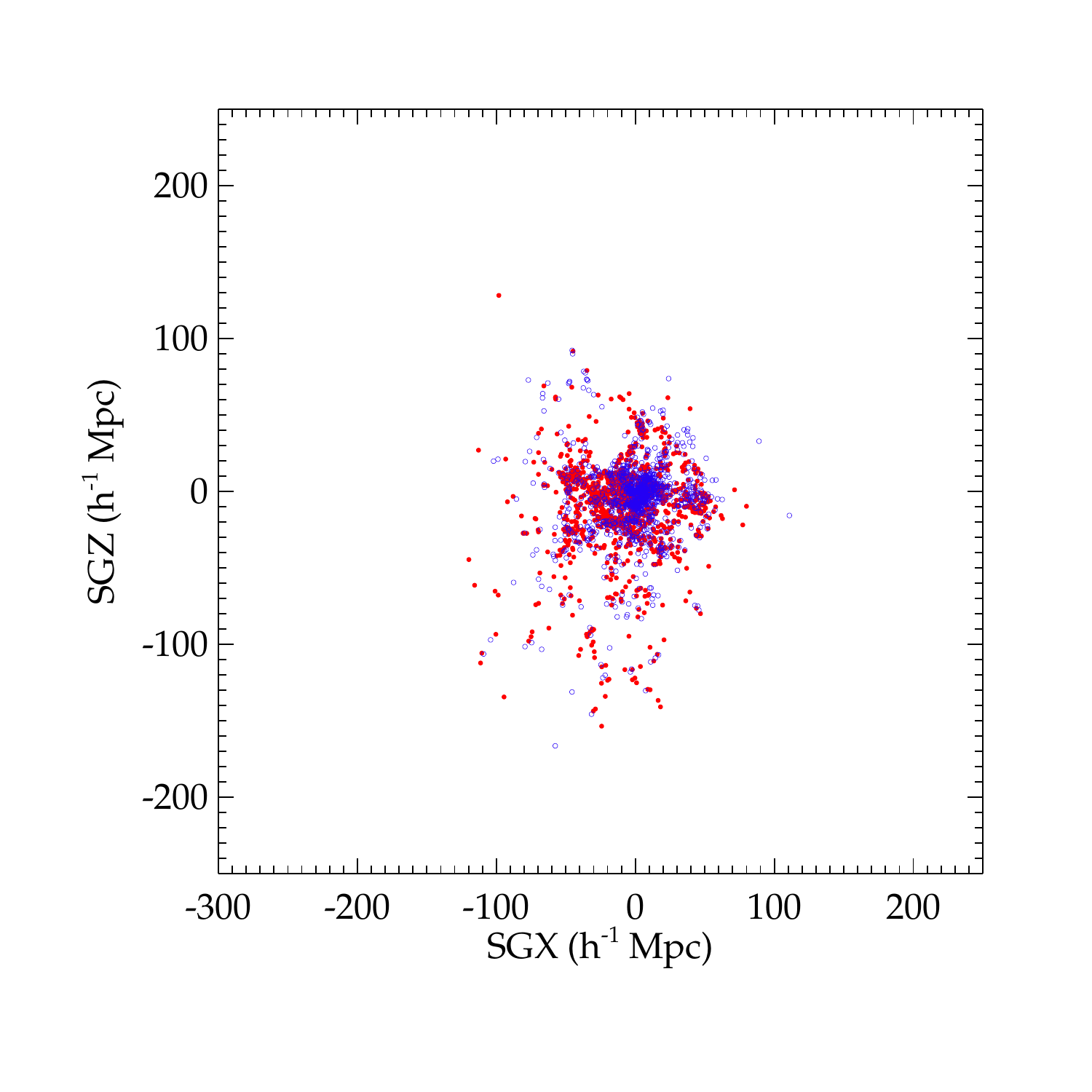}
\hspace{-0.72cm}\includegraphics[width=0.4 \textwidth]{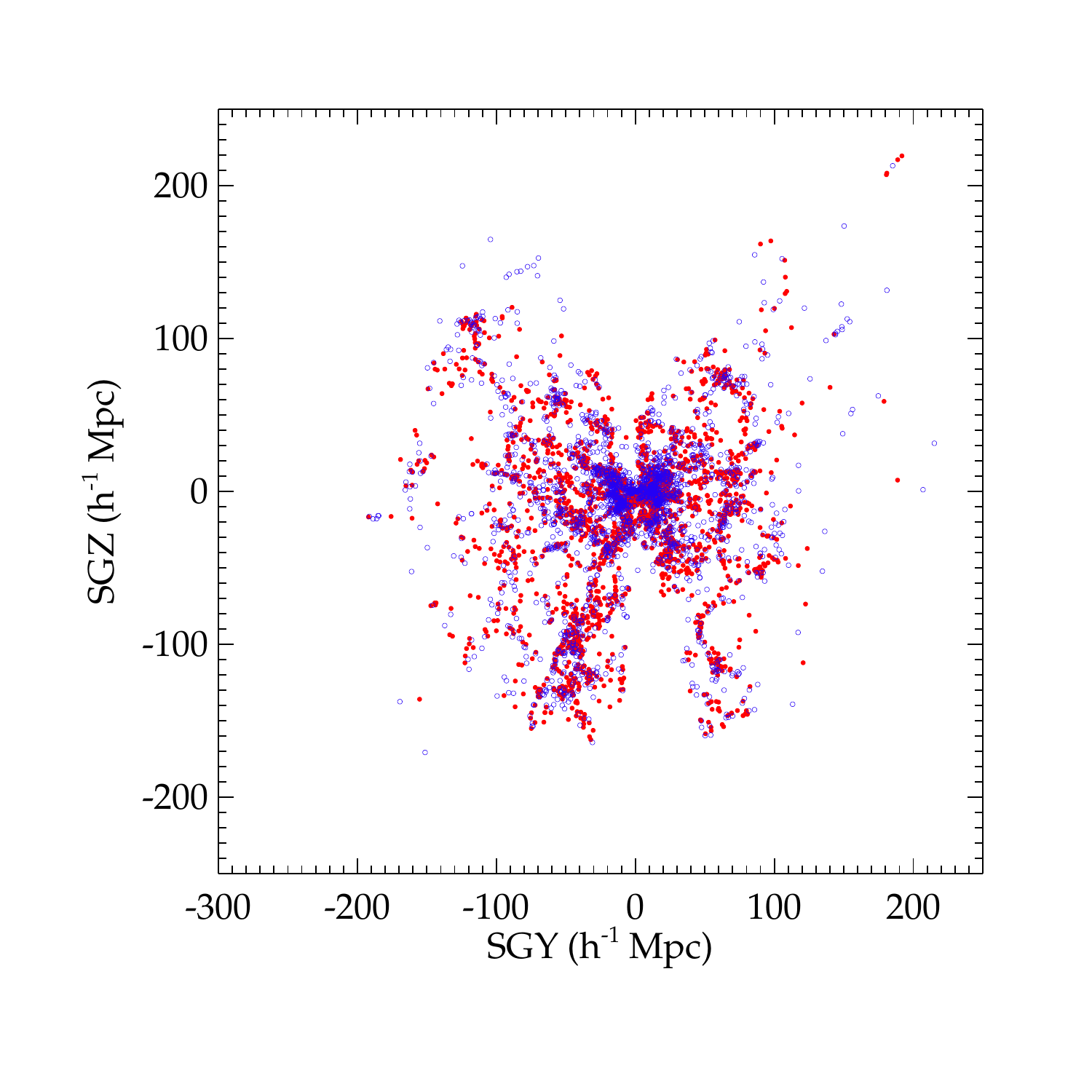}\\

\vspace{-1cm}
\caption{{Same as Fig. \ref{fig:distribvel} but using WMAP7-like cosmological parameter values in the algorithm.}}
\label{fig:distribvel74}
\end{figure*}

\begin{figure*}
 \vspace{-1cm}
\hspace{1.3cm}\includegraphics[width=0.3 \textwidth]{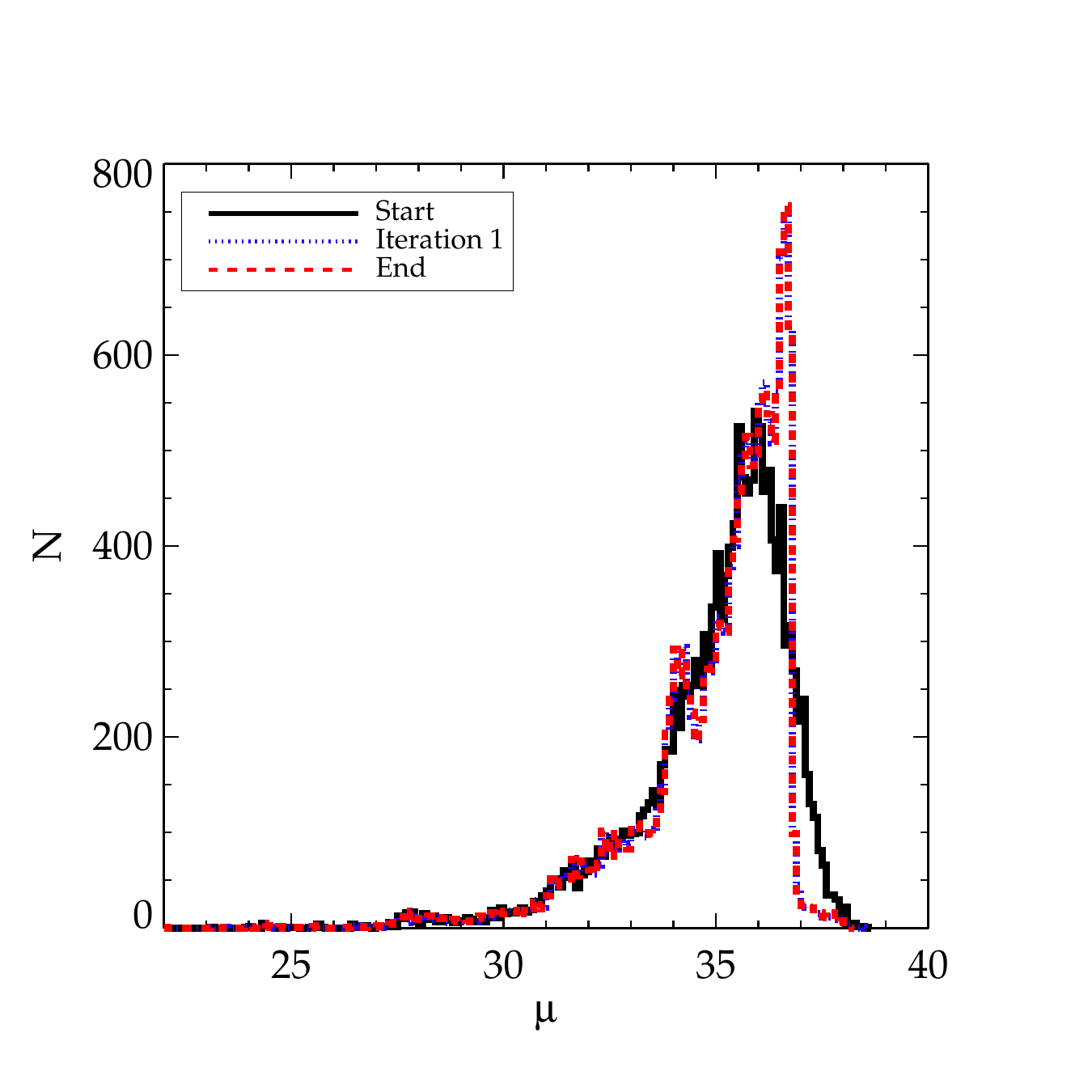}
\hspace{-0.8cm}\includegraphics[width=0.3 \textwidth]{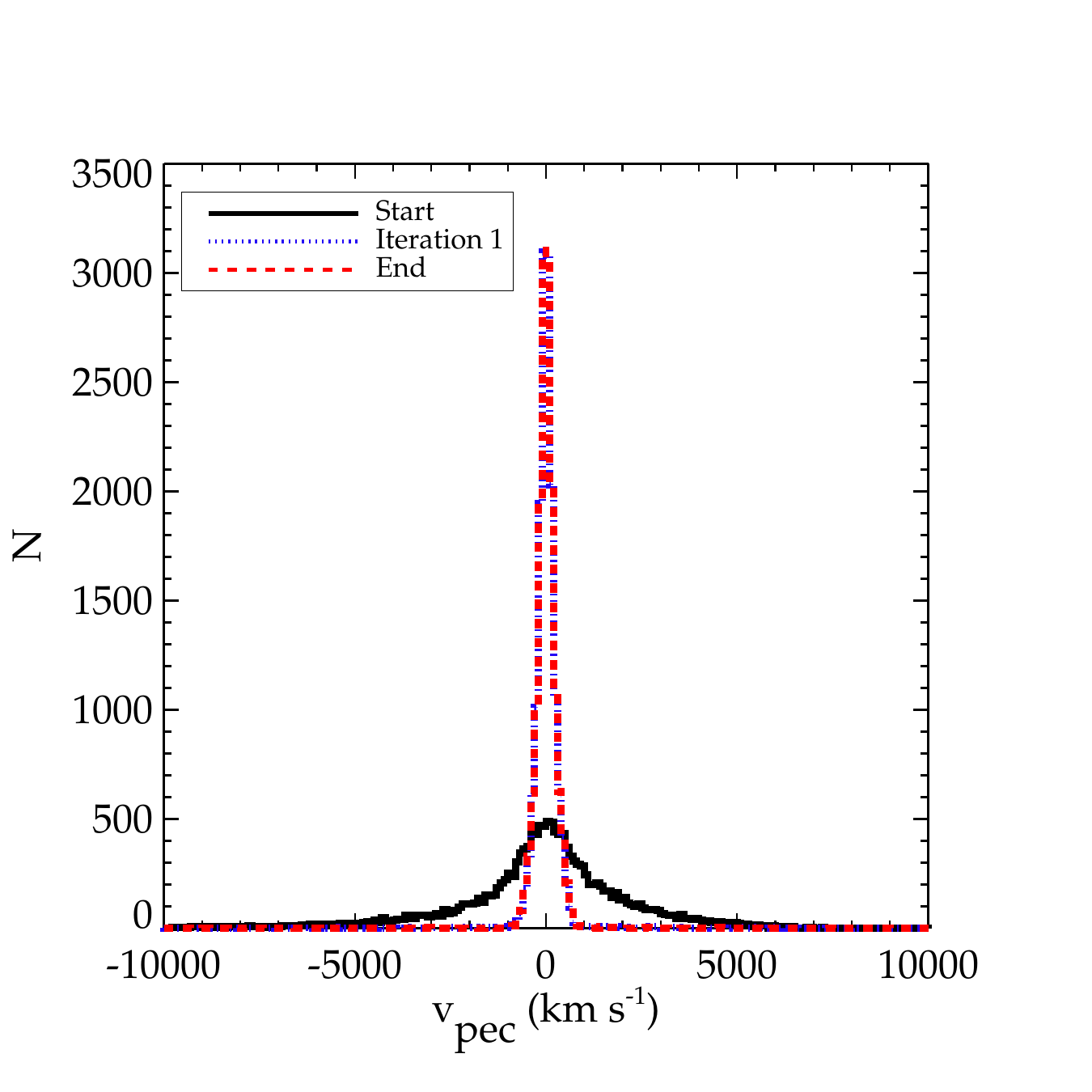}
\hspace{-0.8cm}\includegraphics[width=0.3 \textwidth]{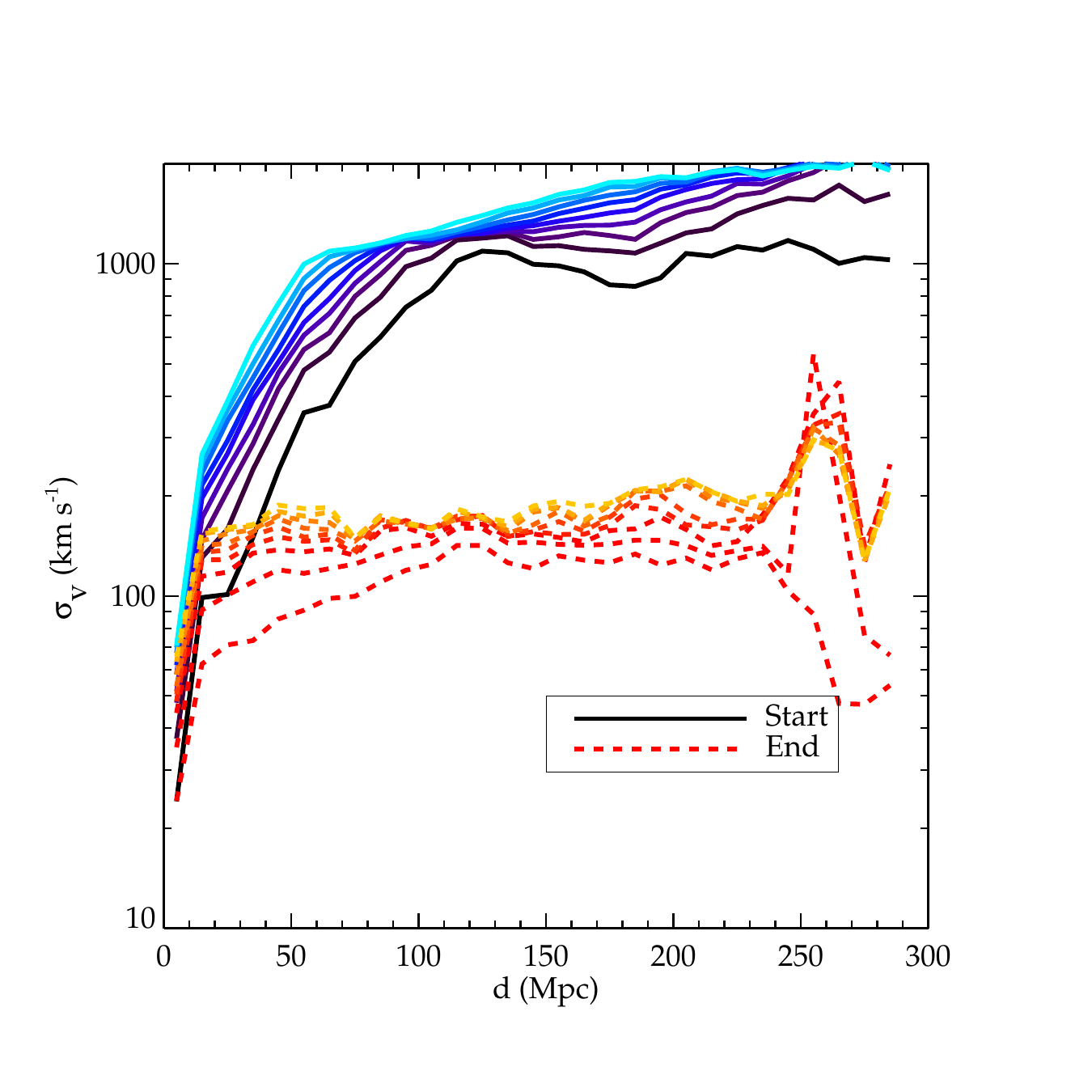}\\

\vspace{-0.8cm}
\hspace{-0.5cm}\includegraphics[width=0.2 \textwidth]{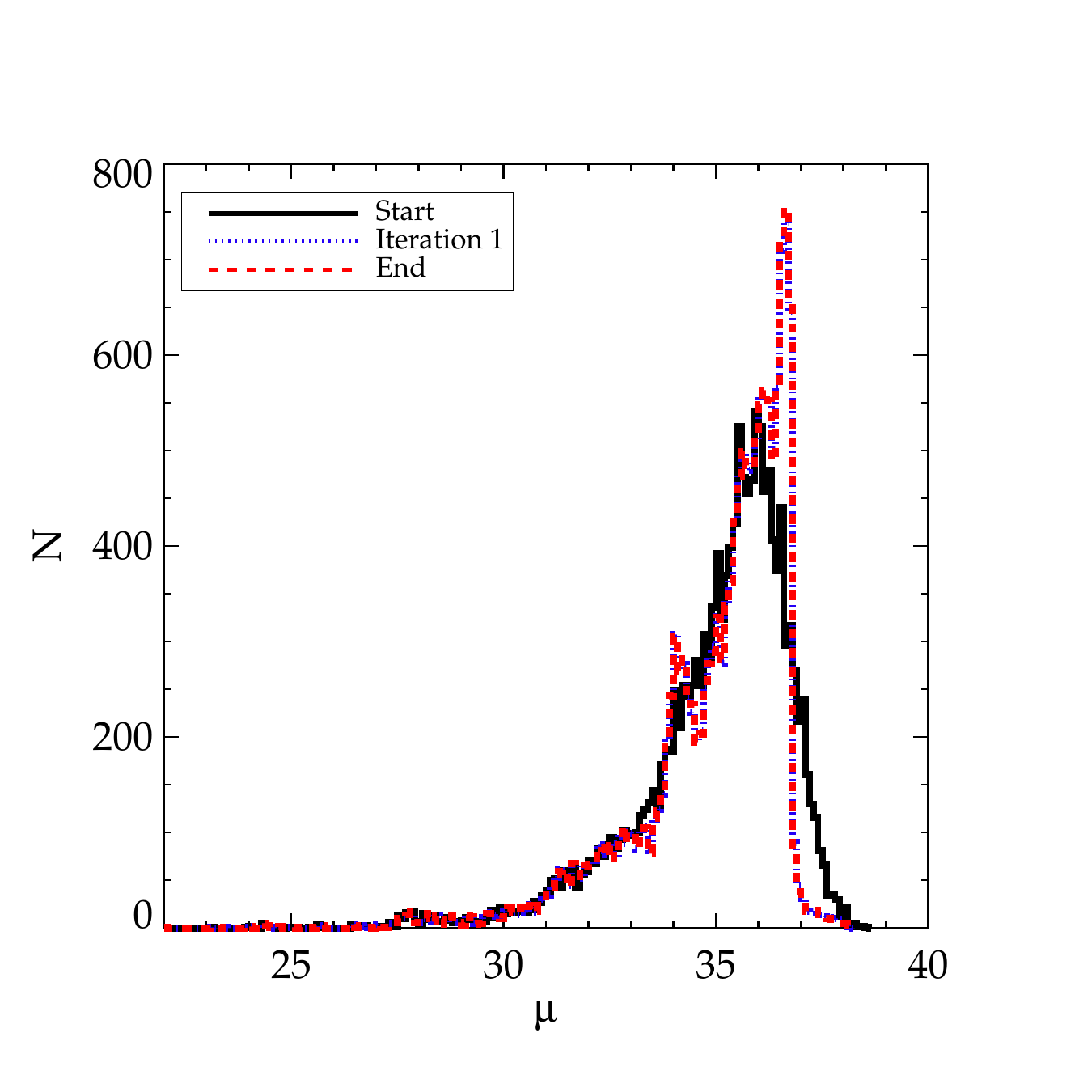}
\hspace{-0.75cm}\includegraphics[width=0.2 \textwidth]{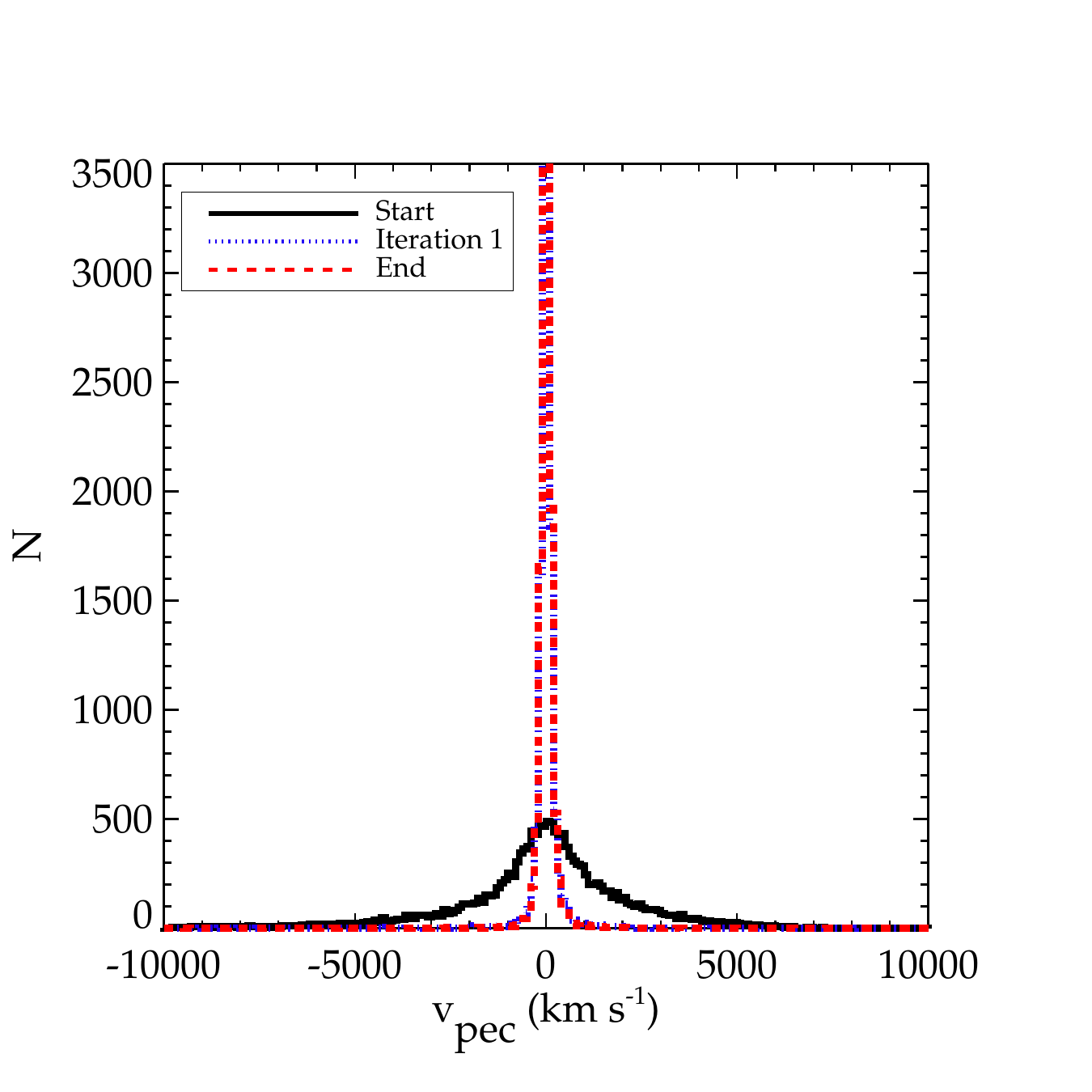}
\hspace{-0.75cm}\includegraphics[width=0.2 \textwidth]{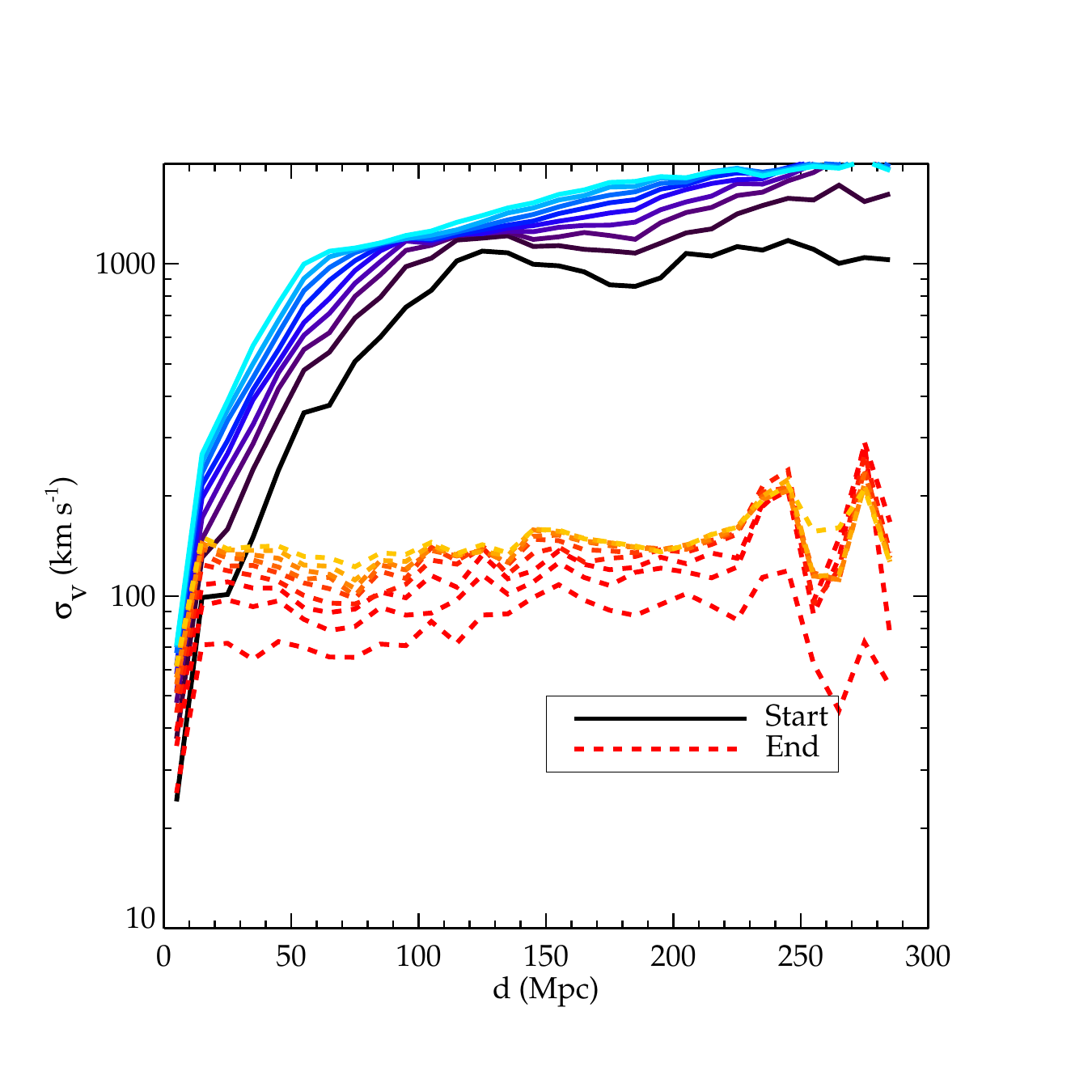}
\hspace{-0.6cm}\includegraphics[width=0.2 \textwidth]{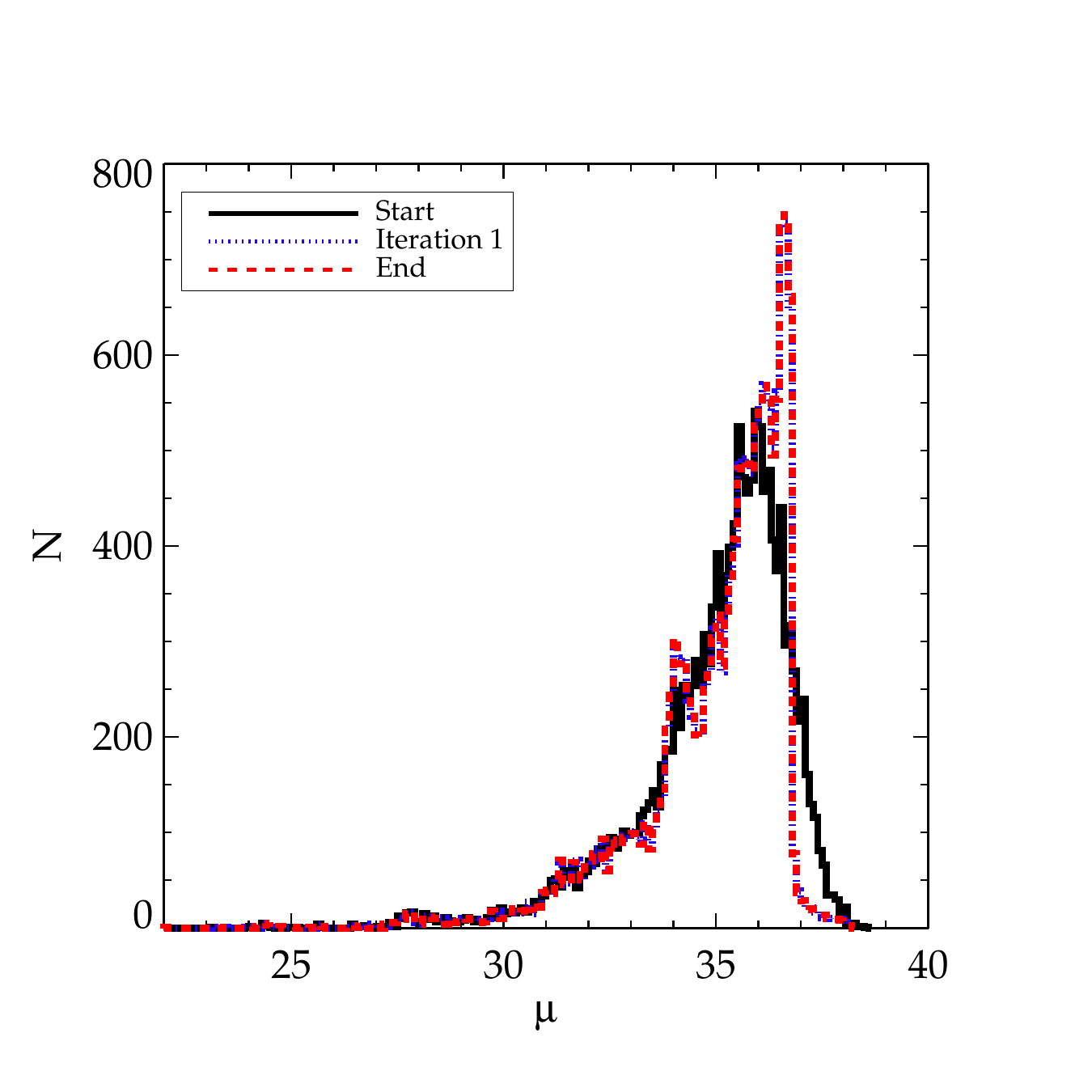}
\hspace{-0.75cm}\includegraphics[width=0.2 \textwidth]{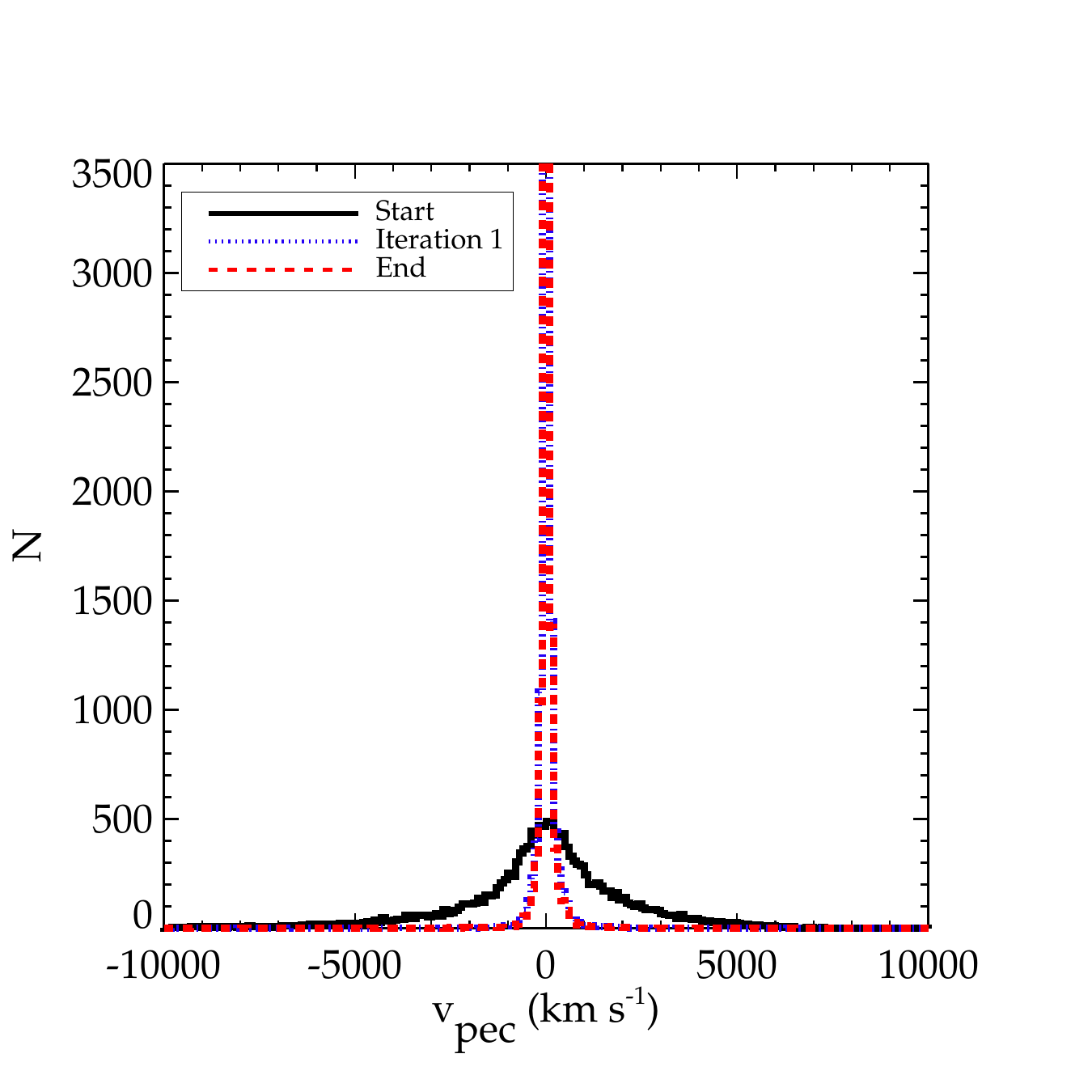}
\hspace{-0.75cm}\includegraphics[width=0.2 \textwidth]{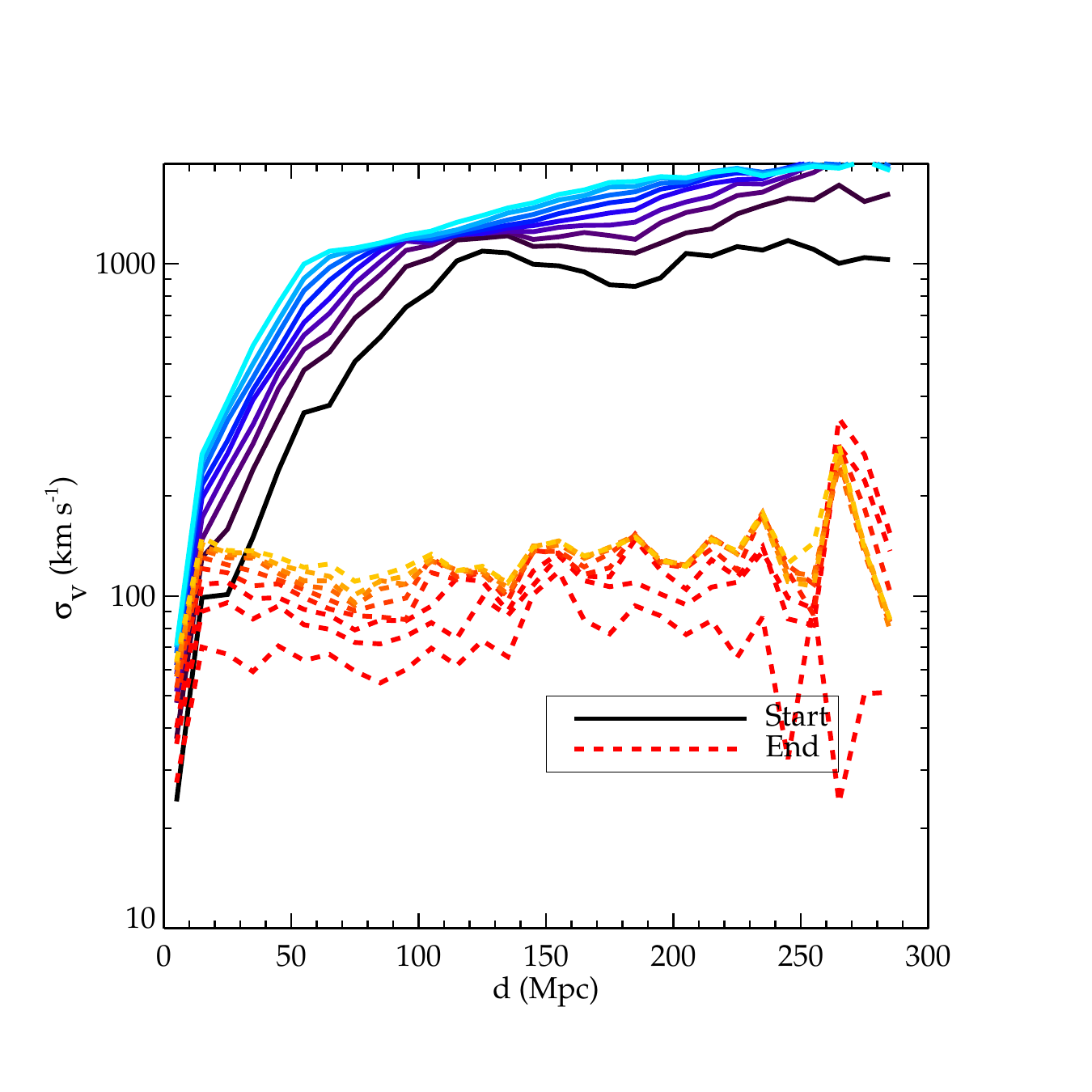}
\caption{{Same as Fig. \ref{fig:cf3corr} but using WMAP7-like cosmological parameter values in the algorithm.}}
\label{fig:cf3corr74}
\end{figure*}

\begin{figure*}
 \vspace{-0cm}\centering
\includegraphics[width=0.9\textwidth]{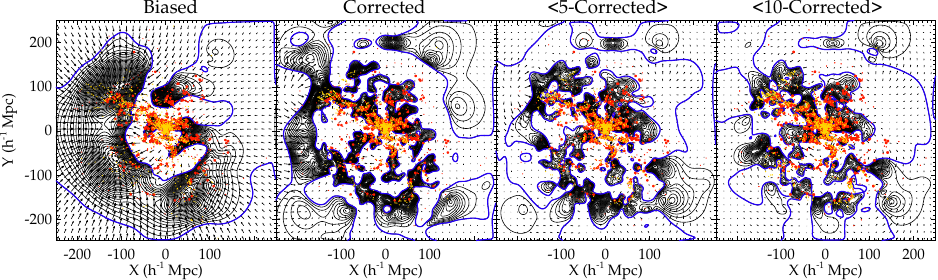}
\includegraphics[width=0.9\textwidth]{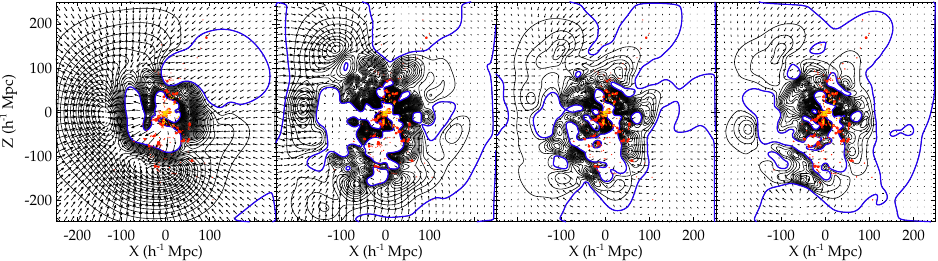}
\includegraphics[width=0.9\textwidth]{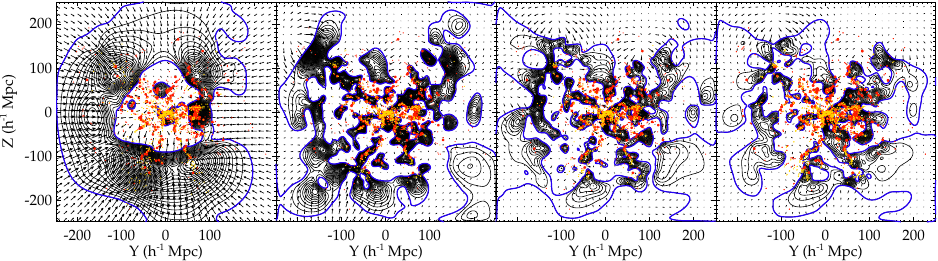}
\vspace{-0.1cm}

\caption{{Same as Fig. \ref{fig:cf3wf} but using WMAP7-like cosmological parameter values in the algorithm.}}
\label{fig:cf3wf74}
\end{figure*}

\begin{figure*}
 \vspace{-0cm}\centering
\includegraphics[width=0.9\textwidth]{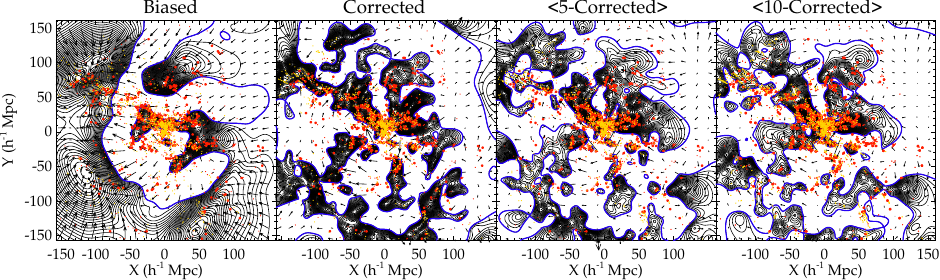}
\includegraphics[width=0.9\textwidth]{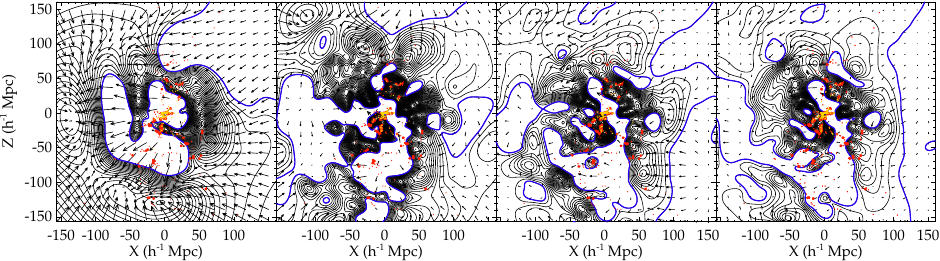}
\includegraphics[width=0.9\textwidth]{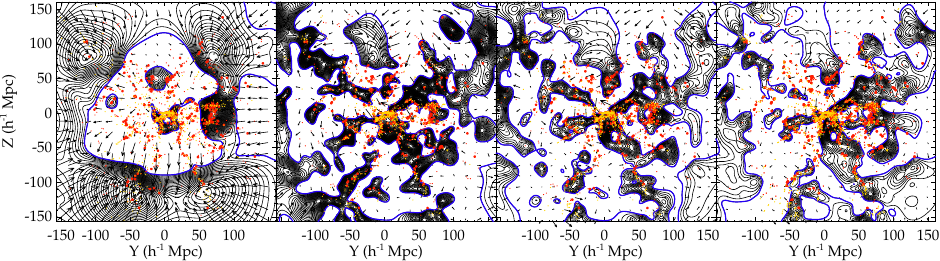}
\vspace{-0.1cm}

\caption{{Same as Fig. \ref{fig:cf3wfzoom} but using WMAP7-like cosmological parameter values in the algorithm.}}
\label{fig:cf3wfzoom74}
\end{figure*}

This appendix reports the results obtained for the third catalog of the Cosmicflows project when using WMAP7-like cosmological parameter values rather than Planck values. Indeed, \citet{2016AJ....152...50T} estimate the third dataset of the Cosmicflows project to be compatible with H$_0$=75$\pm$2~km~s$^{-1}$~Mpc$^{-1}$, this appendix thus uses  WMAP7-like parameters: H$_0$=74~km~s$^{-1}$~Mpc$^{-1}$, $\Omega_m=0.27$ and $\Omega_\Lambda=0.73$. Figures presented in the paper core for the observational catalog are reproduced in this appendix (Fig. \ref{fig:distribvel74} to Fig. \ref{fig:cf3wfzoom74}). No drastic changes appear when using either set of cosmological parameter values. It confirms that this kind of bias minimization techniques seems prone to smooth any effect due to the Hubble constant value choice. Further work will consist in determining whether it could also permit estimating the Hubble constant value that is a best fit to the data.

 \label{lastpage}
\end{document}